\documentclass[12pt]{iopart}
\expandafter\let\csname equation*\endcsname\relax
\expandafter\let\csname endequation*\endcsname\relax

\usepackage{graphicx,enumerate,bm,amsbsy}
\usepackage{amsmath,amssymb,wasysym,graphicx,capt-of,ifthen,calc}
\usepackage{subfigure}
\usepackage{verbatim} 
\usepackage{float}
\usepackage{url}
\usepackage{hyperref}

\newcommand{\gl}{\mathrel{\raise0.6ex\hbox{$>$\kern-.75em\lower1ex\hbox{$<$}}}}

\begin{document}
\title{Zero-Temperature Coarsening in the 2d Potts Model} 
\author{J~Olejarz$^1$, P~L~Krapivsky$^2$, and S~Redner$^1$}
\address{$^1$Center for Polymer Studies and Department of Physics, Boston University, Boston, MA 02215, USA}
\address{$^2$Department of Physics, Boston University, Boston, MA 02215, USA}

\begin{abstract}
  We study the fate of the 2d kinetic $q$-state Potts model after a sudden
  quench to zero temperature.  Both ground states and complicated static
  states are reached with non-zero probabilities. These outcomes resemble
  those found in the quench of the 2d Ising model; however, the variety of
  static states in the $q$-state Potts model (with $q\geq 3$) is much richer
  than in the Ising model, where static states are either ground or stripe
  states.  Another possibility is that the Potts system gets trapped on a set
  of equal-energy blinker states where a subset of spins can flip \emph{ad
    infinitum}; these states are similar to those found in the quench of the
  3d Ising model.  The evolution towards the final energy is also
  unusual---at long times, sudden and massive energy drops may occur that are
  accompanied by macroscopic reordering of the domain structure.  This
  indeterminacy in the zero-temperature quench of the kinetic Potts model is
  at odds with basic predictions from the theory of phase-ordering kinetics.
  We also propose a continuum description of coarsening with more than two
  equivalent ground states.  The resulting time-dependent Ginzburg-Landau
  equations reproduce the complex cluster patterns that arise in the quench
  of the kinetic Potts model.

\end{abstract}
\pacs{64.60.My, 05.40.-a, 05.50.+q, 75.40.Gb}

\section{Introduction}

What happens when a system with degenerate ground states is suddenly quenched
from an initial supercritical temperature, $T_i>T_c$, to a final subcritical
temperature, $0\leq T_f<T_c$?  The answer to this question depends on the
temperatures $T_i$ and $T_f$, the spatial dimension $d$, and the dynamics.
Numerous studies of systems with two degenerate ground states, particularly
the Ising model, revealed that qualitative behaviors are remarkably universal
and depend only on basic features of the dynamics (see,
e.g.,~\cite{gunton_dynamics_1983,bray_review, KRB10} and references therein).
For spin systems, this universality allows us to consider the simplest
microscopic single spin-flip dynamics, leading to what is known as the
kinetic Ising model. A more macroscopic approach is based on the Landau
free-energy functional with a potential that has two degenerate minima.  The
corresponding non-conservative dynamics is the time-dependent Ginzburg-Landau
(TDGL) equation for a scalar order
parameter~\cite{gunton_dynamics_1983,bray_review, KRB10}.

Both for the kinetic Ising model and the TDGL equation, an intricate
coarsening domain mosaic emerges shortly after the quench.  This mosaic has a
characteristic length scale that grows in time as $t^{1/2}$.  Because the
dynamics is universal for all subcritical quenches, we focus on quenches to
zero temperature, $T_f=0$.  In this limit, the early-time evolution is more
rapid and the emergent scaling behavior is more crisp.  Moreover, the kinetic
Ising model considerably simplifies at zero temperature, and for the TDGL
equation the simplification is even more drastic as the governing equation
turns into a {\em deterministic} non-linear partial differential equation.

Recent studies of the fate of the kinetic Ising model that is quenched to
zero temperature gave results that contradict the naive expectation that the
system merely reaches one of the two ground states~\cite{Adam99,
  spirin_fate_2001, OKR2011,ONSS06, BKR09, OKR12, Leticia07,Picco13}.  For $d\geq 3$,
the kinetic Ising model gets trapped on a set of topologically-complex and
temporally fluctuating spongy states, so that the ground state is never
reached in the thermodynamic limit~\cite{spirin_fate_2001,OKR2011}.  The
topological structure of the accessible state space at long times has been
found to be replete with many saddle points~\cite{KL1996}, which may explain
some of the strange features of the 3d system.  In $d=2$, the kinetic Ising
model reaches either ground states or states with straight stripes (all
vertical or horizontal)~\cite{spirin_fate_2001, ONSS06}. There is also a
tantalizing connection between the fate of the 2d kinetic Ising model and the
equilibrium critical behavior in the seemingly unrelated system of critical
continuum 2d percolation~\cite{BKR09,OKR12}.  This percolation mapping allows
one to extract apparently exact predictions for the probabilities of the 2d
kinetic Ising model to freeze into the ground states or stripe
states~\cite{BKR09}, and even the probabilities to reach states with inclined
straight stripes (with arbitrary winding numbers in horizontal and vertical
directions) that occur for the 2d TDGL equation.  There are additional
intriguing connections to 2d critical percolation that give exact predictions
for geometric properties of the kinetic Ising model in the coarsening
regime~\cite{Leticia07}.

Our main goal here is to investigate the long-time phase-ordering kinetics of
systems with more than two degenerate ground states after a quench to zero
temperature.  There are two natural examples of such systems.  The first is
the $q$-state kinetic Potts model (with $q\geq 3$) on the square lattice, in
which evolution occurs by single spin-flip dynamics.  We also study a
complementary continuum description that is based on the TDGL equation.  The
latter provides a more general description of phase-ordering kinetics because
it is heavily based on symmetry considerations rather than fidelity to a
particular microscopic system.

The two-dimensional $q$-state kinetic Potts model that is quenched to zero
temperature was investigated in the 80s, which predates many studies of the
corresponding kinetic Ising model.  One of reasons for this earlier focus on
the Potts model is that kinetic trapping phenomena are much more
pronounced~\cite{SSG82,SSG83}, and unusual relaxation
properties~\cite{SSGAS83} become evident already for small systems.
Numerical work, especially simulations of quenches to positive subcritical
temperatures, was additionally driven by applications, e.g., to coarsening in
soap froths~\cite{soap} and magnetic grains~\cite{grain}, and to the
description of emergent structures of cellular tissues~\cite{cells}.
Simulations also indicated that general sub-critical quenches exhibit
standard coarsening for early times~\cite{GAS88,FC07}.  This continuous
evolution stops because the system gets trapped in a disordered state and
escape from this state is slow because the process is thermally
activated~\cite{deOliveira}.  When the final temperature is zero, it was
noticed previously that the ground state may not be reached~\cite{OP02,P03}.
It was further suggested that for $q\geq d+1$, glassy evolution can
occur~\cite{FC07,deOliveira,Lifshitz,Safran} because the system gets pinned
in a disordered state and the ultimate relaxation to equilibrium proceeds via
thermally-activated processes~\cite{FC07,deOliveira,VG1986,VG87,BFCLP07}.
From a more geometrical perspective, recent work~\cite{Leticia12} determined
the distribution of hull enclosed areas of the 2d kinetic Potts model during
the coarsening regime.

\begin{figure}[ht]
\begin{center}
\includegraphics*[width=0.3\textwidth]{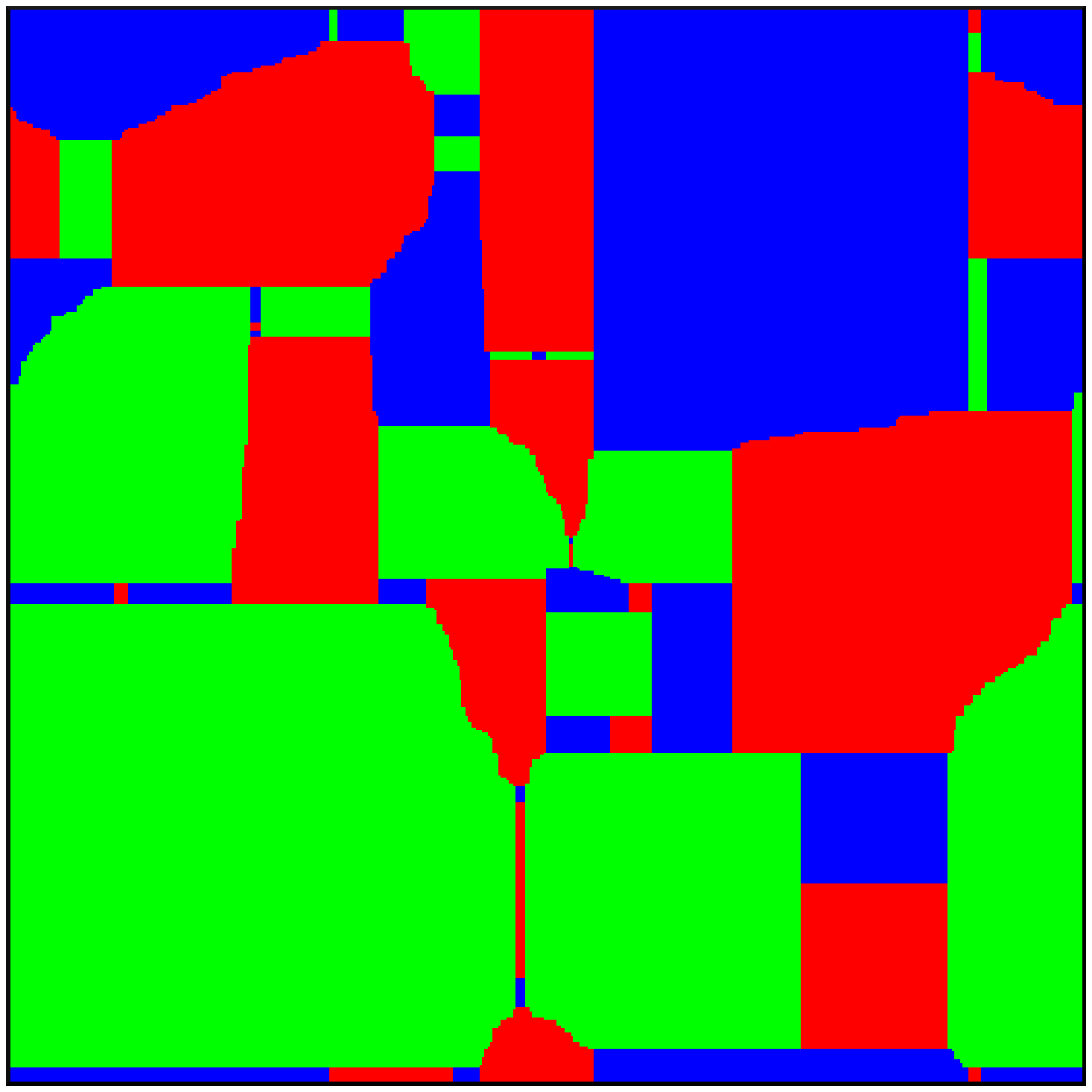}\quad
\includegraphics*[width=0.3\textwidth]{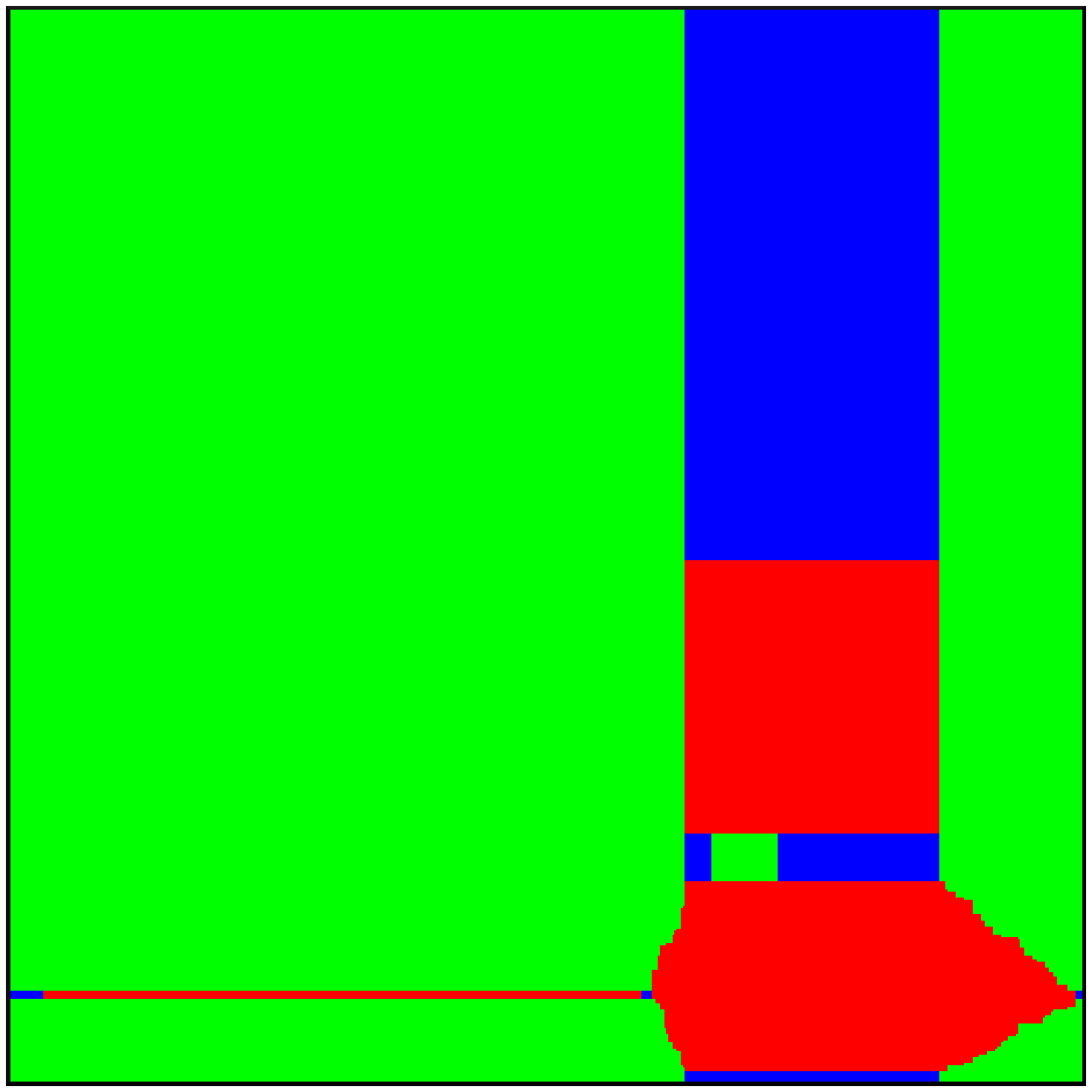}\quad
\includegraphics*[width=0.3\textwidth]{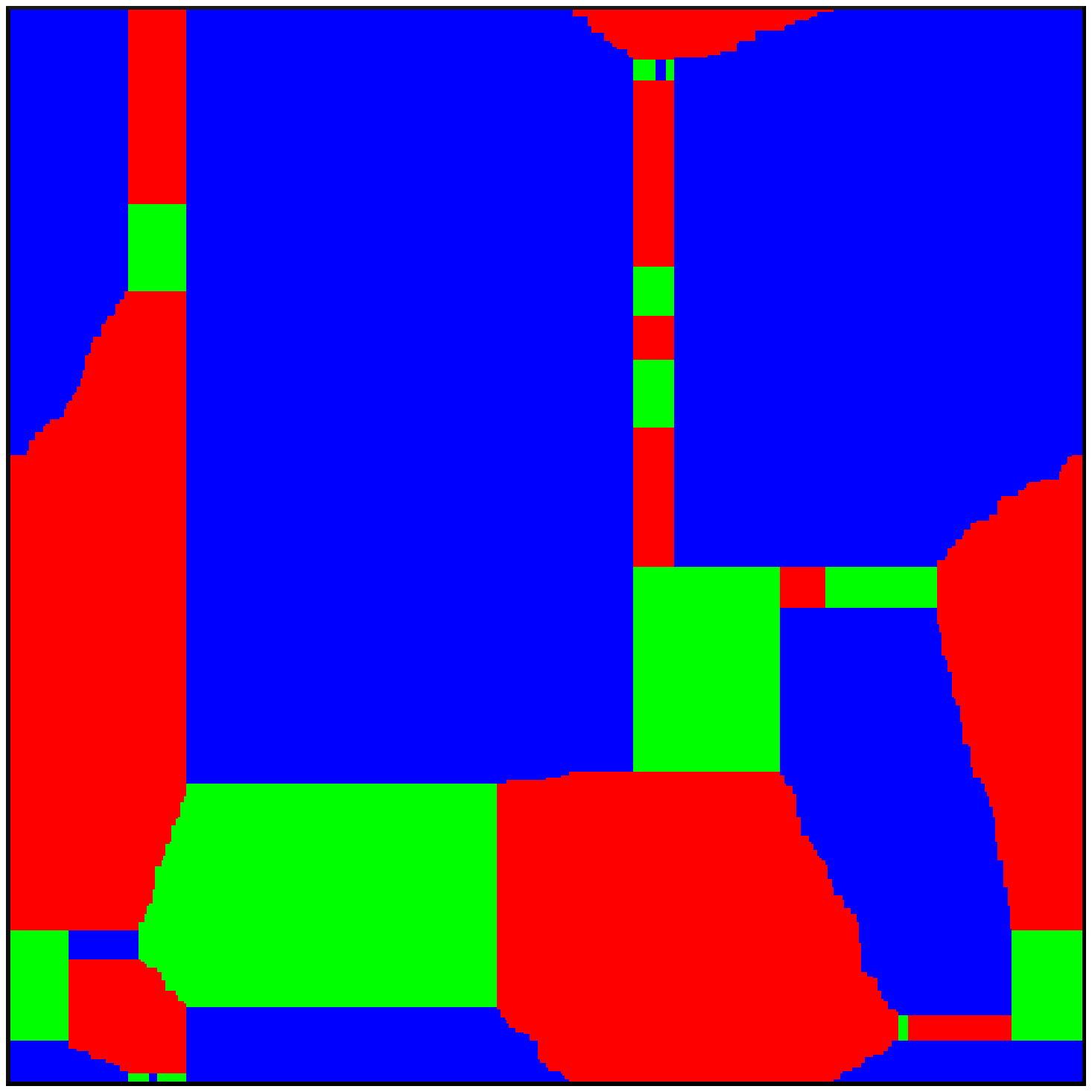}
\vskip 0.1in
\includegraphics*[width=0.3\textwidth]{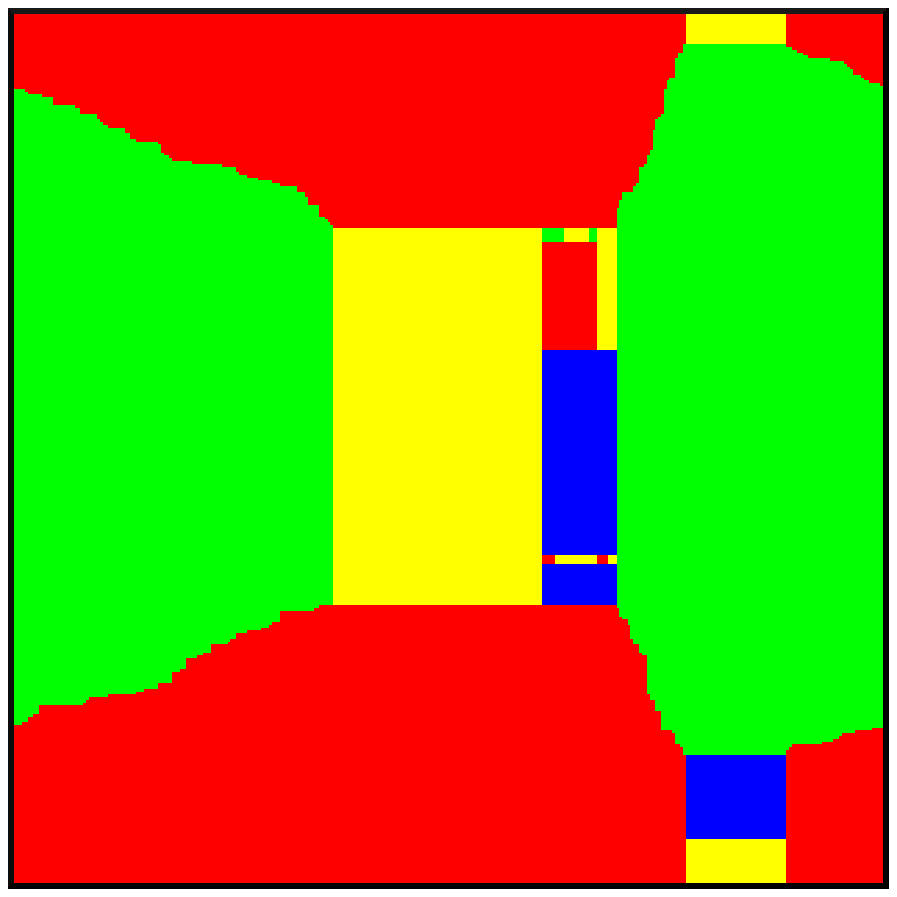}\quad
\includegraphics*[width=0.3\textwidth]{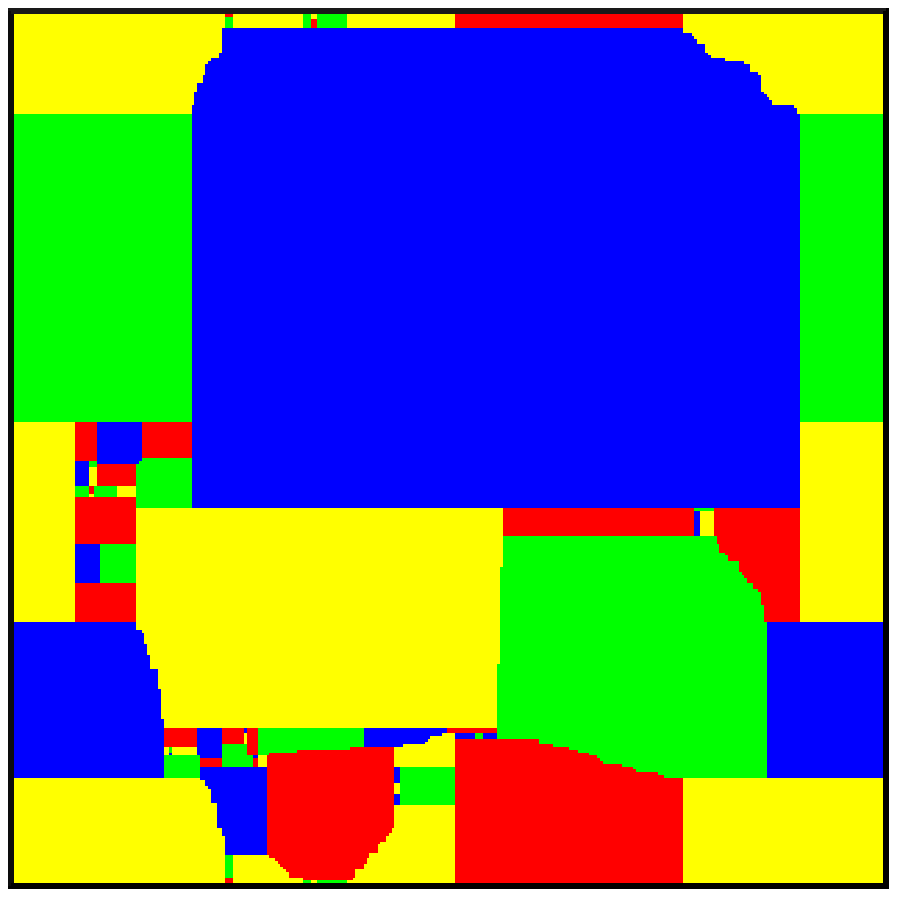}\quad
\includegraphics*[width=0.3\textwidth]{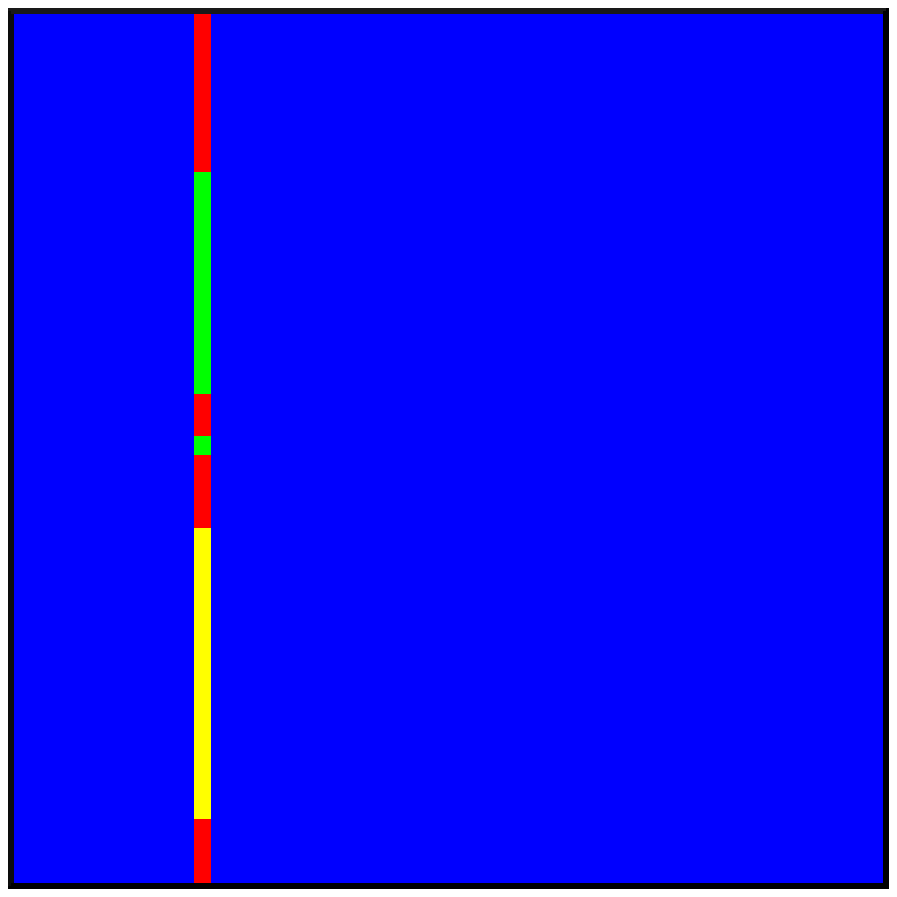}
\vskip 0.1in
\includegraphics*[width=0.3\textwidth]{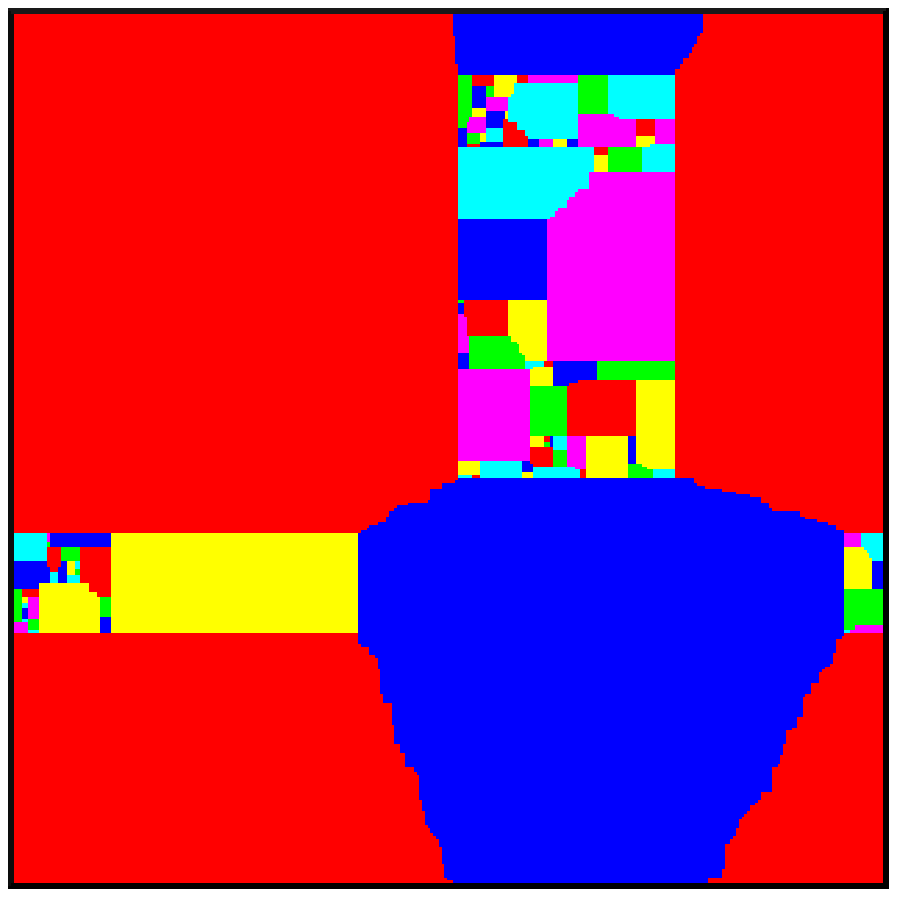}\quad
\includegraphics*[width=0.3\textwidth]{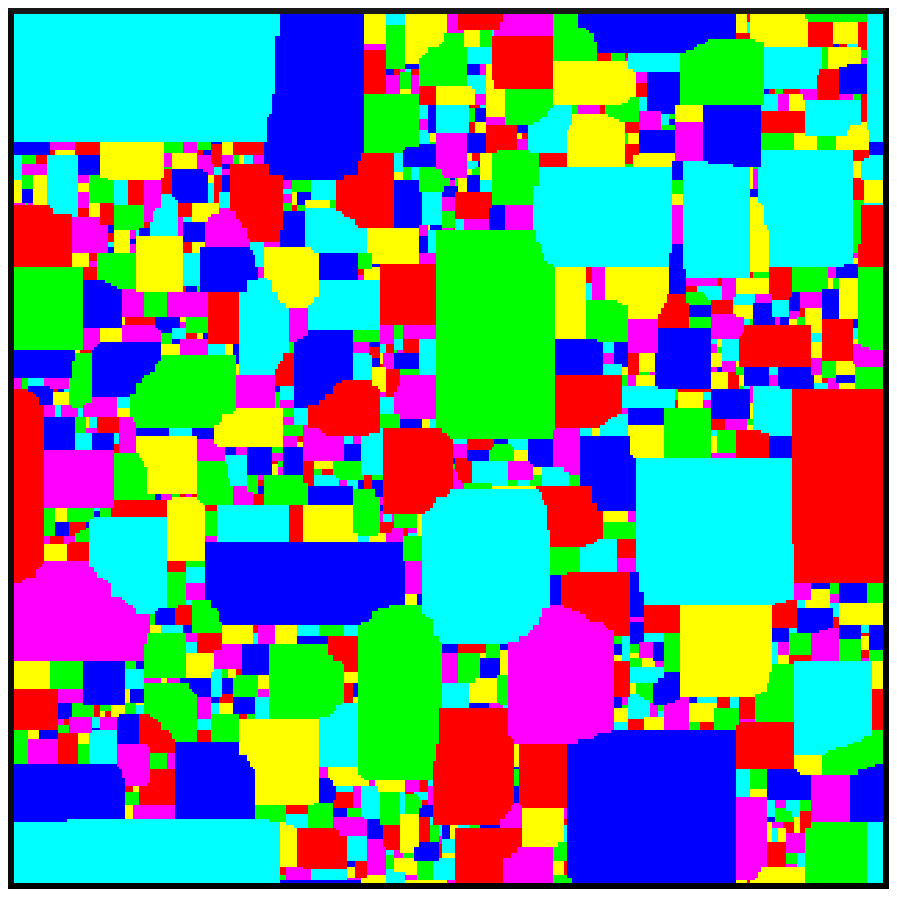}\quad
\includegraphics*[width=0.3\textwidth]{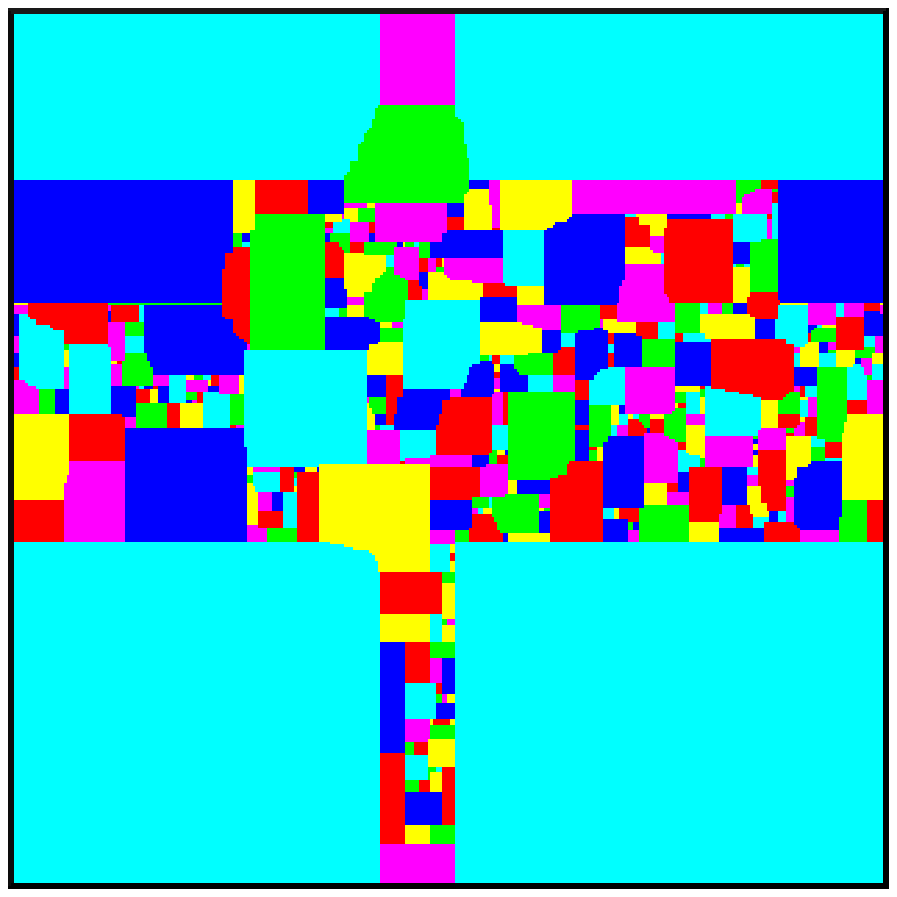}
\caption{Representative long-time states of the 3-state (top) 4-state
  (middle) and 6-state (bottom row) Potts model on a square lattice of linear
  dimension $L=384$ after quench from $T=\infty$ to $T=0$.  Notice that only
  the configuration in the middle right, with only vertical or horizontal
  interfaces, is static.  All others, with non-straight interfaces, are still
  evolving, and some may evolve ad infinitum.  }
\label{fig:clusters}
  \end{center}
\end{figure}

With the huge increase of computing power since the 80s, one can try to
settle the fate of the 2d kinetic Potts model that is quenched to zero
temperature.  When the system is sufficiently isotropic, e.g., when the
underlying lattice is triangular\footnote{Similar behaviors have been
  observed \cite{SSGAS83} on the square lattice with additional
  next-nearest-neighbor ferromagnetic interactions.} the relaxation appears
to be rapid~\cite{SSGAS83}, while on the square lattice the system can get
pinned in a disordered state.  We therefore limit ourselves to the most
anisotropic (even-coordinated) 2d lattice, the square lattice, and we ask
about the possible outcomes in the thermodynamic limit (representative
long-time states are shown in Fig.~\ref{fig:clusters}).  Our simulations lead to
the following conclusions:
\begin{itemize}
\itemsep -0.1ex
\item Ground states are reached with non-zero probability.  
\item More complicated static states are also reached with non-zero probability.  
\item The system may wander \emph{ad infinitum} on an iso-energy subspace of
blinker states.
\end{itemize}

The latter possibility has not been seen in earlier work, and it has come as
a surprise to us.  Indeed, the common lore in phase-ordering kinetics is that
in following an energy-decreasing path, the system can fall into a metastable
state, not into a collection of such states.  We are aware of one
counterexample to the above rule, namely, the fate of the Ising model that is
quenched to zero temperature, but only in greater than two
dimensions\footnote{More precisely, the probability to reach a static state
  is finite for the 3d kinetic Ising model on a torus, but this probability
  very quickly decays as the system size is
  increased~\cite{spirin_fate_2001,OKR2011}.}; in $d=2$, static states are
always reached.

The 2d kinetic Potts model undergoes an extremely slow relaxation that, in a
finite fraction of realizations, is reminiscent to the 3d Ising
model~\cite{OKR2011}.  In the Ising case, slow relaxation is driven by rare
events in which diffusively fluctuating interfaces with an additional
entropic repulsive bias ultimately merge.  The rarity of such coalescence
events causes a logarithmically-slow decay in the energy of the system.  The
ultra-slow evolution in the Potts model has the same origin.  We shall see,
however, that interface merging can sometimes lead to a rapid and drastic
reordering of the global domain structure in the 2d Potts model.  This
avalanching is unpredictable---small stochastic fluctuations in the domain
coarsening at long times can generate substantially different long-time
cluster geometries.

We also emphasize that the most `obvious' outcome, namely, that the 2d kinetic
Potts model reaches the ground state with non-zero probability, is by no
means obvious from simulations.  For the 3-state Potts model, the probability
to reach the ground state initially decreases as the linear dimension $L$ of
system grows and a naive extrapolation of the small-$L$ data to $L\to\infty$
suggests that this probability vanishes in the thermodynamic limit.  For
larger system sizes, however, the probability starts to grow.  Such a
non-monotonic behavior is more clean for the kinetic Potts model with $q>3$,
leading to the conclusion that the ground state is reached with positive
probability.

\begin{figure}[ht]
\begin{center}
\subfigure[]{\includegraphics[width=0.25\textwidth]{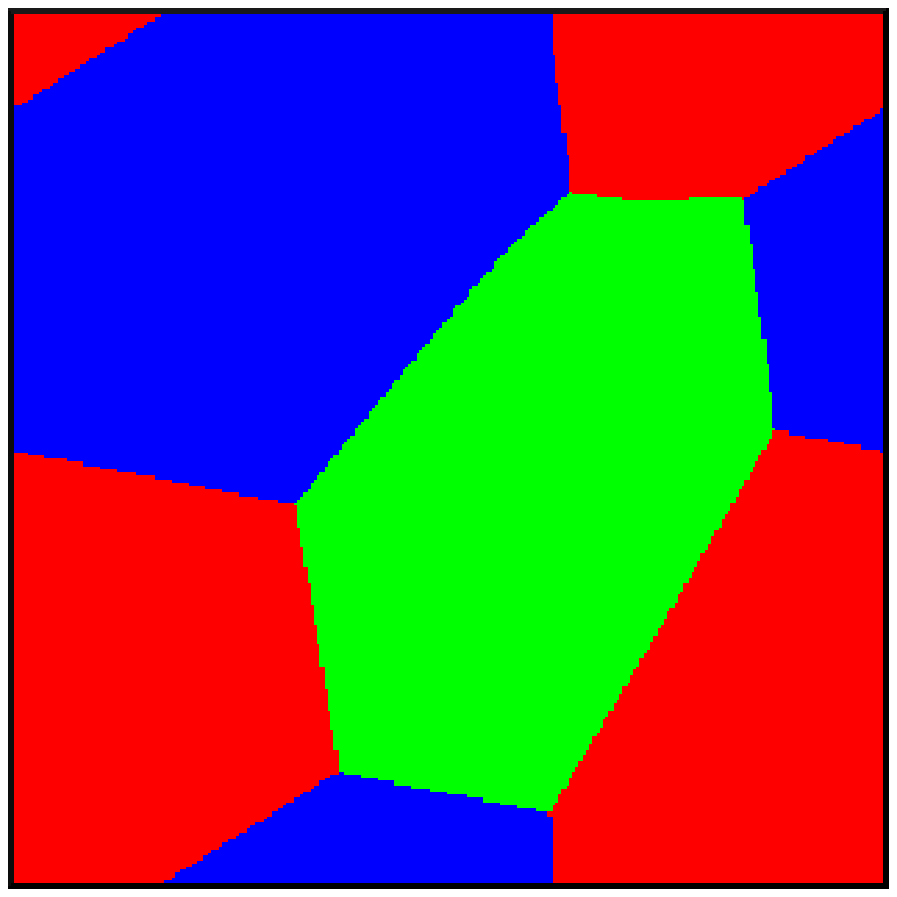}}\qquad
\subfigure[]{\includegraphics[width=0.25\textwidth]{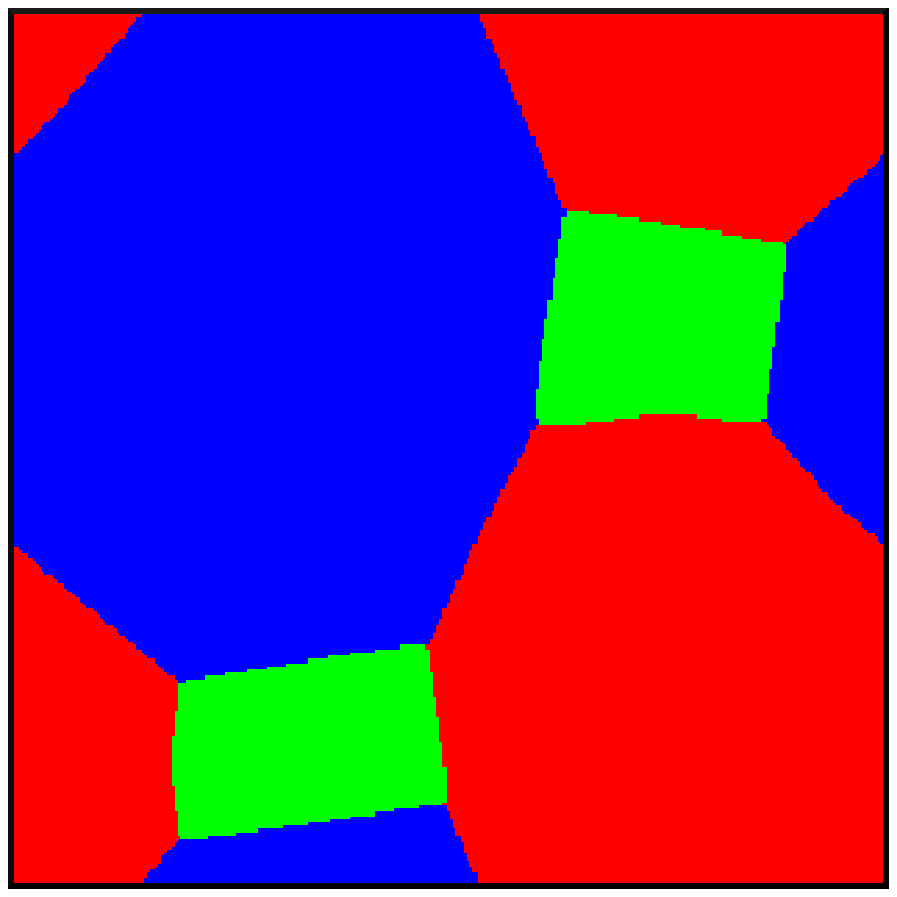}}\qquad
\subfigure[]{\includegraphics[width=0.25\textwidth]{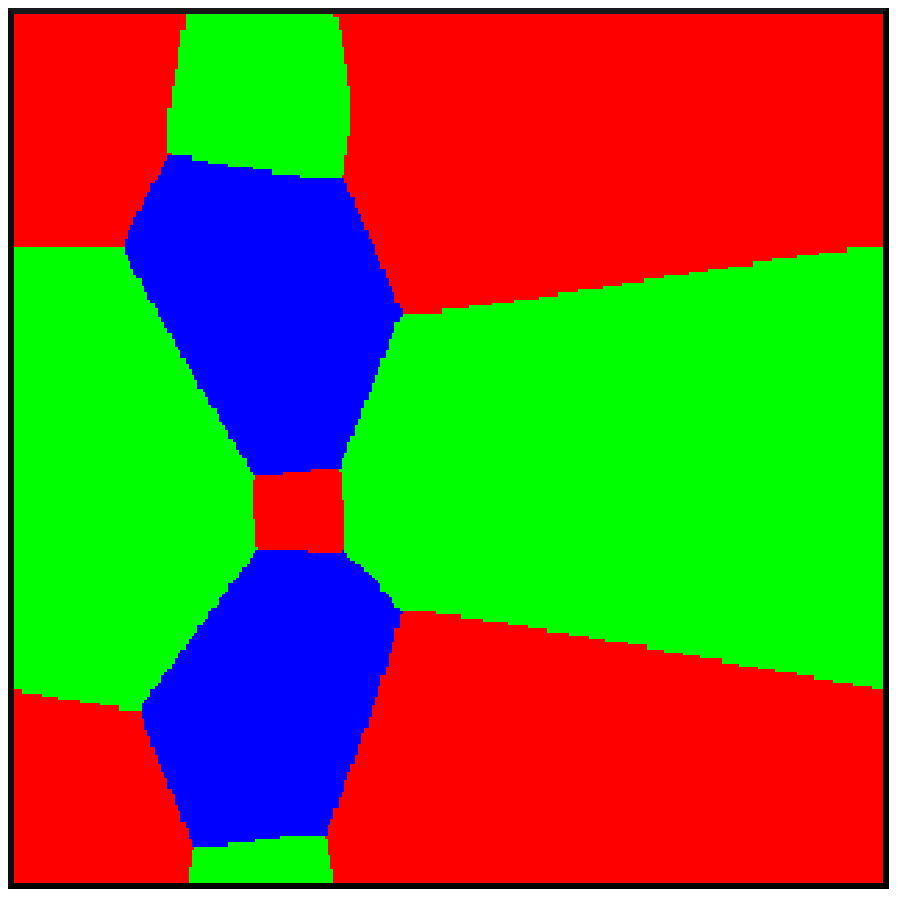}}
\vskip 0.05in
\subfigure[]{\includegraphics[width=0.25\textwidth]{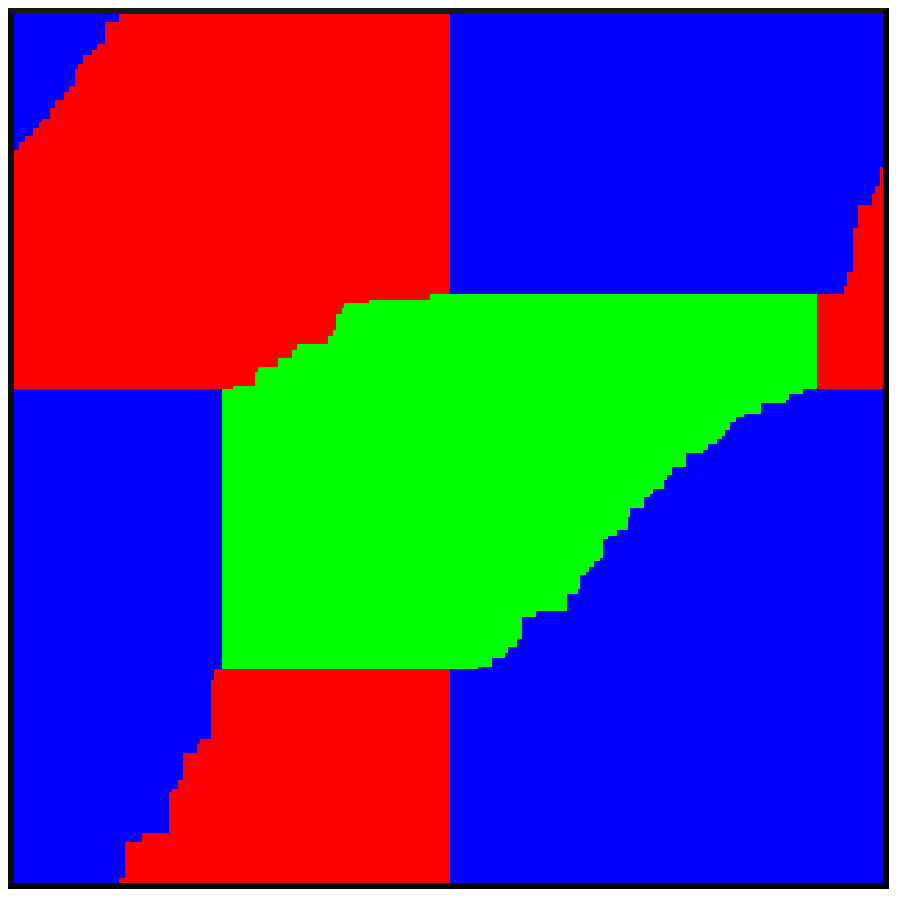}}\qquad
\subfigure[]{\includegraphics[width=0.25\textwidth]{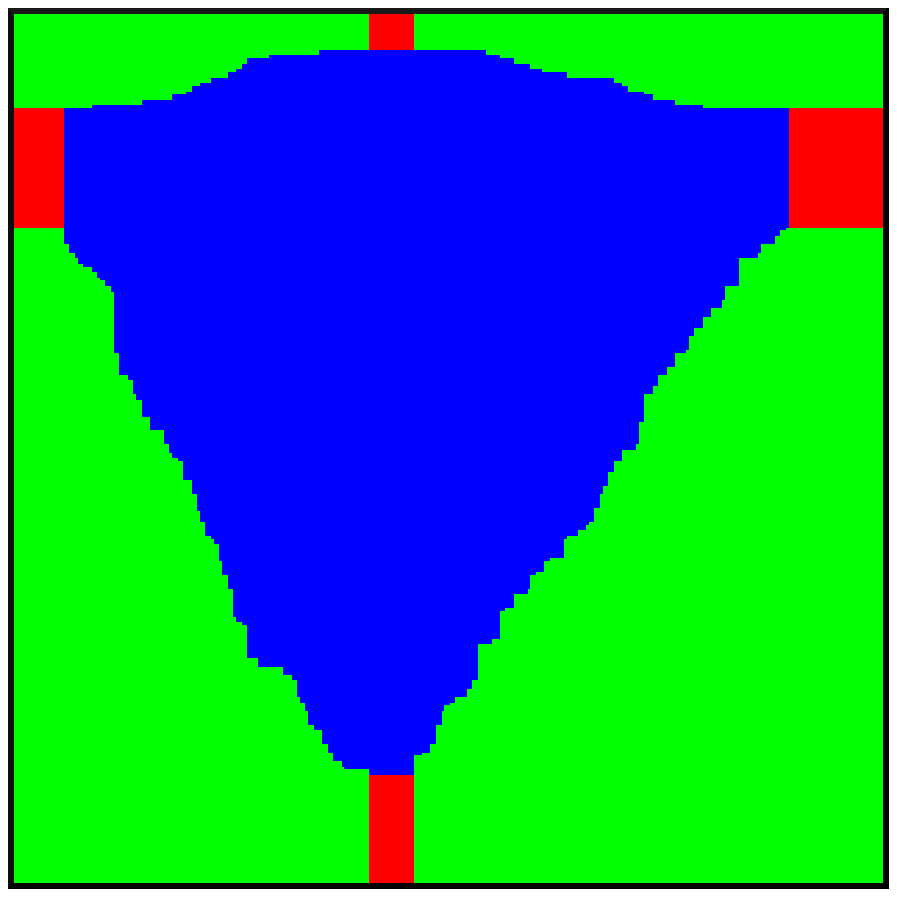}}\qquad
\subfigure[]{\includegraphics[width=0.25\textwidth]{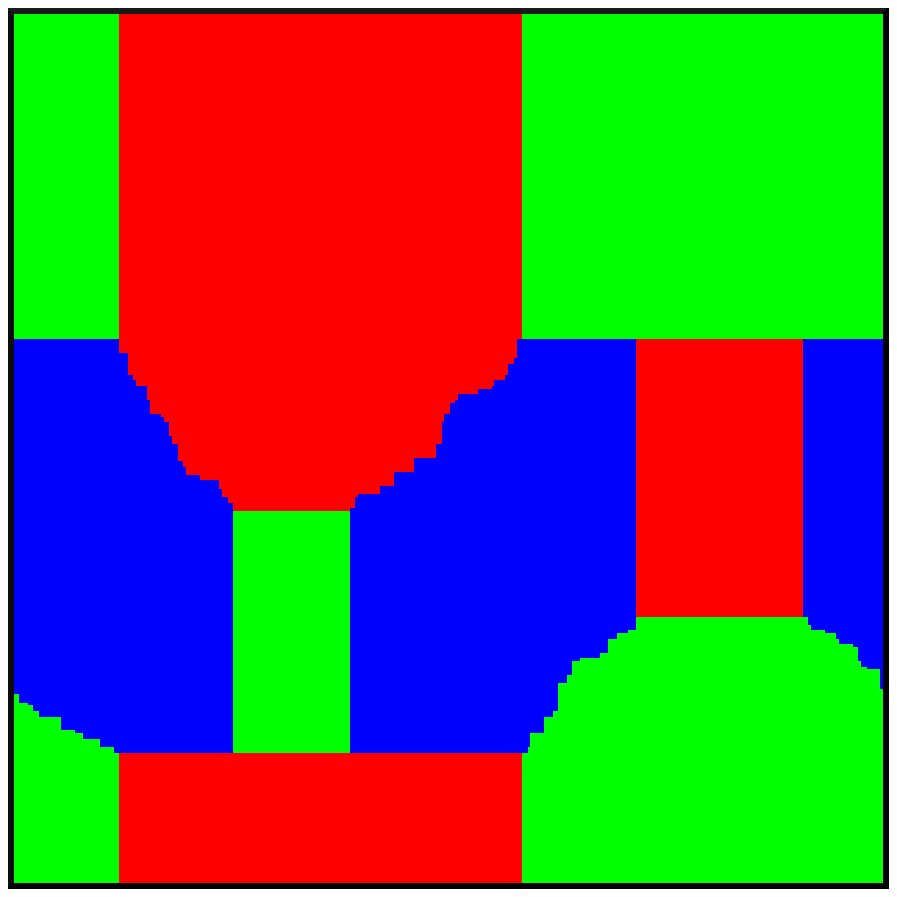}}
\caption{Long-time states from integration of the three-state TDGL equation
  with $L=256$ (a-c) and simulations of the 3-state Potts model with $L=192$
  (d-f).}
\label{fig:patterns}
  \end{center}
\end{figure}

The second major part of our work is the study the evolution of the 2d Potts
system that is based on the TDGL equation. This is a manifestly isotropic
framework and the results are conceptually easier to interpret because the
dynamics are not accompanied by the anomalous discreteness effects that
characterize the behavior of the kinetic Potts model.  Thus in the TDGL
equation with three degenerate ground states, the final states are always
static.  However, these final states are much more rich than those that arise
in the TDGL equation with two degenerate ground
states~(Fig.~\ref{fig:patterns}).  For example, we observed states with three
hexagons (in roughly $11\%$ of realizations for $L=128$), states with four
clusters---two squares and two octagons---(roughly $8\%$ of realizations),
and six-cluster states (roughly $0.2\%$ of all realizations).  In spite of
the diversity of geometrically complex states, the probability of reaching
the ground state has the approximate value 0.7 for the largest systems that
we simulated, and appears to remain positive in the thermodynamic limit.

Although we extensively simulate the kinetic Potts model with $q=3,4,5,6$, it
is still difficult to extract reliable estimates for basic observables, such
as the probability to reach a ground state, the probability to end in blinker
states, etc.  The reasons for this difficulty are twofold. First, the
relaxation is extremely slow and many realizations are still evolving, with
their ultimate fate still undecided, at times that are orders of magnitude
longer than the coarsening time, which scales as $L^2$. (The situation is
more clear for $q=3$ where one can often resolve that a system is in a
genuine blinker state.)~ If we stop simulations at the coarsening time, the
time to simulate a system of linear dimension $L$ scales as $L^4$.  A second
difficulty is that for the system sizes that we are able to simulate,
substantial finite-size effects arise that obscure the thermodynamic
behavior.  Collection of all of our data took roughly 10 days of computing
time on a 32-CPU cluster, with each CPU running at 2 GHz.

Our simulations for the 2d kinetic Potts model
with $q=3,4,5,6$ lead to the conclusion that the probabilities $\Pi_q$ to
reach the ground state remain positive in the thermodynamic limit:
\begin{equation}
\label{GS_yes}
\Pi_q>0\qquad\text{for all}\quad q\geq 2.
\end{equation}
Even for the 2d kinetic Ising model this statement is conjectural, although
in this case the connection with 2d critical continuum percolation leads to
the precise conjecture~\cite{BKR09} 
\begin{equation}
\label{GS_Ising}
\Pi_2 = \frac{1}{2}+ \frac{\sqrt{3}}{2\pi}\ln\left(\frac{27}{16}\right)=0.64424...,  
\end{equation}
for free boundary conditions and $\Pi_2 = 0.661169...$ for periodic boundary
conditions; these are well-supported heuristically and
numerically~\cite{BKR09,OKR12}.

\section{Models}

\subsection{Kinetic Potts Model}
\label{discrete-potts}

The Hamiltonian of the Potts model is
\begin{equation}
\mathcal{H}=-\sum_{\langle ij\rangle}\delta(\sigma_i,\sigma_j)
\end{equation}
where $\delta(\sigma_i,\sigma_j)$ is the Kronecker delta function, each spin
$\sigma_i$ takes one of the $q$ integer values $1,2,\ldots,q$, and the sum
$\langle ij\rangle$ is over all nearest-neighbor pairs.  We are interested in
the long-time state of the system following a quench from infinite
temperature to zero temperature.  Thus we initialize the lattice with equal
fractions of each of the $q$ spin states; the assumption that the initial
temperature is infinite implies that the initial spin values in different
sites are uncorrelated.

We employ zero-temperature Glauber dynamics \cite{Glauber} in which each spin
flip event that would decrease the energy occurs with rate $1$, while each
flip event that would conserve the energy occurs with rate $1/2$.  At long
times, the number of flippable spins (those whose flips would conserve or
decrease the energy) is small, since all evolution occurs on interfaces of
large domains.  For efficiency, we therefore track all flippable spins.  For
each such spin $\sigma_i$ the corresponding flip rate $R_i$ is determined
from all of its possible transitions.  Each update event then consists of
selecting a flippable spin at random, flipping it, and incrementing the time
by $1/(\Sigma_i R_i)$~\cite{BKL1975}.  After each such spin-flip event, the
list of flippable spins is also updated.

Our simulations are performed on square $L\times L$ lattices with periodic
boundary conditions, and our goal is to extract the results in the
$L\to\infty$ thermodynamic limit.  One could consider different boundary
conditions, e.g., free boundary conditions, and different geometries, e.g.,
$L\times rL$ rectangles and investigate the $L\to\infty$ limit, with the
aspect ratio $r$ held constant. The influences of these modifications have
been investigated in the Ising case~\cite{BKR09,OKR12}, and these extensions
have provided a crucial support in favor of the connection to 2d critical
continuum percolation.  The boundary conditions and the aspect ratio
similarly affect quantitative results in the Potts case, but for simplicity
we limit ourselves to periodic boundary conditions and to square systems.

We tacitly assume (when not stated otherwise) that the densities for all $q$
spin states are equal.  In simulations of finite systems we always start with
equal number of spins of each species; thus for $q=3$, we choose $L$ to be
divisible by 3, so that there initially are $\frac{1}{3} L^2$ spins of each
type.  This choice of equal initial densities for the $q$ states is
physically natural because it corresponds to an initial supercritical
temperature, $T_i>T_c$.  It is possible, however, to organize a disparity in
the densities in each state; in the Ising case, for example, an applied
external magnetic field can be turned off when the quench occurs.  Our
results for the general case when the initial densities in each state are
unequal are briefly discussed in Sec.~\ref{non-equal}.

\subsection{Time-Dependent Ginzburg-Landau (TDGL) Equation}
\label{TDGL}

Because discrete models of kinetic spin systems are too complex to solve
analytically, especially in more than one dimension, much effort has focused
on phenomenological continuum models~\cite{KRB10}.  The latter are often
better suited for analytical and numerical work; additionally, continuum
models can better mimic the dynamics of real systems.  For the Ising model
($q=2$), the simplest continuum description is based on a scalar
magnetization field $\phi({\bf x},t)$ and the free-energy functional
\begin{equation}
F_2[\phi]=\int \left[(\nabla\phi)^2 + (\phi-1)^2(\phi+1)^2 \right] d{\bf x}
\end{equation}
that evolves by gradient descent 
\begin{equation}
  \frac{\partial\phi}{\partial t}=-\frac{\delta F_2}{\delta\phi} = 2\nabla^2\phi + 4\phi(1-\phi^2)\,.
\end{equation}
The double-well potential $V(\phi)= (\phi-1)^2(\phi+1)^2$ has two degenerate
minima that correspond to the ground states.  The field $\phi$ flows to one
of the two potential minima, and the Laplacian term captures the effects of
the surface tension between neighboring clusters.  While there is no direct
connection to the Ising model, this TDGL equation description accurately
describes coarsening in the Ising model and related
systems~\cite{BKR09,OKR12,ABCS07}.
  
\begin{figure}[ht]
\begin{center}
\includegraphics*[width=0.4\textwidth]{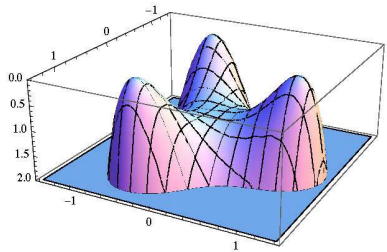}
\caption{The negative of the potential,
  $-(\bm{\phi}-\bm{A})^2(\bm{\phi}-\bm{B})^2(\bm{\phi}-\bm{C})^2$, that
  corresponds to the TDGL-equation description of the 3-state Potts model. }
\label{fig:potts-V}
  \end{center}
\end{figure}

We now formulate an analogous TDGL-equation description of non-conserved
coarsening for systems with more than two degenerate ground states.  For
three degenerate states, we postulate the free-energy functional:
\begin{equation}
  F_3[\bm{\phi}]=\int \left[(\nabla\phi_1)^2 +  (\nabla\phi_2)^2 
    + (\bm{\phi}-\bm{A})^2(\bm{\phi}-\bm{B})^2(\bm{\phi}-\bm{C})^2 \right] d{\bf x}\,,
\end{equation}
with a two-component order parameter $\bm{\phi}({\bf x}, t)=\{\phi_1({\bf x},
t), \phi_2({\bf x}, t)\}$, and a potential that has three
symmetrically-located minima in the plane (Fig.~\ref{fig:potts-V}).  We may
choose $\bm{A},\bm{B},\bm{C}$ to be at the corners of the equilateral
triangle
\begin{equation*}
  \bm{A}=(1,0), \quad \bm{B}=\tfrac{1}{2}\big(-1,\sqrt{3}\big),\quad 
  \bm{C}=\tfrac{1}{2}\big(-1,-\sqrt{3}\big)\,.
\end{equation*}
Gradient descent now leads to the coupled equations of motion
\begin{subequations}
\label{potts-tdgl}
\begin{equation}
\label{eqn:Potts_TDGL_x}
\frac{\partial\phi_1}{\partial t} = 2\nabla^2\phi_1-\frac{\delta}{\delta\phi_1}\!
\left[(\bm{\phi}-\bm{A})^2(\bm{\phi}-\bm{B})^2(\bm{\phi}-\bm{C})^2\right]\,,
\end{equation}
\begin{equation}
\label{eqn:Potts_TDGL_y}
\frac{\partial\phi_2}{\partial t} = 2\nabla^2\phi_2 - \frac{\delta}{\delta\phi_2}\!\left[(\bm{\phi}-\bm{A})^2(\bm{\phi}-\bm{B})^2(\bm{\phi}-\bm{C})^2\right]\,.
\end{equation}
\end{subequations}
For the continuum system, there are several natural choices for the initial
condition: (i) $\bm{\phi}$ uniformly distributed on a circle of unit radius,
(ii) $\bm{\phi}$ uniformly distributed within a ball of radius $\epsilon\ll
1$, (iii) $\bm{\phi}$ that equiprobably takes the values $\bm{A}$, $\bm{B}$,
or $\bm{C}$.  We find that the long-time behavior is largely independent of
the initial conditions.

\section{Non-Symmetric Initial Conditions}
\label{non-equal}

The initial state where the densities of the different spin states are equal
is the most physically relevant as it corresponds to quenching from a
supercritical temperature.  The more general case is still interesting and it
arises, e.g., when the system was in a magnetic field before the quench.  We
first study what happens when not all initial densities are equal in the
simplest non-trivial case of the 3-state Potts model.

\begin{figure}[ht]
\begin{center}
\subfigure[][t=200]{\includegraphics[width=0.24\textwidth]{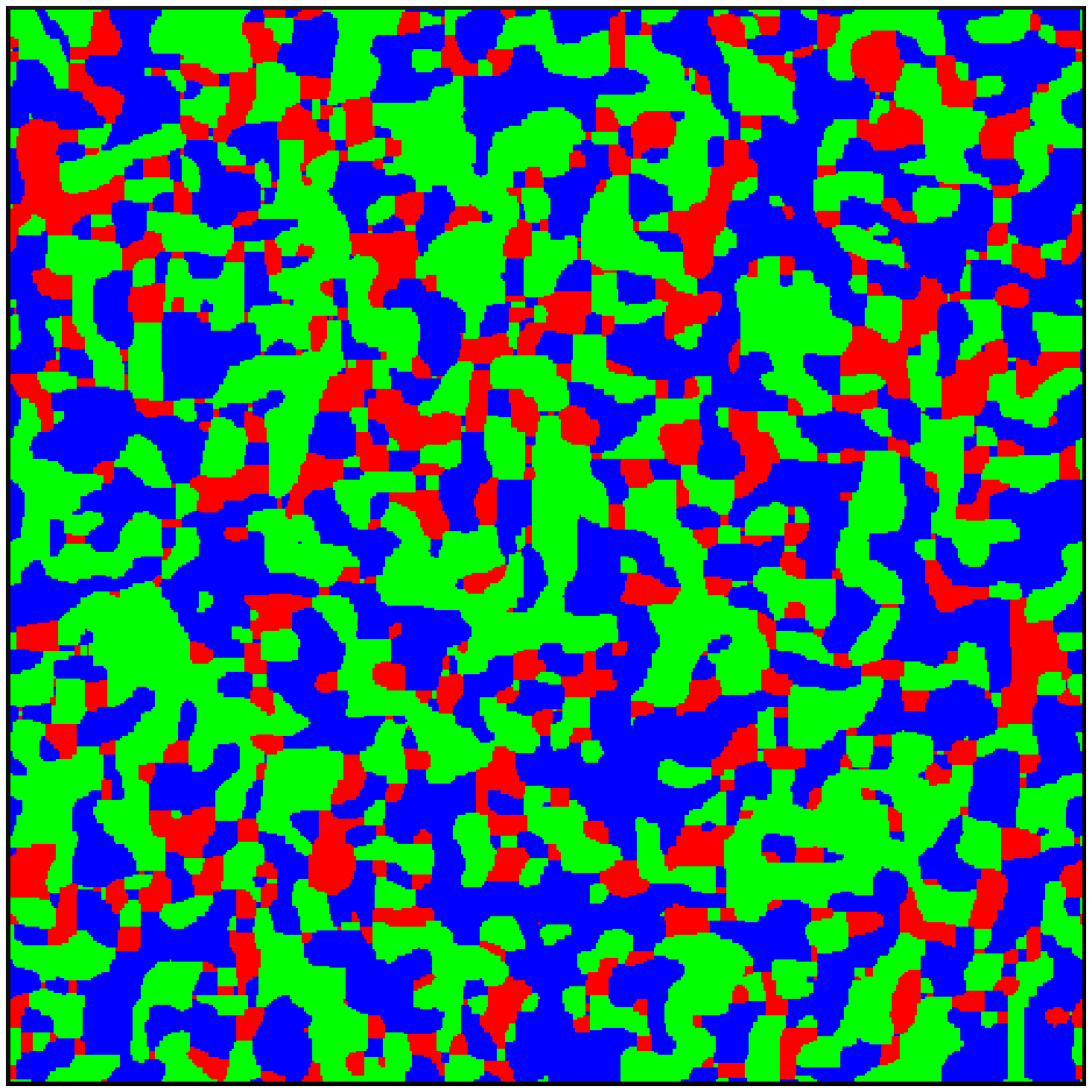}}
\subfigure[][t=1000]{\includegraphics[width=0.24\textwidth]{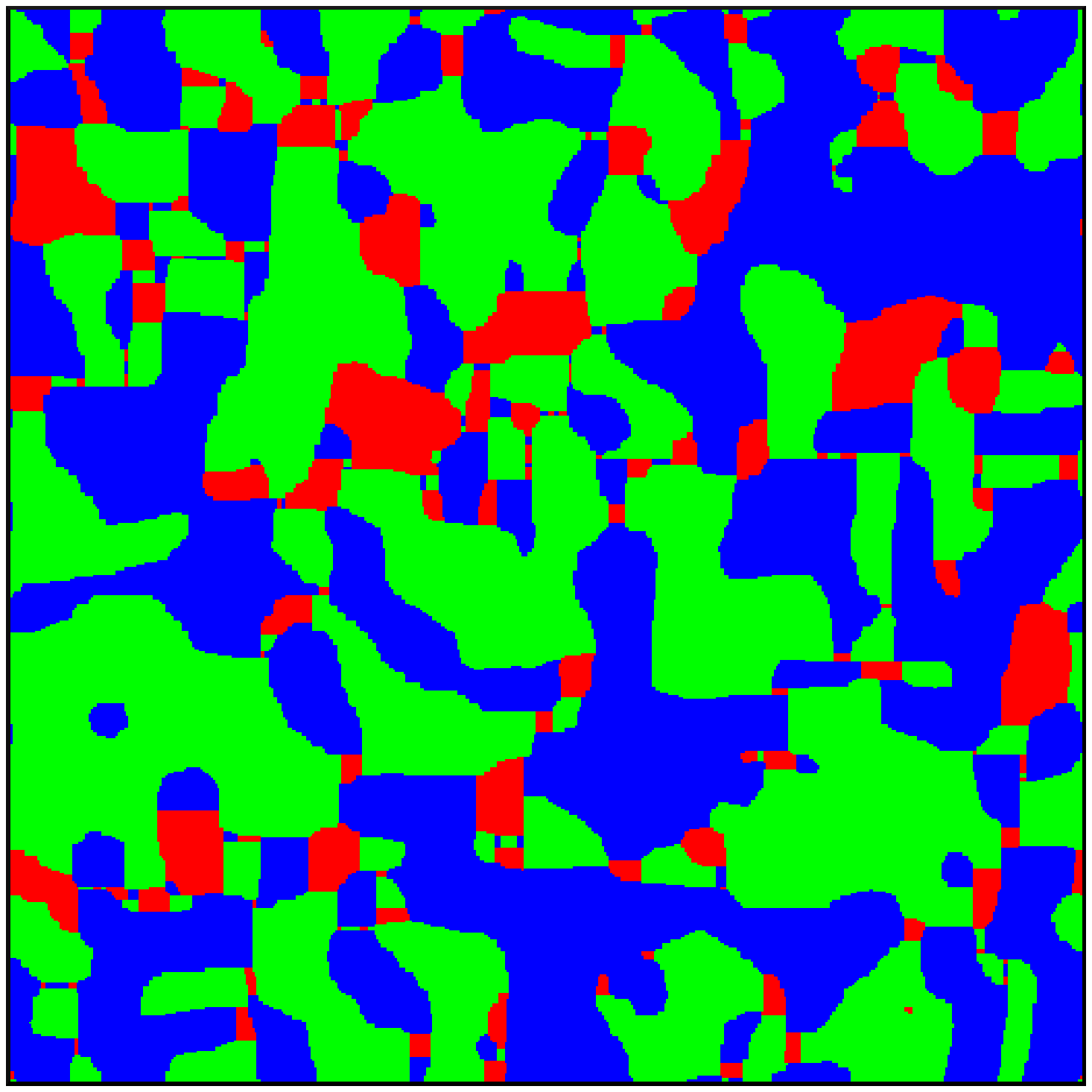}}
\subfigure[][t=5000]{\includegraphics[width=0.24\textwidth]{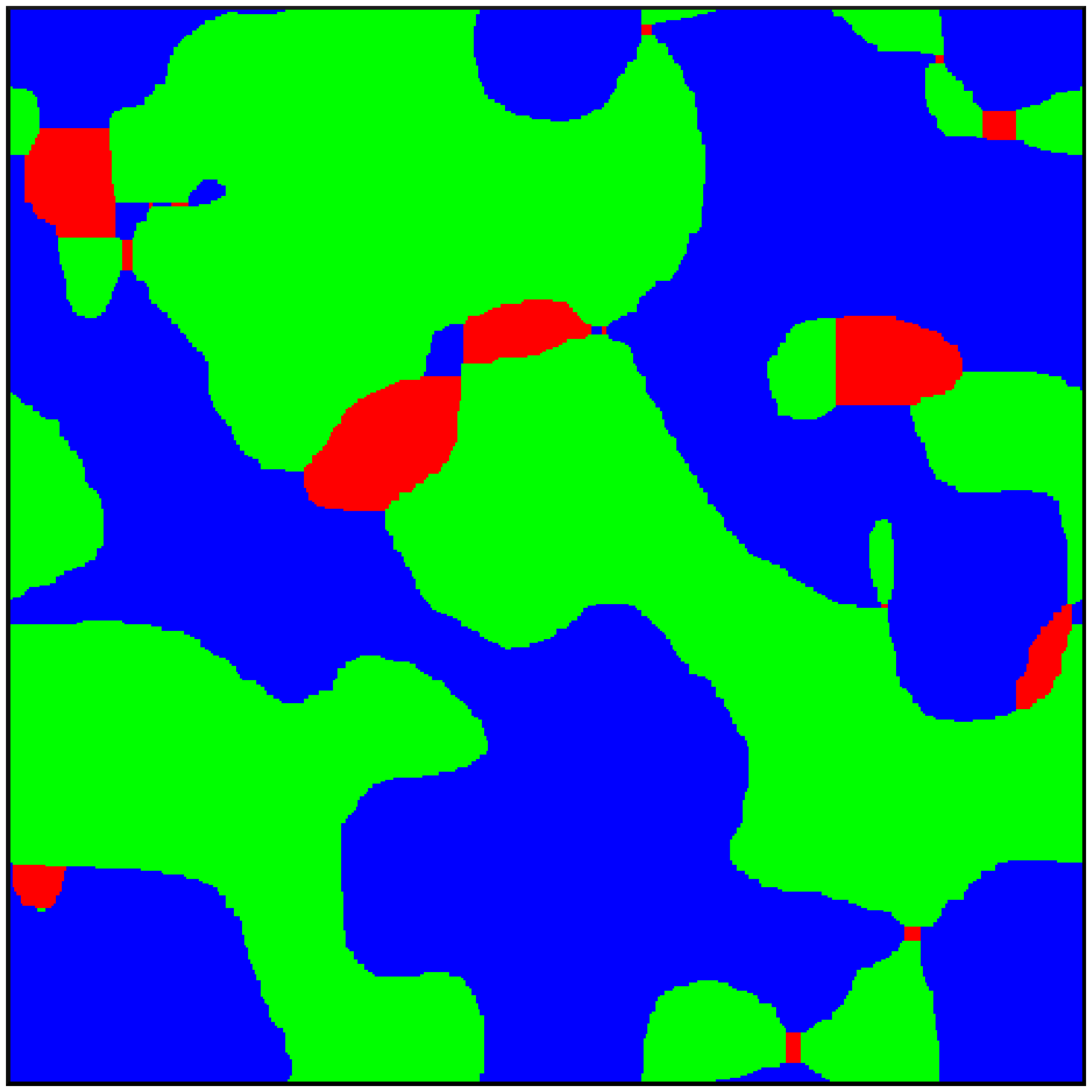}}
\subfigure[][t=50000]{\includegraphics[width=0.24\textwidth]{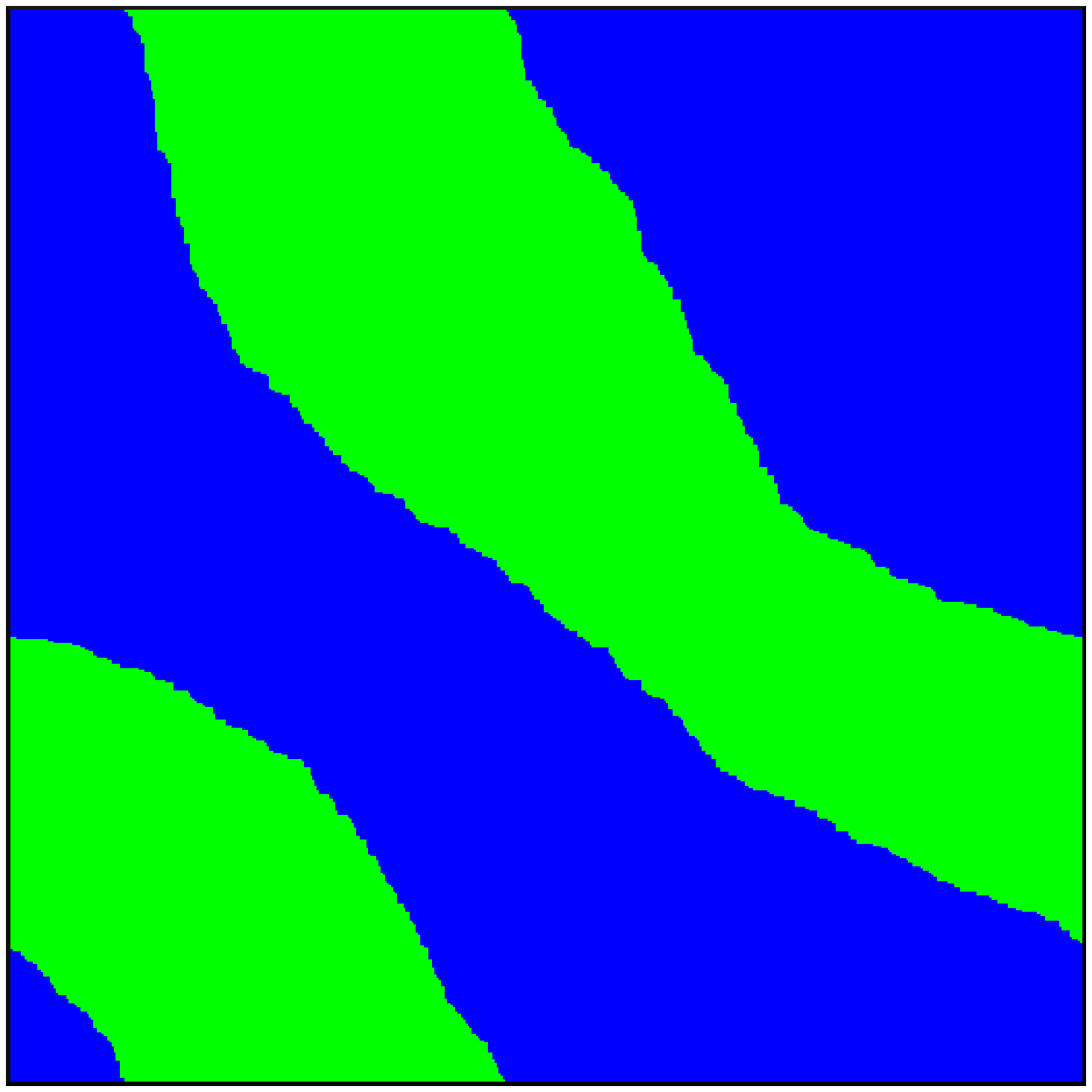}}
\caption{Coarsening of the 3-state Potts model to a diagonal stripe for
  linear dimension $L=768$, with respective initial densities of red, green,
  and blue $\frac{10}{32}$, $\frac{11}{32}$, and $\frac{11}{32}$.
  See~\cite{SM} for an animation of this evolution.}
\label{fig:diagonal}
  \end{center}
\end{figure}

There are two natural types of non-symmetric initial conditions: (i) one
species in the majority, and (ii) one species in the minority, while the
other two more abundant species have equal concentrations. In case (i),
simulations indicate that the final outcome is always the ground state of the
initial majority species.  In case (ii), simulations show that the minority
species disappears relatively quickly, so that the long-time dynamics is
Ising like.  Figure~\ref{fig:diagonal} shows the evolution of the 3-state
Potts model on a square lattice of linear dimension $L=768$ when the initial
densities of the three states are $\frac{1}{3}+\epsilon$ (blue),
$\frac{1}{3}+\epsilon$ (green), and $\frac{1}{3}-2\epsilon$ (red), with
$\epsilon=\frac{1}{96}$.  The important feature is that the evolution is
Ising like in the long-time limit.  For this example, the two persisting
species evolve into a very long lived (1,1) stripe topology, an event that
occurs with roughly 4\% probability\footnote{In the Ising model with
  nearest-neighbor interactions, this (1,1) stripe eventually disappears, but
  this topology is stable when  second-neighbor interactions exist;
  see~\cite{OKR12}.} in the kinetic Ising model~\cite{OKR12}.

\begin{figure}[ht]
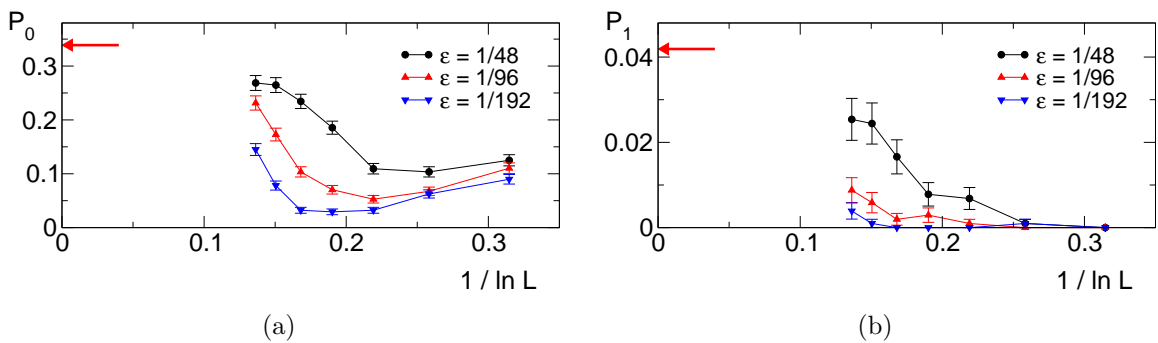

\begin{center}
\subfigure[]{\includegraphics*[width=0.47\textwidth]{figs/stripe10}}
\quad
\subfigure[]{\includegraphics*[width=0.47\textwidth]{figs/stripe11}}
\caption{Probabilities of freezing into: (a) a (1,0) stripe or (b) a (1,1)
  stripe versus $1/\ln L$ for the 3-state Potts model for different
  $\epsilon$ values.  The arrows indicate the theoretical predictions from
  the Ising model~\cite{BKR09,OKR12}. }
\label{fig:stripe_prob}
  \end{center}
\end{figure}

As a more stringent test of whether this initial condition corresponds to the
kinetic Ising model, we study the probability that the system reaches a
$(1,0)$ (equivalently (0,1)) or a $(1,1)$ stripe state when the initial
densities of the three spin states are $\frac{1}{3}+\epsilon$,
$\frac{1}{3}+\epsilon$, and $\frac{1}{3}-2\epsilon$ for varying $\epsilon$
(Fig.~\ref{fig:stripe_prob}).  Here the notation $(m,n)$ denotes a domain
that winds $m$ times around the torus in one coordinate direction and $n$
times in the orthogonal direction.  The freezing probabilities into these
distinct $(m,n)$ stripe topologies on periodic $L\times L$ lattices was
previously found in the kinetic Ising model~\cite{BKR09,OKR12}.  For the
3-state Potts model, it appears that the probabilities of reaching the (1,0)
and (1,1) stripe states are slowly approaching the Ising model predictions,
even for the smallest value of $\epsilon = \frac{1}{192}$ that we studied.
These results therefore suggest that dynamical behavior that is intrinsically
associated with the presence of all states in the 3-state Potts model occurs
only when the initial densities of the three spin states are equal.  
  
We now briefly discuss the fate of the Potts model with arbitrary initial
densities for general $q$. For the Ising model with initial magnetization
$m_0\ne 0$, it is believed that on an even-coordinated lattice in $d\geq 2$,
the majority ground state is reached with probability one in the
thermodynamic limit. In other words, the final magnetization is $m_\infty =
\text{sgn}(m_0)$.\footnote{The one-dimensional case is exceptional, both
  ground states can be reached, e.g., the plus ground state is reached with
  probability $(1+m_0)/2$.} This intuitively appealing property has been
proved \cite{M11} only in the $d\to\infty$ limit. Thus it is still a
conjecture even in two dimensions (where the connection with 2d continuum
percolation \cite{BKR09} `explains' why the majority ground state should be
reached).

For the Potts model, we denote initial densities by $m_1,\ldots,m_q$.  We can
always relabel the spin types to ensure that these densities satisfy $m_1\geq
m_2\geq\cdots\geq m_q$, and the normalization constraint $m_1+\ldots+m_q=1$.
Our numerical simulations (Sec.~\ref{non-equal}) of the 3-state kinetic Potts
model indicate three possibilities:
\begin{enumerate}
\item $m_1>m_2\geq m_3$. The majority ground state is reached.
\item $m_1=m_2> m_3$. The fate of this 3-state kinetic Potts model is the
  same as that in the Ising model: The minority phase disappears, and the
  system either reaches the ground state of one of the two majority phases
  (this occurs with probability \eqref{GS_Ising}), or a stripe state with
  stripes composed of spins of the majority phases is reached (which occurs
  with the complementary probability $1-\Pi_2$).
\item $m_1=m_2=m_3$. This is the only point where the fate of the system is
  novel, and it faithfully represents the 3-state Potts model.
\end{enumerate}

Based on the above findings, we conjecture that the $q$-state Potts model
with initial densities $m_1=\cdots=m_Q>m_{Q+1}\geq\cdots\geq m_q$ exhibits
the same behavior as the $Q$-state Potts model with equal initial
densities. It is feasible that this conjecture is actually valid not only on
the square lattice, but for other (even-coordinated) lattices of arbitrary
spatial dimension $d\geq 2$.

In the remainder of this paper, we concentrate on the symmetric initial
condition for the general $q$-state Potts model.

\section{Cluster Geometry at Long Time}
\label{cluster}

\subsection{Kinetic Potts Model}

For the kinetic Potts model, fundamental questions that we investigate are:
(i) What is the nature of the long-time state?  Is the ground state
eventually reached or is there freezing into more complex final states?  (ii)
What is the time dependence of the long-time relaxation?  The examples in
Fig.~\ref{fig:clusters} suggest that the ground state is likely not reached,
but rather the final state typically consists of a domain mosaic with
characteristic length scale $L$ and with `breathing' domain walls; such
states are the analogs of blinker states that arise in quenches of the 3d
Ising model to zero temperature.  Our simulations indicate, however, that
this outcome is just one of the three possibilities mentioned in the
introduction.  Namely, in the thermodynamic limit the system falls into a
ground state, or a more complicated static state, or it wanders on a set of
equal-energy blinker states (see also Ref.~\cite{OP02}).  As already
mentioned in the introduction, it is hard to provide quantitative predictions
for these probabilities (see also Sec.~\ref{ultra}).

\begin{figure}[ht]
\begin{center}
  \includegraphics*[width=0.1\textwidth]{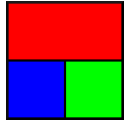}
  \caption{A T-junction in the 3-state Potts model.}
\label{fig:T_vertex}
  \end{center}
\end{figure}

For the 3-state Potts model, the fundamental reason for the increased cluster
complexity compared to the Ising model is that an interface between domains
may have a beginning and an end.  The termini of these interfaces occur at
T-junctions where three distinct spin states appear in the elemental
plaquette that circumscribes the junction (Fig.~\ref{fig:T_vertex}).  Each of
these spins has at least 2 neighbors in the same state and thus cannot flip
at zero temperature; the importance of these T-junctions was first noted
in~\cite{SSGAS83,OP02}.  These T-junctions therefore pin domain boundaries
and provide a multitude of ways for the 3-state Potts model to get stuck in
an infinitely long-lived metastable state (at zero temperature), such as the
examples on the top line of Fig.~\ref{fig:clusters}.  For larger $q$, more
complex interface junctions can arise, which suggest that pinning of domain
walls is even more likely to occur.

\begin{figure}[ht]
\begin{center}
\includegraphics*[width=0.4\textwidth]{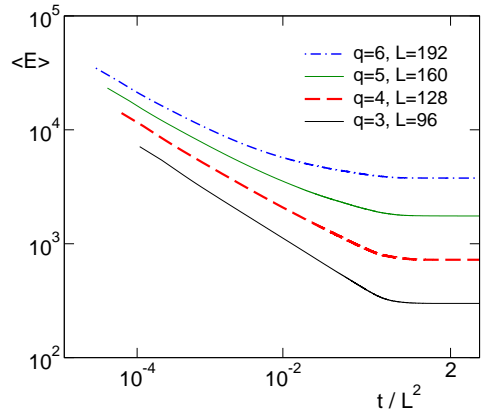}
\caption{Average energy versus $t/L^2$ for the $q$-state Potts model for
  $q=3$--6.  The average is over $4096$ realizations for $q=3$ and $512$
  realizations for larger $q$.}
\label{fig:avg_E_t}
  \end{center}
\end{figure}

To understand the long-time state of the $q$-state Potts model, consider the
time dependence of the energy for $q$ between 3 and 6
(Fig.~\ref{fig:avg_E_t}).  At early times, the energy decays in a manner that
is consistent with a power-law time dependence, as expected from
curvature-driven coarsening~\cite{bray_review,F1997}.  This apparent
power-law decay defines the coarsening regime.  By time $t \approx L^2$, the
decrease in the energy has essentially stopped (see Ref.~\cite{P03} for a
more quantitative discussion of this feature), although we shall see that
there are subtle relaxation mechanisms that arise at much longer times.  For
both tractability and concreteness, most of our observations are based on
measurements at time $\tau_c=2L^2$, which we define here as the Potts
coarsening time.  We choose this value of $\tau_c$ because it is safely
beyond the initial coarsening regime, so that a typical realization of the
system is pinned in a quasi-final state.

\begin{figure}[ht]
\begin{center}
\vspace{0.5in}
\subfigure[]{\includegraphics[width=0.31\textwidth]{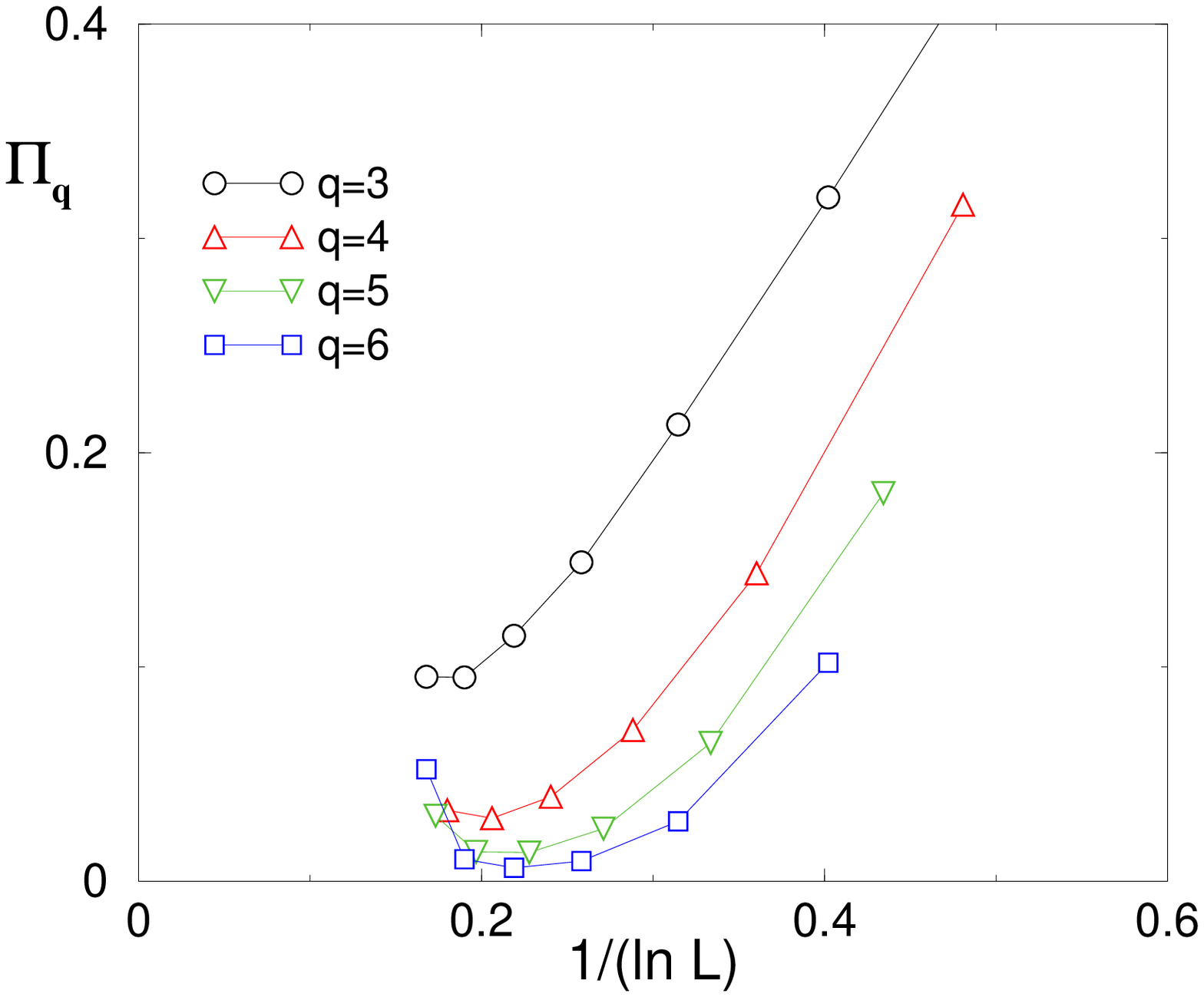}}\quad
\subfigure[]{\includegraphics[width=0.32\textwidth]{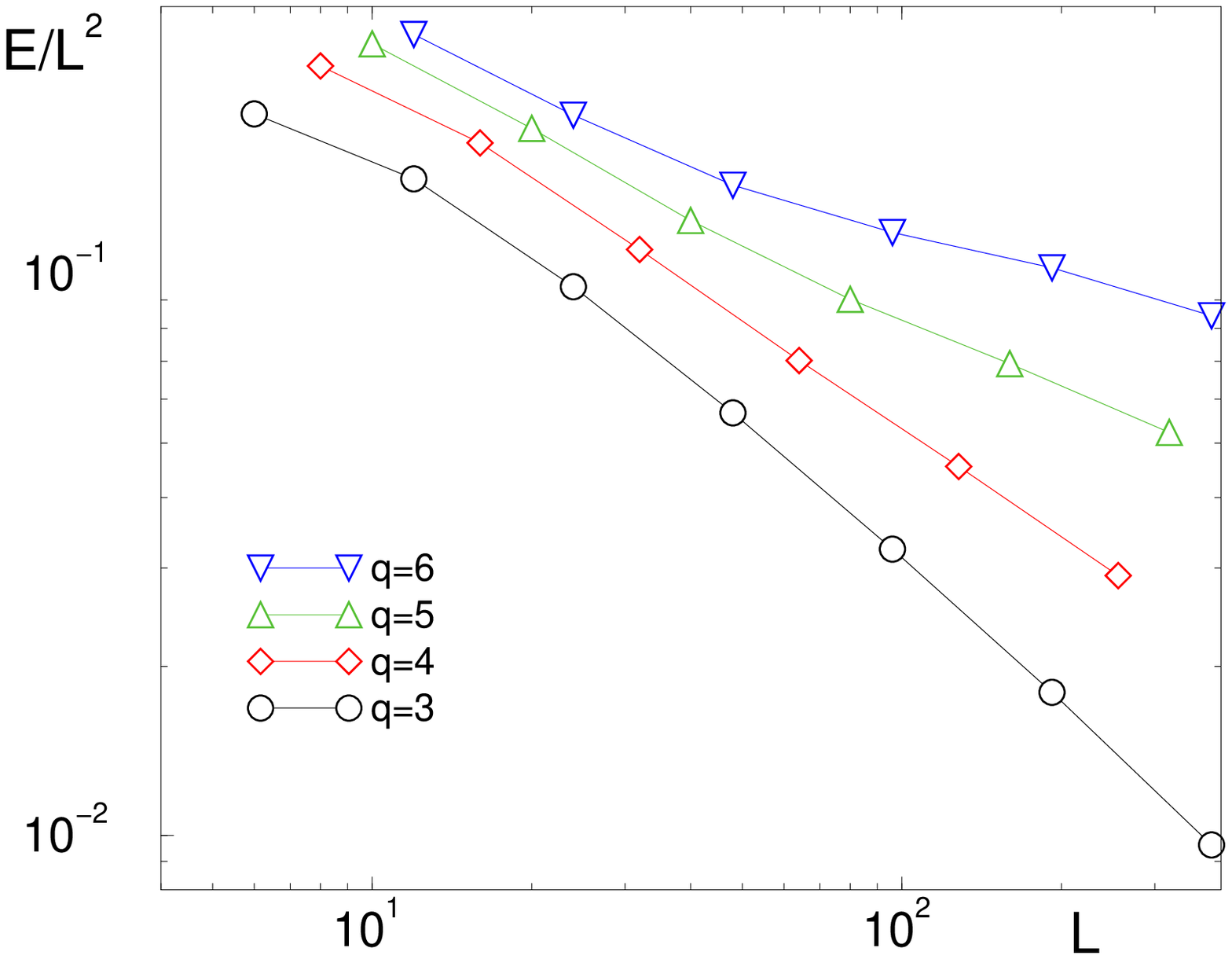}}\quad
\subfigure[]{\includegraphics[width=0.3\textwidth]{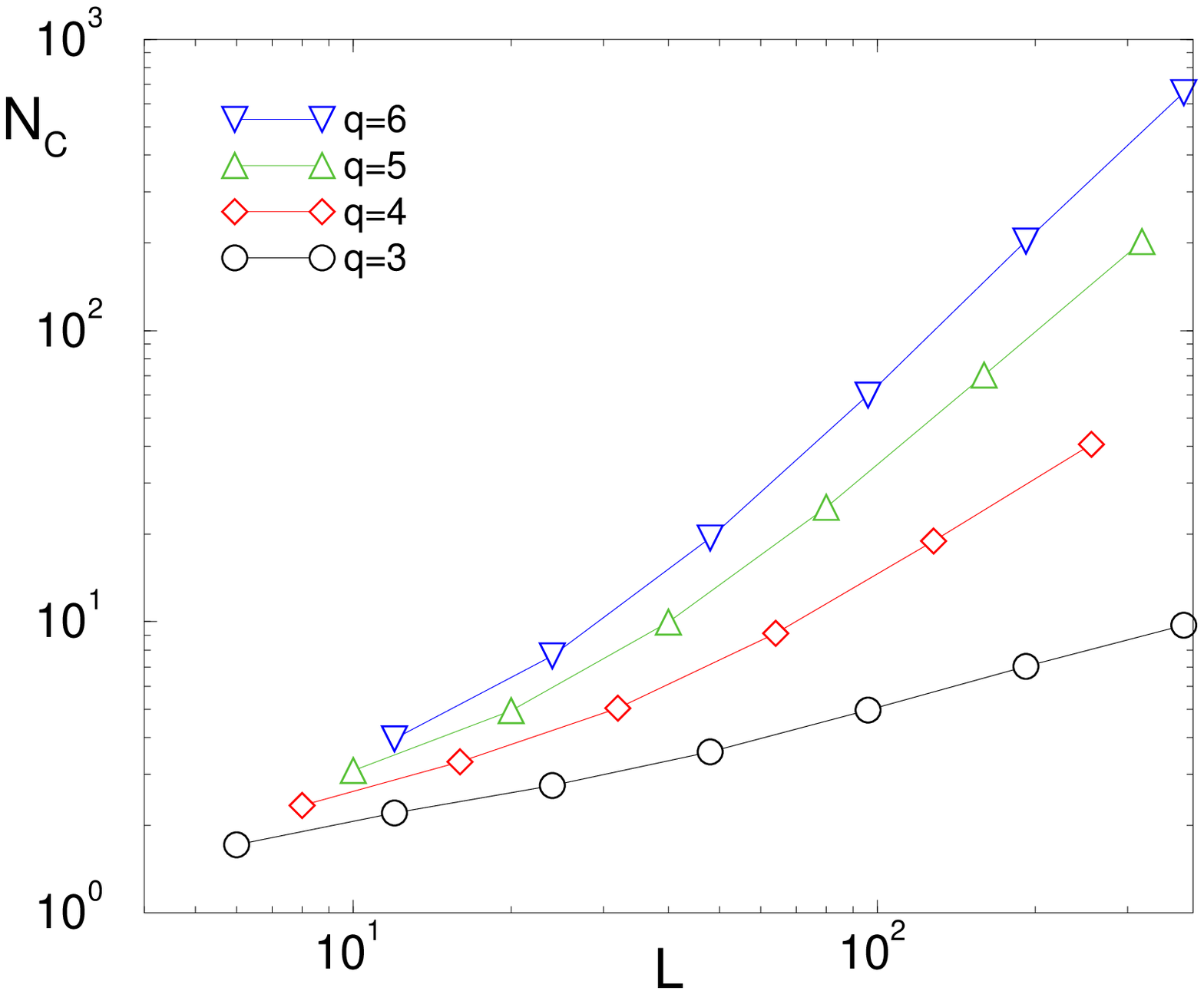}}
\caption{ (a) Probability $\Pi_q$ for the $q$-state Potts model to reach the
  ground state by time $2L^2$ following a quench to zero temperature.  The
  data for the largest two values of $L$ for $q=3$ and for the largest
  value of $L$ for $q=4,5,6$ are based on $2^{14}$ realizations; all
  other data are based on $2^{17}$ realizations.  Error bars are
  smaller than the size of the symbols.  (b) Average energy per spin
   and (b) average number of clusters versus $L$ for the $q$-state 
   Potts model with $q=3$--6.
   \label{at-tc}}
  \end{center}
\end{figure}

We now focus on the systematic $q$ dependences of basic observables in the
$q$-state Potts model at $\tau_c=2L^2$ (Fig.~\ref{at-tc}).  Perhaps the most
surprising feature is the behavior of the probability $\Pi_q$ to reach the
ground state in the limit of $L\to\infty$.  For all $q\geq 3$, $\Pi_q$ varies
non-monotonically with $L$, and the data suggest that the probability of
reaching the ground state will be greater than zero as $L\to\infty$ for any
$q\geq 3$.  It is paradoxical that as the number of Potts states is increased,
the data also suggest that it is more likely that a finite system will reach
the ground state (Fig.~\ref{at-tc}(a)).  We will provide more details about
this intriguing phenomenon below.

Next, we consider the $L$ dependence of the energy per spin.  If one fits
these data to a power law (which becomes more dubious for larger $q$), this
normalized energy decays roughly as $L^{-0.8}$ for $q=3$ and decays
progressively more slowly with $L$ for larger $q$.  As one might naively
anticipate, as the number of Potts states increases, a typical realization at
the coarsening time consists of progressively smaller patches of single-spin
states.  Concomitantly, there would be more interfaces for larger $q$ and the
normalized energy should decay therefore progressively more slowly with $L$.

Finally, let us examine the average number of clusters $N_c$. Strictly
speaking, we can not cleanly probe $N_c$ for sufficiently large $L$, as the
true asymptotic regime is beyond the reach of the simulations for many
realizations.  Nevertheless, we can still infer some basic $q$-dependent
properties: For $q=6$, the average number of clusters grows roughly as $L^2$.
This result indicates that for $q=6$ (and larger), domains typically have a
size comparable to the lattice spacing, so that their number grows linearly
with the area of the system.  In contrast, for $q<6$, $N_c$ grows more slowly
than $L^2$, indicating that in these cases a characteristic domain length
increases with system size.

Interestingly, as the number of spin states $q$ increases, the probability to
reach the ground state $\Pi_q$ becomes more likely while the average number
of clusters $N_c$ grows more rapidly with $L$. These seemingly incompatible
behaviors are not necessarily contradictory, however.  Even if it becomes
more likely to end up with a single cluster (the ground state) for larger
$q$, the {\em average} number of clusters can still grow linearly in the area
of the system.  Importantly, this growth rate increases with $q$.

\subsection{TDGL Equation}

For the case of three degenerate ground states, we start with
$\bm{\phi}(x,y,t)$ in one of the symmetric and disordered initial
configurations (i)--(iii) given in subsection~\ref{TDGL} and numerically
integrate the TDGL equations~\eqref{potts-tdgl} in time on $L\times L$
lattices, with $L=16,32,64,128$.  We stop the integration at time $t=L^2$,
which is at the end of the coarsening regime for the 3-state TDGL.
Figure~\ref{fig:patterns} (top) shows three nontrivial outcomes from these
integrations.  To make this plot, we map the continuous order parameter in
the TDGL equation to one of three discrete values that depend on whether
$\bm{\phi}$ lies inside the wedge with angular range
$(\frac{\pi}{3},-\frac{\pi}{3})$ that is centered on the potential minimum at
$\bm{A}=(1,0)$, the wedge $(\frac{\pi}{3},\pi)$ that is centered on
$\bm{B}=\tfrac{1}{2}\big(-1,\sqrt{3}\big)$, or the wedge
$(-\pi,\frac{5\pi}{3})$ that is centered on
$\bm{C}=-\tfrac{1}{2}\big(1,\sqrt{3}\big)$.  After this discretization, we
identify the resulting clusters by the multi-labeling method~\cite{labeling}.

We find that the 3-state TDGL can become trapped in unexpected types of
pinned metastable arrangements as shown in Fig.~\ref{fig:patterns}.  The
outcome of three hexagons (Fig.~\ref{fig:patterns}(a)) occurs in roughly
$11\%$ of all realizations, four pinned clusters (Fig.~\ref{fig:patterns}(b))
occur in roughly $8\%$ of realizations, while a state that contains six
hexagonal clusters (Fig.~\ref{fig:patterns}(c)) occurs in less than 1\% of
all realizations.  In addition, the probability that the system gets stuck in
a state that consists of two straight stripes (as in the TDGL with two ground
states) occurs roughly 10\% of the time, while roughly 70\% of the time the
ground state is reached.  The probabilities of reaching these different
long-time states do depend on the system size, but the data clearly indicate
that in the thermodynamic limit the probability of reaching the ground state
is less than 1, while the probabilities of reaching 3- and 4-cluster states
are both greater than zero.

\begin{figure}[ht]
\begin{center}
\subfigure[][t=0]{\includegraphics[width=0.22\textwidth]{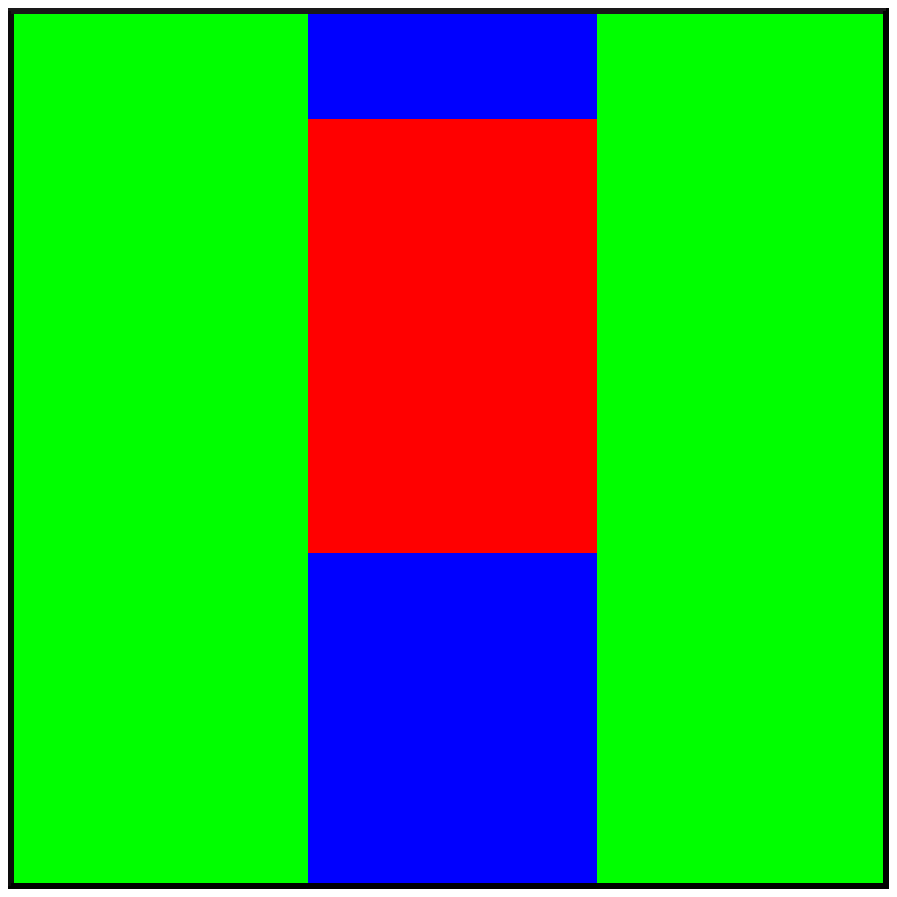}}\quad
\subfigure[][t=1000]{\includegraphics[width=0.22\textwidth]{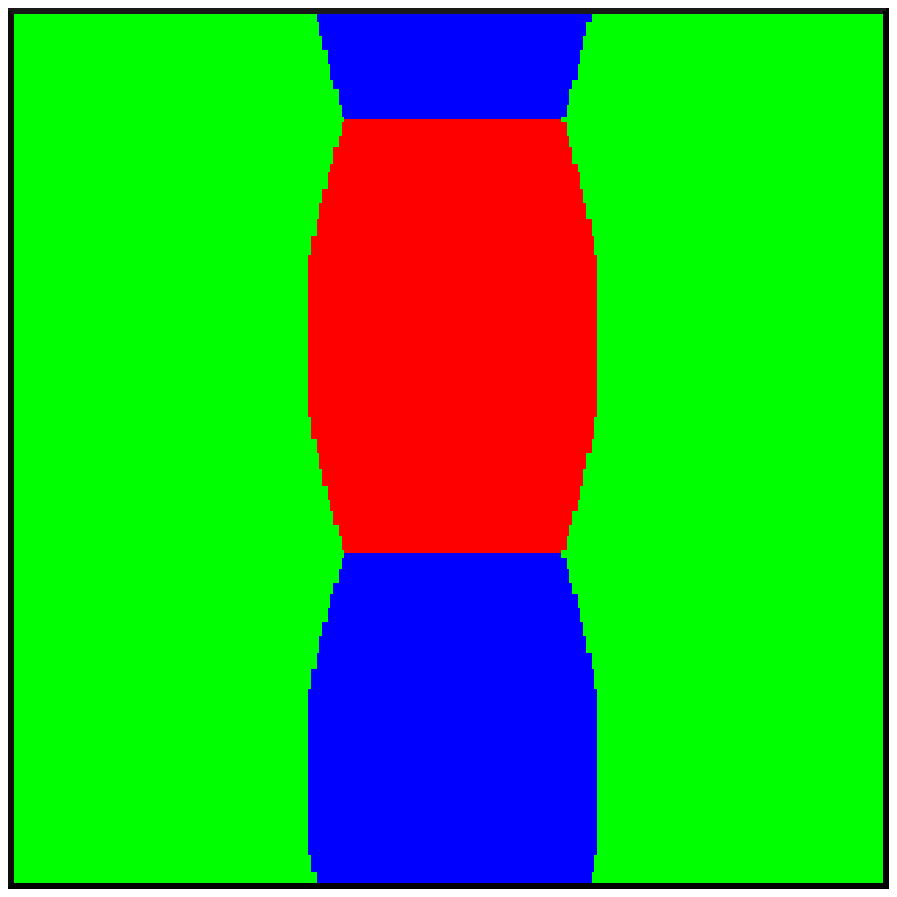}}\quad
\subfigure[][t=7000]{\includegraphics[width=0.22\textwidth]{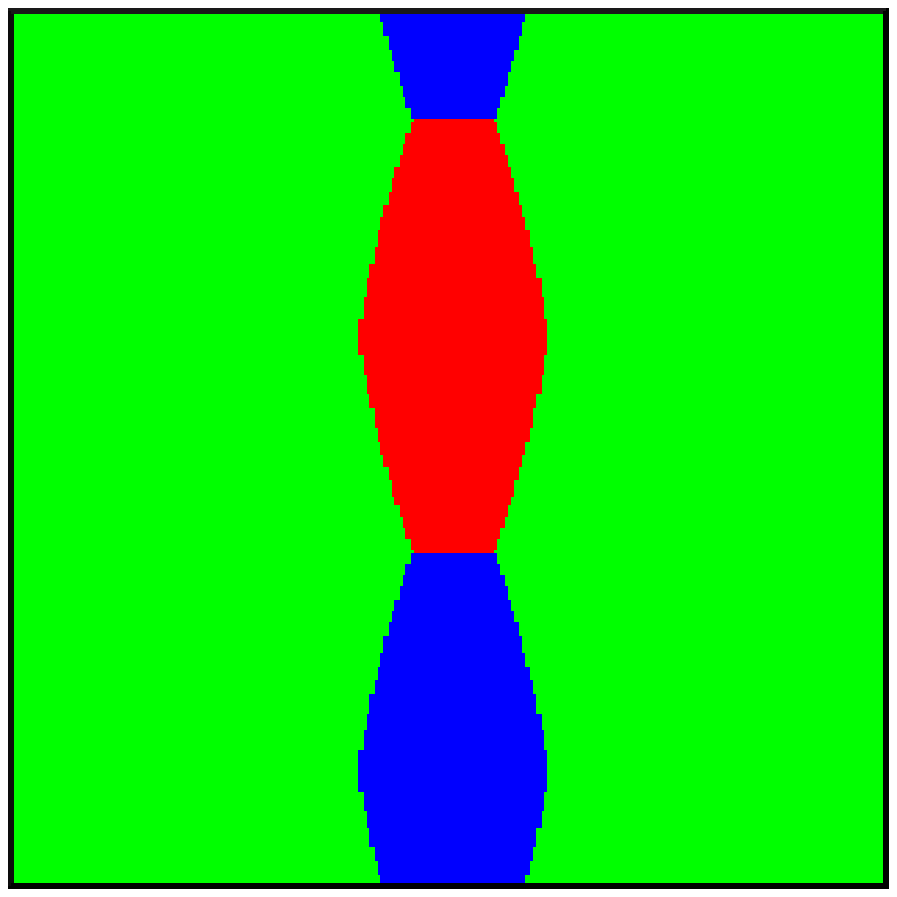}}\quad
\subfigure[][t=12000]{\includegraphics[width=0.22\textwidth]{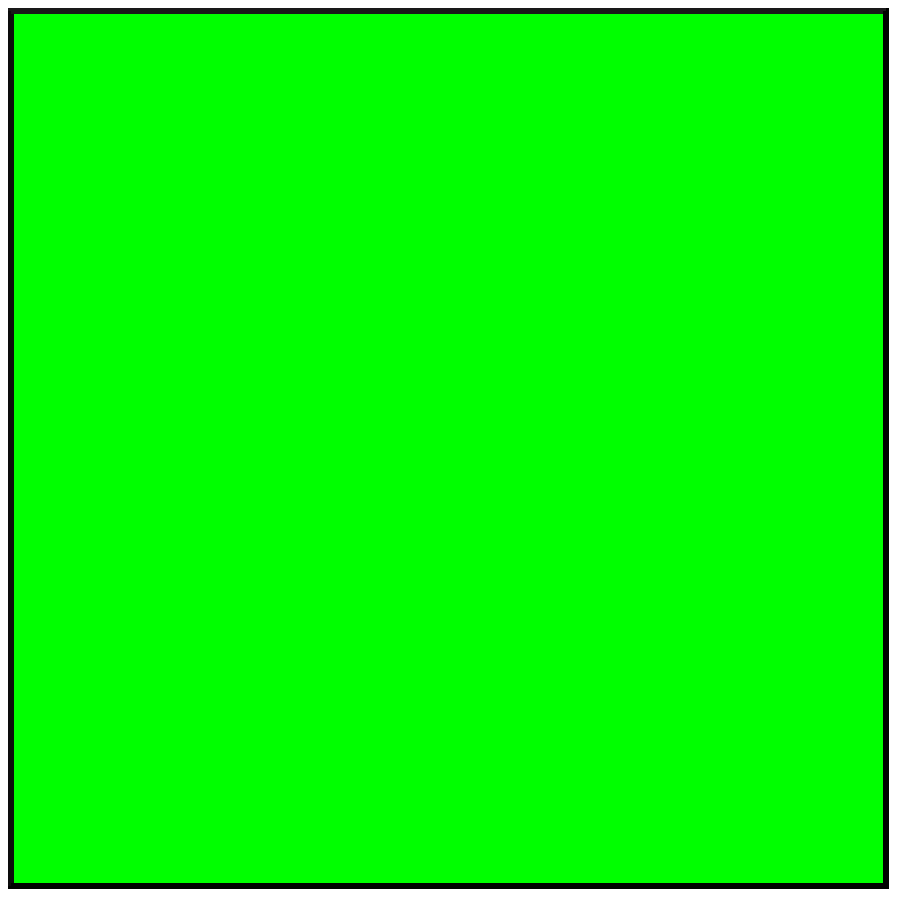}}
\caption{A multi-state stripe is stable in the discrete Potts model
  (a) but is unstable under TDGL evolution (b-c).  }
\label{fig:bar}
\end{center}
\end{figure}

Finally, it is worth noting an important difference between the discrete
3-state Potts model and the TDGL equation with 3 degenerate minima that has
consequences for the nature of the final states.  In the 3-state Potts model,
T-vertices that pin domain walls always contain right angles.  However, the
TDGL equations are isotropic, so neighboring clusters must meet at
$120^\circ$ angles for the 3-state model.  This difference has profound
implications on the cluster structure at long times.  Consider the stripe
state shown in Fig.~\ref{fig:bar}(a).  This configuration is stable in the
Potts model but is unstable in 3-state TDGL.  Starting with this state and
integrating the TDGL equations \eqref{potts-tdgl}, the angles between domain
walls at the T-junctions begin to equalize.  In so doing, curvature is
generated on the interface that tends to shrink the stripe and the final
outcome is the ground state (Fig.~\ref{fig:bar}(d)).

\section{Ultra-Slow Evolution}
\label{ultra}

\begin{figure}[ht]
\begin{center}
\subfigure[]{\includegraphics[width=0.28\textwidth]{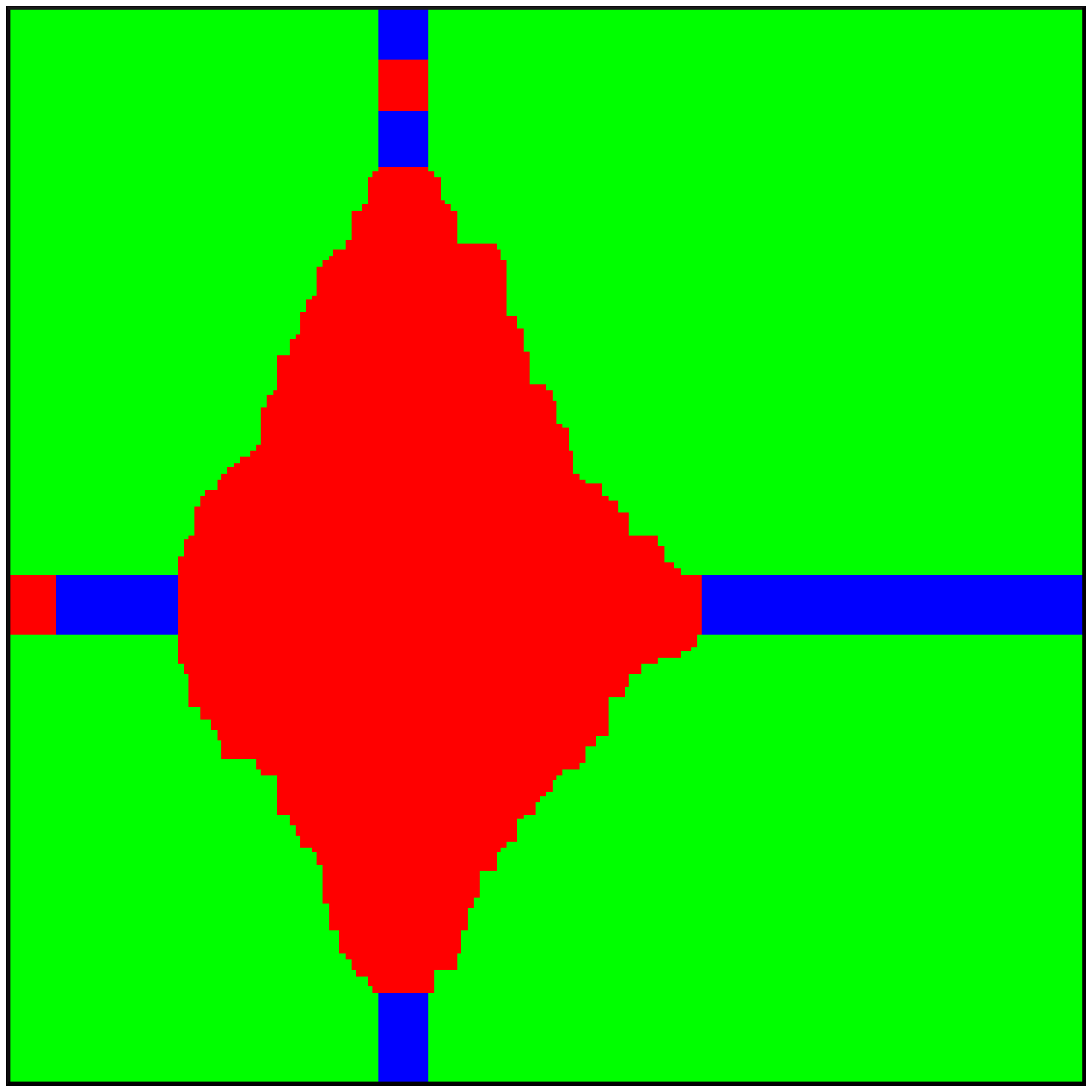}}\qquad
\subfigure[]{\includegraphics[width=0.28\textwidth]{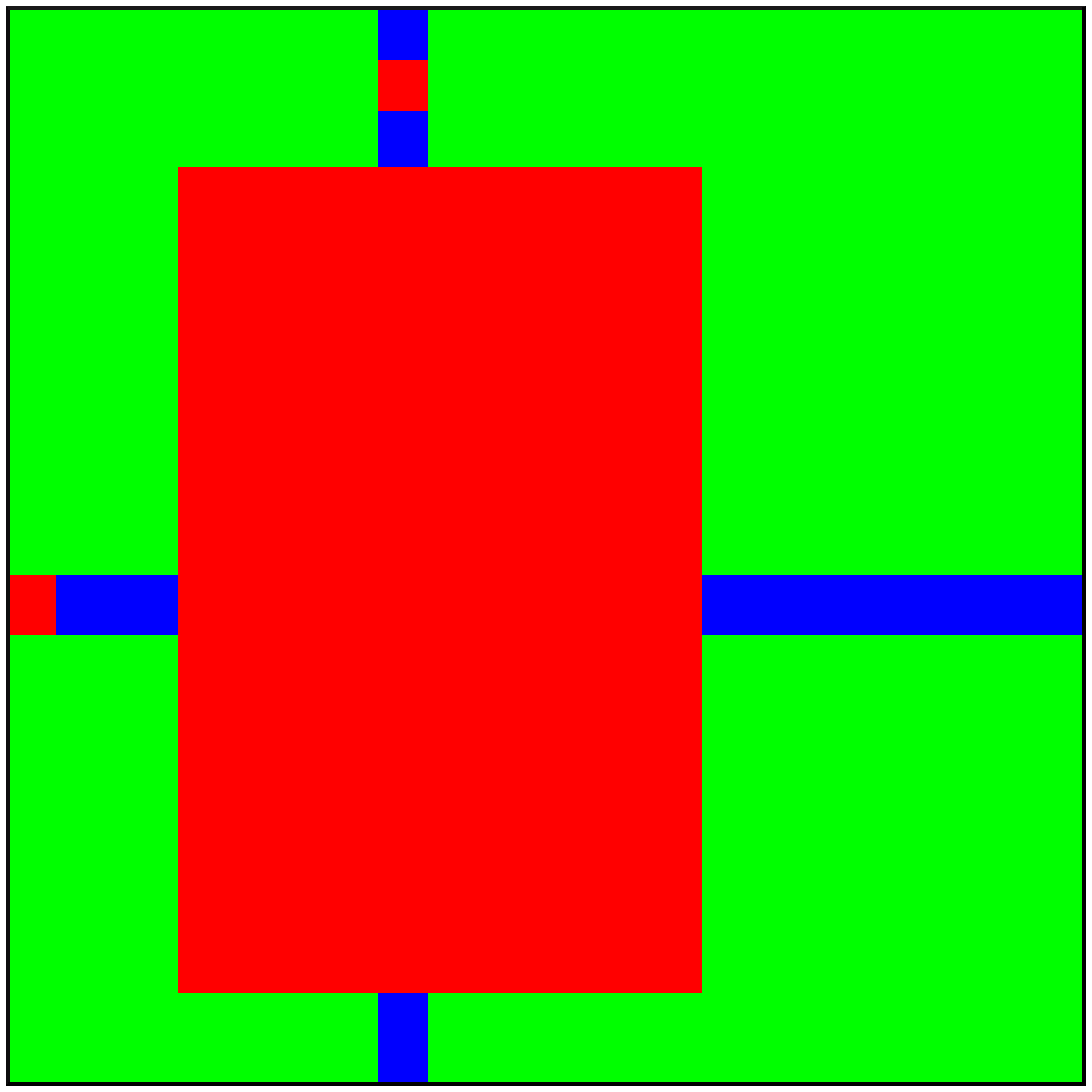}}\qquad
\subfigure[]{\includegraphics[width=0.28\textwidth]{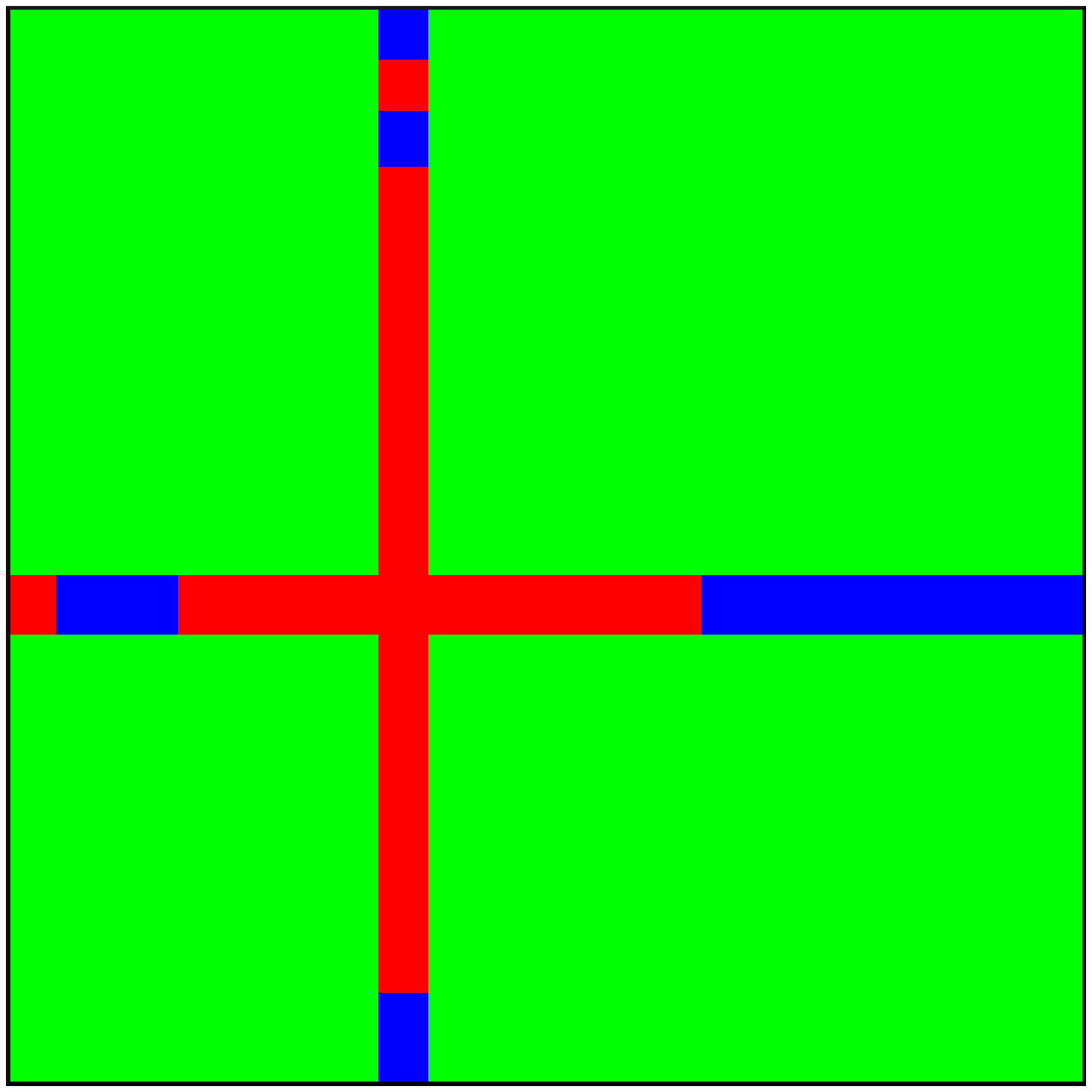}}
\caption{ (a) A blinker state, with four staircase interfaces of blinker
  spins.  The extremes where these interfaces are fully inflated
  (corresponding to the convex envelope of the domain in (a)) or fully
  deflated are shown in (b) and (c).  }
\label{fig:blinker_example}
  \end{center}
\end{figure}

The long-time states of the Potts system are often \emph{not} static, but are
either very long-lived or infinitely long-lived ``blinker'' states.  A
blinker spin is one that can freely flip between various states without any
energy cost, and a blinker state is a configuration that contains multiple
blinker spins.  These blinker states also arise in zero-temperature quenches
of the 3d kinetic Ising model; however, in the 2d Potts system, they are more
readily visualized.  A typical example is shown in
Fig.~\ref{fig:blinker_example}.  Here, the T-junctions are frozen in place at
zero temperature, but the spins on the interfaces that connect two
T-junctions can flip freely with zero energy cost.

\begin{figure}[ht]
\begin{center}
\subfigure[]{\includegraphics[width=0.2\textwidth]{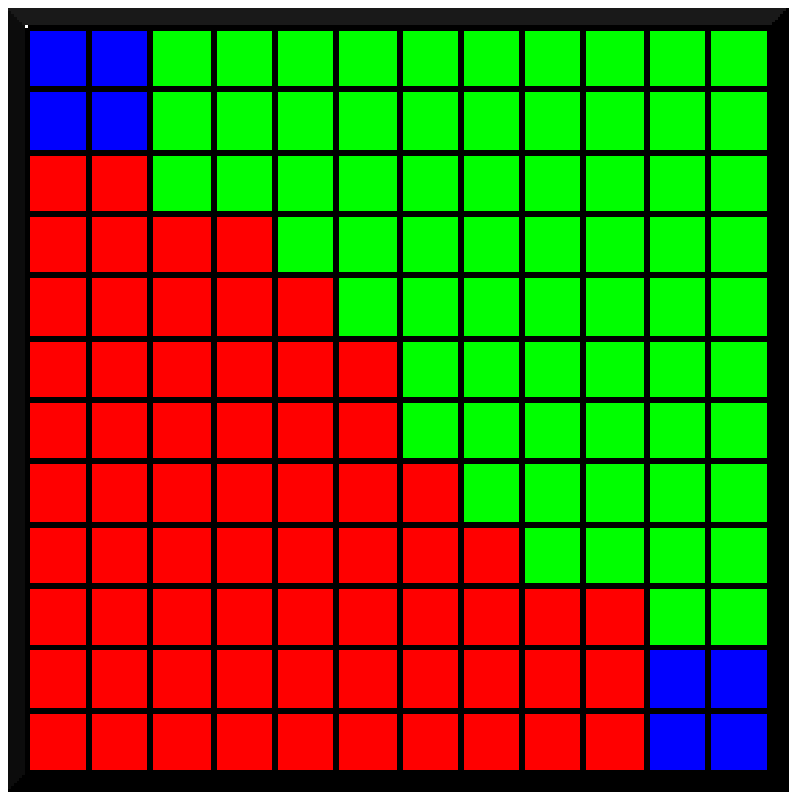}}\qquad\qquad
\subfigure[]{\includegraphics[width=0.2\textwidth]{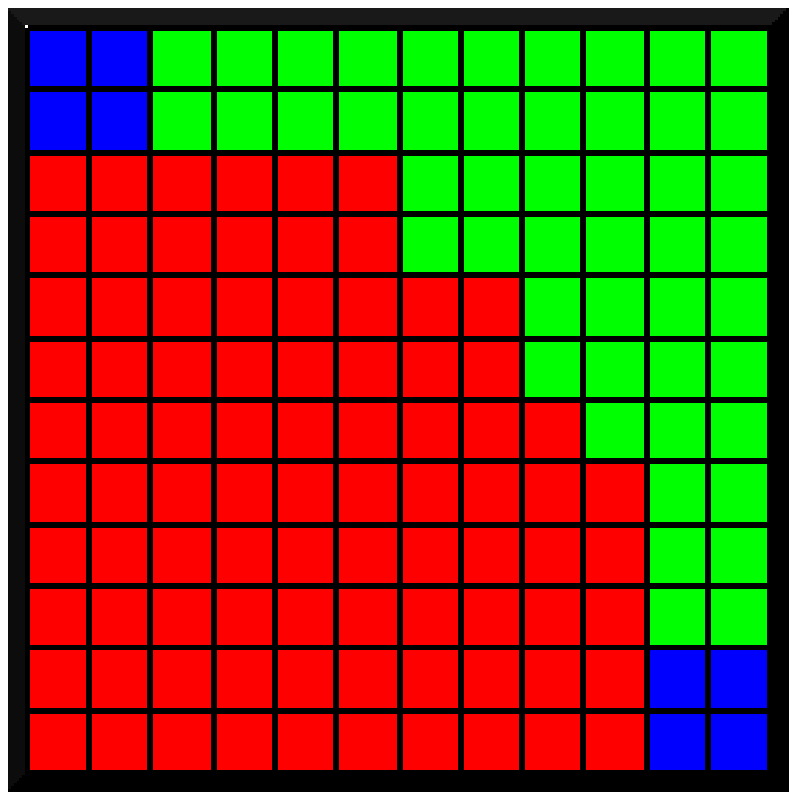}}\qquad\qquad
\subfigure[]{\includegraphics[width=0.2\textwidth]{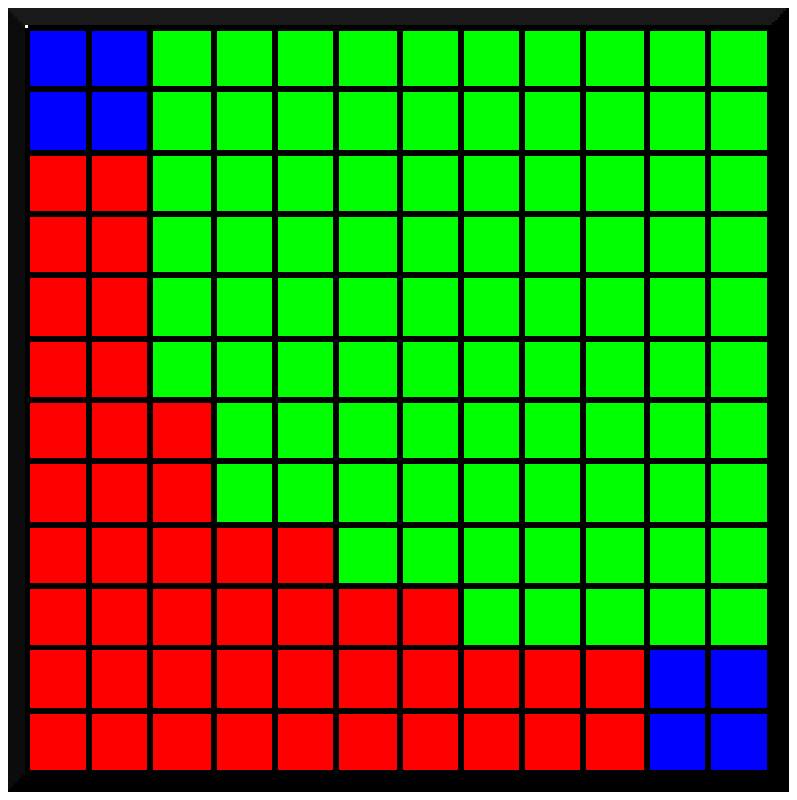}}
\caption{A staircase interface with blinker spins at the interface corners,
  showing the (a) intermediate, (b) partially inflated, and (c) partially
  deflated states.}
\label{fig:corners}
  \end{center}
\end{figure}

A crucial feature of these blinkers is that they predominantly remain in
their intermediate states, as in Fig.~\ref{fig:blinker_example}(a).  This
behavior arises because there exists an effective bias in the diffusive
motion of the staircase interface that pushes it to an intermediate state
(Fig.~\ref{fig:corners}).  In Fig.~\ref{fig:corners}(b), there are four red
and three green blinker spins, so that the red interface tends to recede.
Similarly, in Fig.~\ref{fig:corners}(c), there are four green and three red
blinker spins, so that the green interface tends to recede.  These biases
push the interface to the intermediate state, as in
Figs.~\ref{fig:blinker_example}(a) and \ref{fig:corners}(a).

\begin{figure}[ht]
\begin{center}
\subfigure[]{\includegraphics[width=0.23\textwidth]{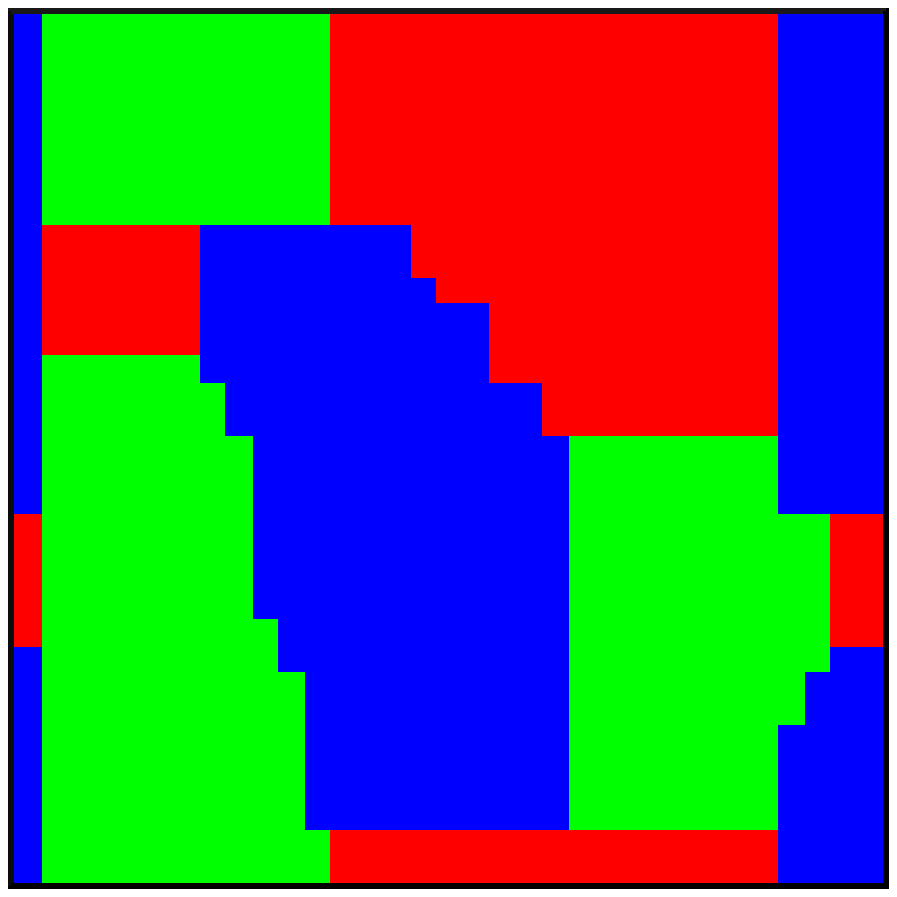}}\quad
\subfigure[]{\includegraphics[width=0.23\textwidth]{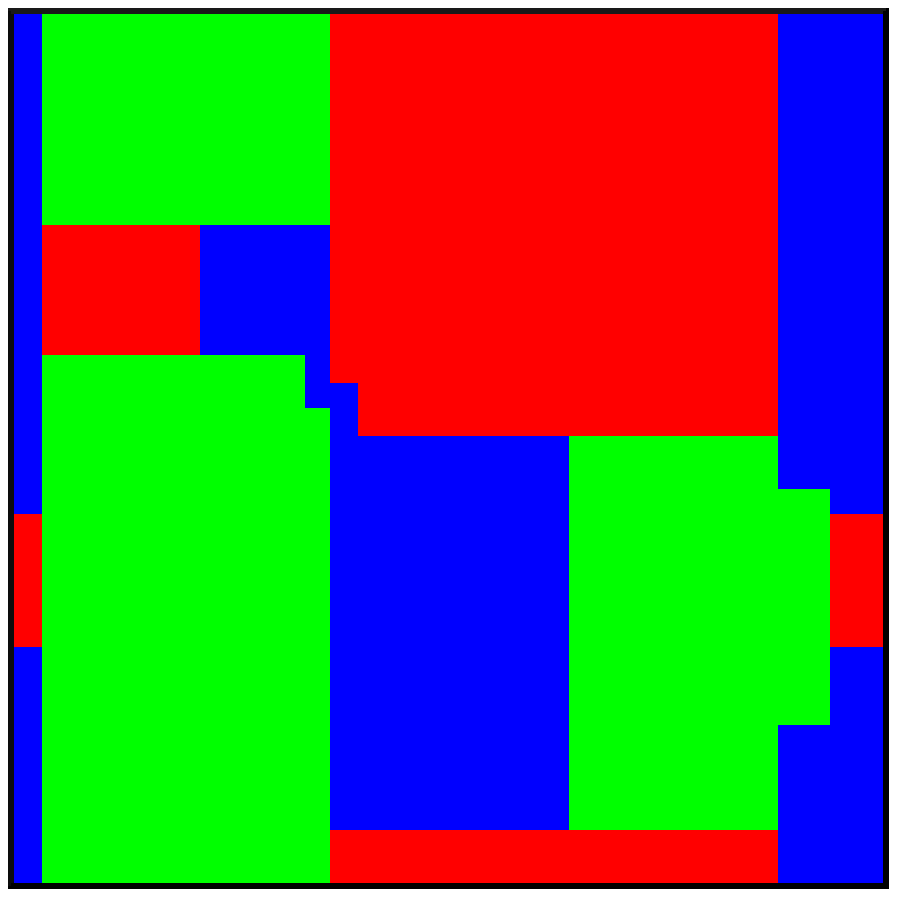}}\quad
\subfigure[]{\includegraphics[width=0.23\textwidth]{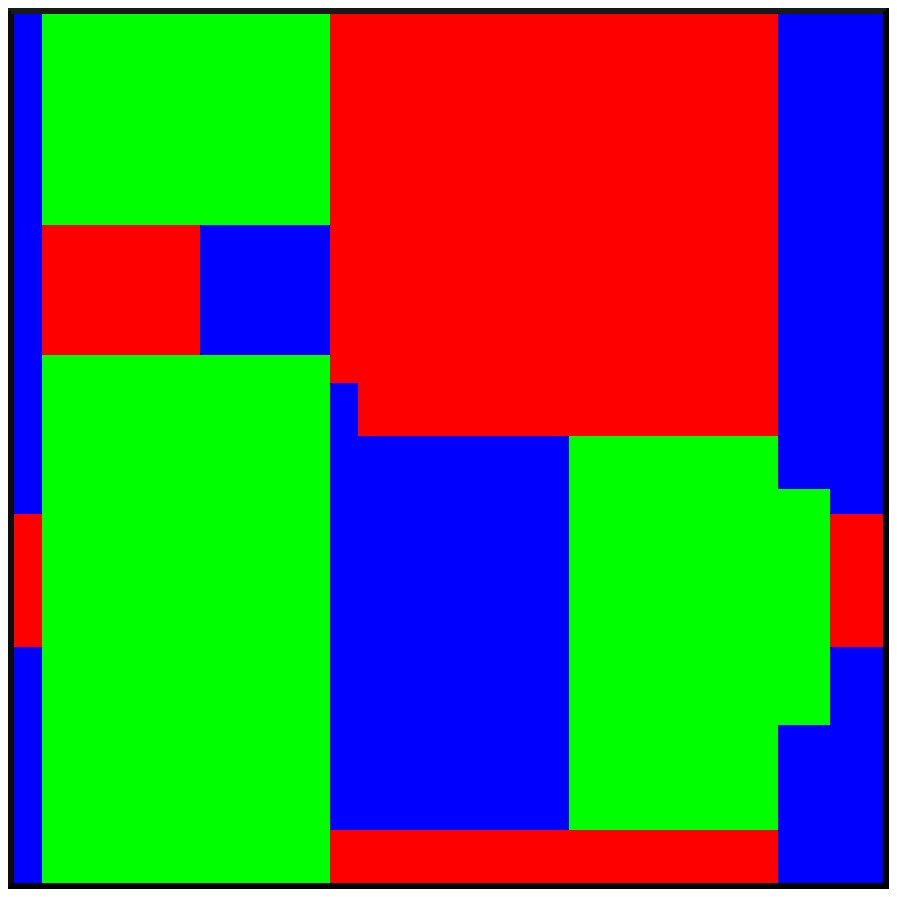}}\quad
\subfigure[]{\includegraphics[width=0.23\textwidth]{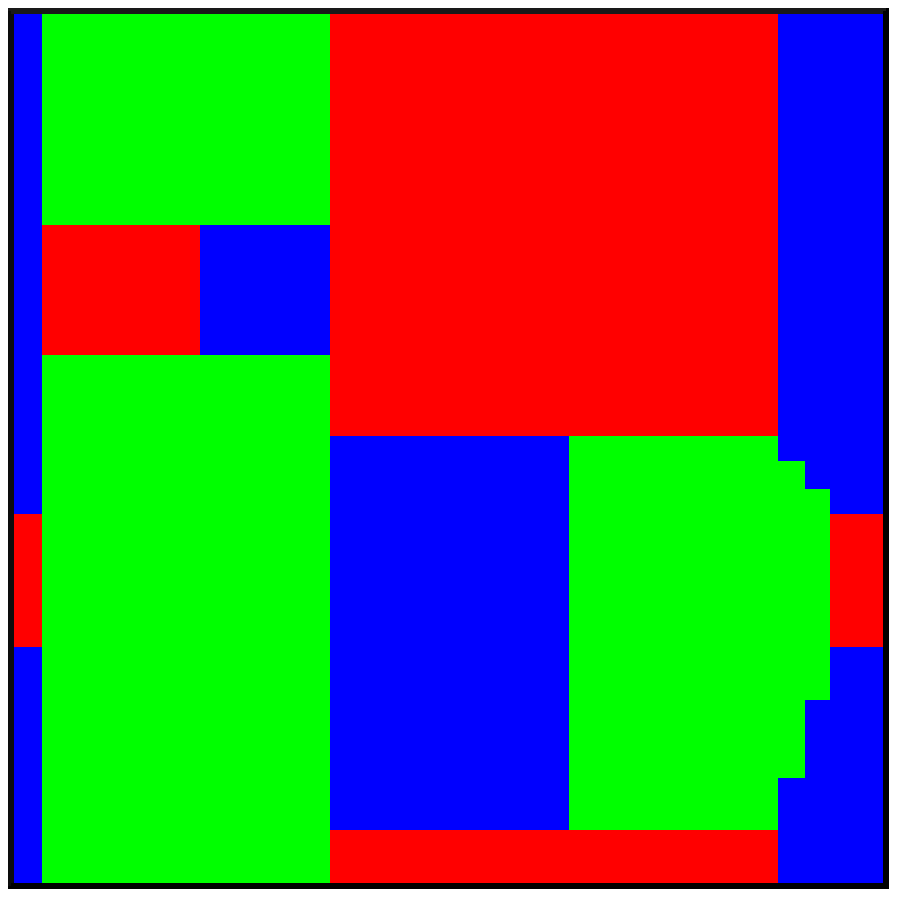}}\quad
\caption{Interface merging in the 3-state Potts model on a $33\times 33$
  square.  (a) Pseudo-blinker state with energy $E=188$, which is first
  reached at $t=251$.  (b) System at $t=16160676.9$ and (c) $t=16160677.0$,
  where the blinker interfaces first touch.  (d) At $t=16160677.7$, the
  merging is complete and the energy is $E=185$.  See~\cite{SM} for an
  animation of this evolution.}
\label{fig:pseudo}
  \end{center}
\end{figure}

This bias towards intermediate states has profound implications for the
long-time relaxation.  While a single isolated interface in a blinker state
can never decrease the energy of the system, interactions between two such
interfaces can give rise to energy-decreasing spin flips at long times
(Fig.~\ref{fig:pseudo}).  Here, each of the two separated blinker interfaces
tends to fluctuate around their respective intermediate states.  However, a
rare event can occur in which these two interfaces touch, leading to
irreversible energy-lowering spin flips that culminate in the red domain
(northeast) and the green domain (southwest) filling their convex envelopes.
The time required for this interface merging scales exponentially in the
length of the domain interface~\cite{OKR2011}, leading to slow relaxation.
We term configurations that consist of two blinker interfaces that eventually
merge at long times as \emph{pseudo-blinkers}.

\begin{figure}[ht]
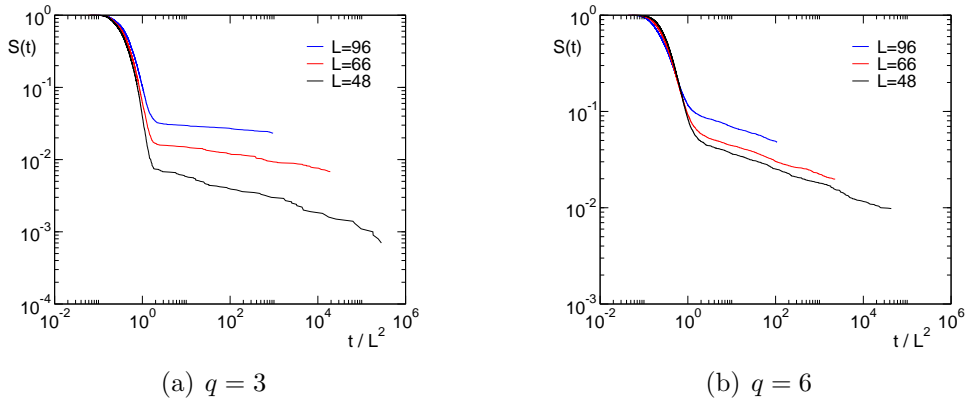

\begin{center}
\subfigure[][$q=3$]{\includegraphics*[width=0.35\textwidth]{figs/S3-scaled}}\qquad\qquad
\subfigure[][$q=6$]{\includegraphics*[width=0.35\textwidth]{figs/S6-scaled}}
\caption{Time dependence of the survival probability $S(t)$, defined as the
  probability that the $q$-state Potts system has not yet reached a
  fixed-energy configuration.  The data are based on $10^4$ realizations.}
\label{fig:S-vs-t}
  \end{center}
\end{figure}

To characterize the slow relaxation, which we argue originates with
pseudo-blinkers, we study the survival probability $S(t)$, the probability
that there are no possible energy-lowering spin flips remaining in the system
at time $t$ after the quench.  This quantity is not easy to determine because
the evolution at long times is characterized by increasingly-long periods of
stasis that are punctuated by very infrequent energy-lowering events
(Fig.~\ref{fig:pseudo}).  Because a direct measurement of $S(t)$ is
prohibitively time consuming, we ``look ahead'' in the simulation to
determine whether an energy-lowering spin flip could occur sometime in the
distant future for a configuration that contains multiple blinker spins.  Our
algorithm is similar to that used to determine the survival probability in
the 3d kinetic Ising model~\cite{OKR2011} and is the following:

\begin{enumerate}
\item Maintain lists of spins for which energy-lowering and energy-conserving
  flips may occur at the current update, $\mathcal{L}_-$ and $\mathcal{L}_0$,
  respectively.  When $\mathcal{L}_-$ becomes empty while $\mathcal{L}_0$ is
  non-empty, then the system may either:
\begin{enumerate}[(a)]
\item be in its lowest-energy state and wander \emph{ad infinitum} on this
  iso-energy set;
\item be in a finite-lifetime metastable state and a lower-energy state is
  reached when two blinker interfaces merge (as in Fig.~\ref{fig:pseudo}).
\end{enumerate}

\item When $\mathcal{L}_-$ first becomes empty while $\mathcal{L}_0$ remains
  non-empty, the configuration $\mathcal{C}_0$ and time $\mathcal{T}_0$ are
  saved.

\item Starting from $\mathcal{C}_0$, an infinitesimal magnetic field is
  imposed in the $-1$ direction:
  
\begin{equation}
  \mathcal{H}=\epsilon\sum_i \delta(\sigma_i,1)-\sum_{\langle
    ij\rangle}\delta(\sigma_i,\sigma_j)\qquad \epsilon\ll 1.
\end{equation}

  This Hamiltonian drives the state-space motion along the iso-energy surface
  by forcing blinker spins in state 1 to change to either state 2 or 3.
  If an energy-lowering move occurs during this driving, then the system is
  not in its lowest-energy state.  The system is returned to state
  $\mathcal{C}_0$ and is evolved with unbiased Glauber dynamics from time
  $\mathcal{T}_0$ until the next energy-lowering move occurs and then step
  (ii) is repeated.

\item If an energy-lowering spin flip does not occur in step (iii), then this
  step is repeated with a field applied in the $-2$ direction and then in the
  $-3$ direction.  If the energy does not decrease after application of all
  three fields, then $\mathcal{C}_0$ is the lowest-energy state of the system
  and its survival time is $\mathcal{T}_0$.
\end{enumerate}

Figure~\ref{fig:S-vs-t} shows the time dependence of the survival probability
for the 3-state and 6-state Potts model.  There are several intriguing and
yet-to-be-explained features in these plots.  First, in the early-time
coarsening regime, $S(t)$ shows good but not excellent data collapse when the
time is rescaled by the coarsening time.  Better collapse arises for the
early-time 3-state Potts data when the time is rescaled by $L^z$, with
$z\approx 2.3$.  For the 6-state Potts model, the small deviation from data
collapse cannot be removed by using a value for $z$ that is different than 2.
Second, relaxation at extraordinarily long times can occur.  For example, for
$q=3$ and $L=48$, there are energy-lowering spin-flip events a factor $10^5$
beyond the coarsening time!

\section{Avalanche Dynamics for Large $q$}

\begin{figure}[ht]
\begin{center}
\subfigure[][t=630000]{\includegraphics[width=0.188\textwidth]{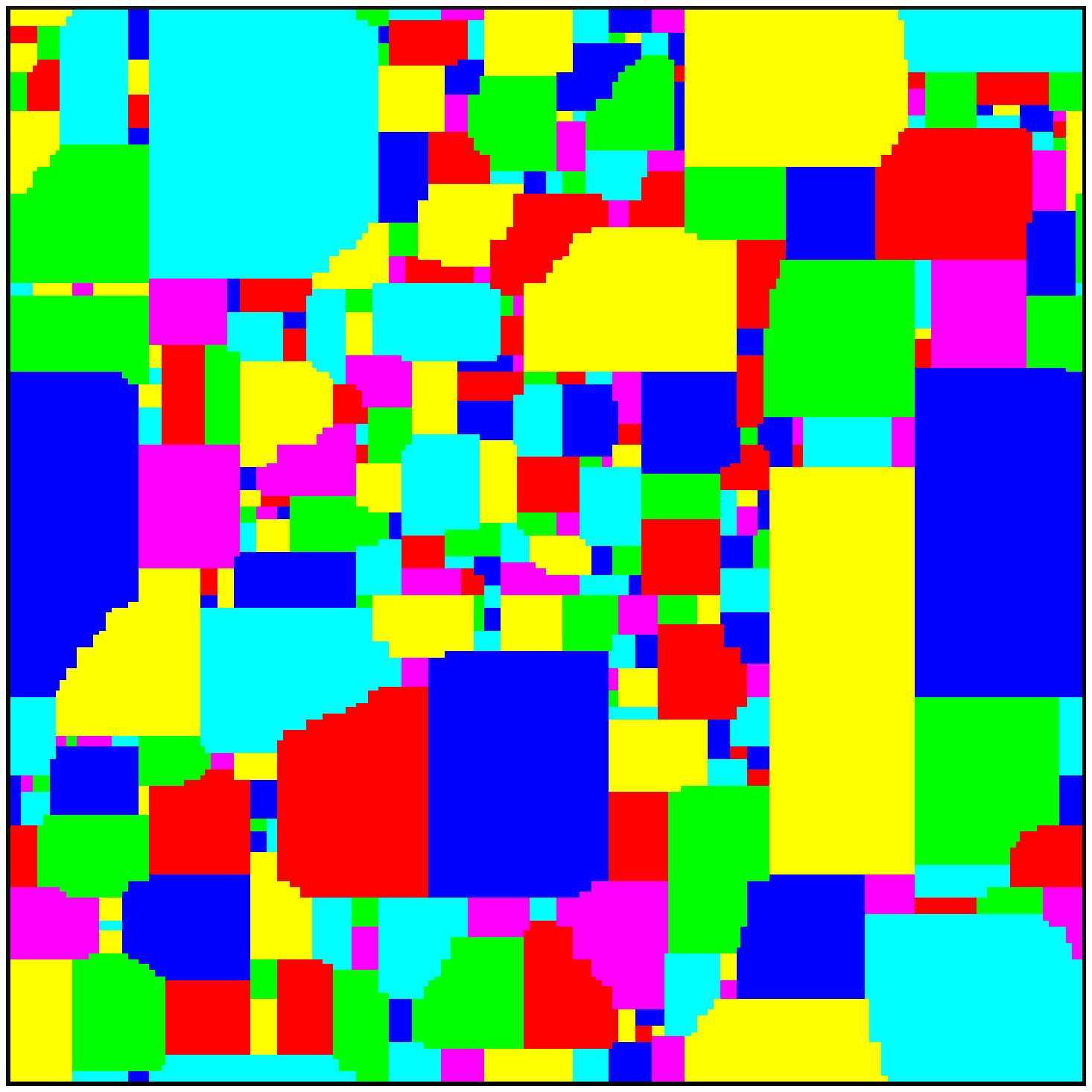}}\,
\subfigure[][t=633000]{\includegraphics[width=0.188\textwidth]{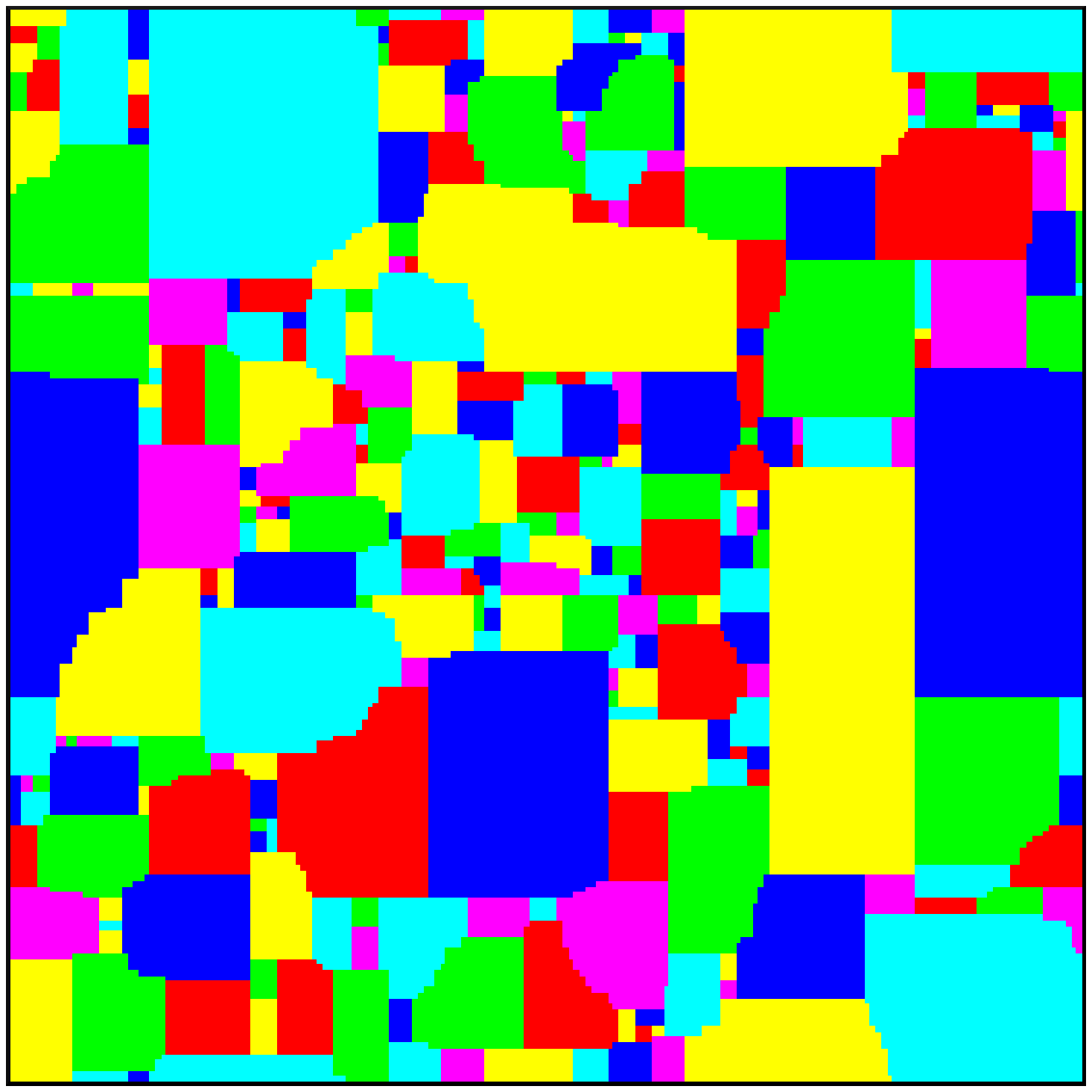}}\,
\subfigure[][t=633500]{\includegraphics[width=0.188\textwidth]{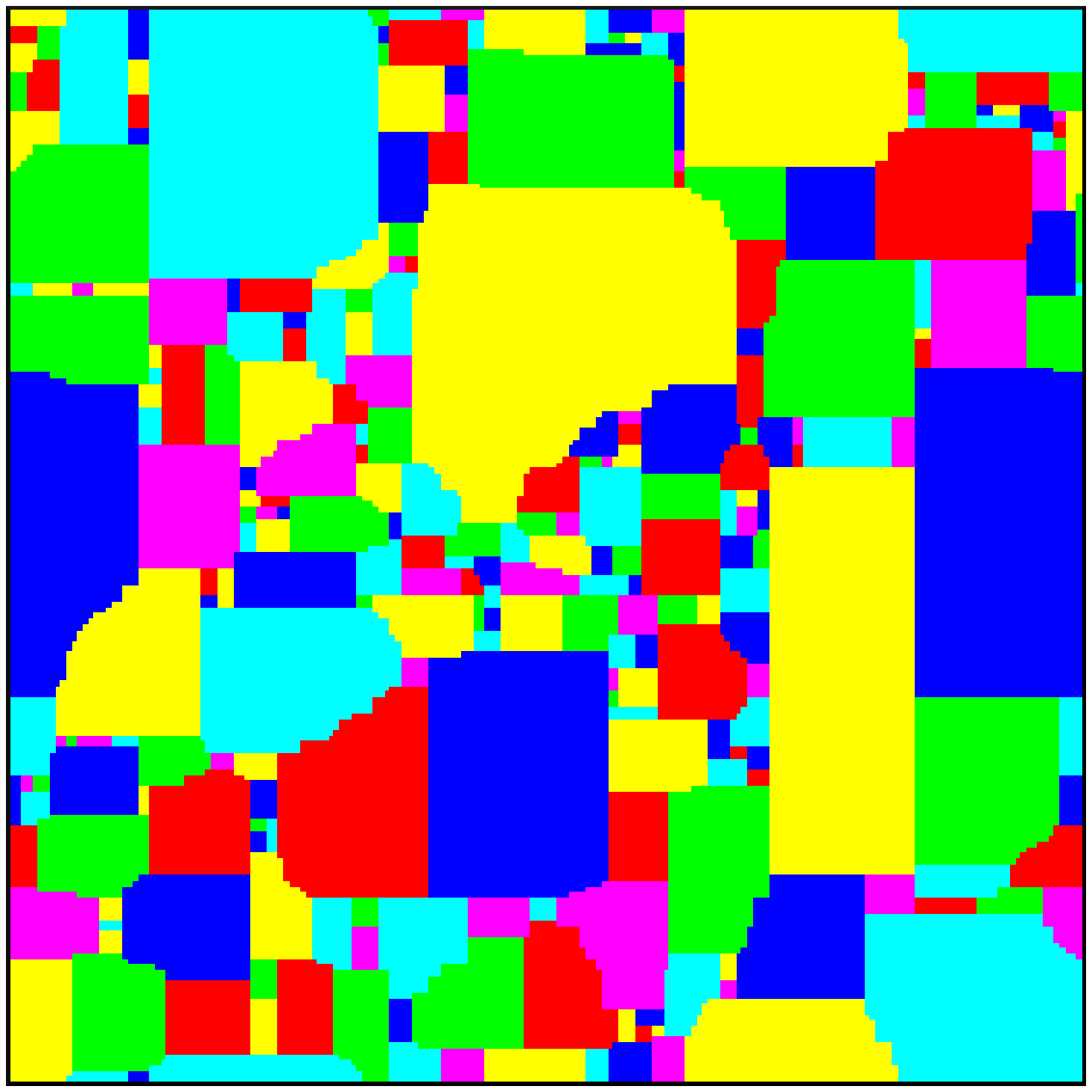}}\,
\subfigure[][t=634500]{\includegraphics[width=0.188\textwidth]{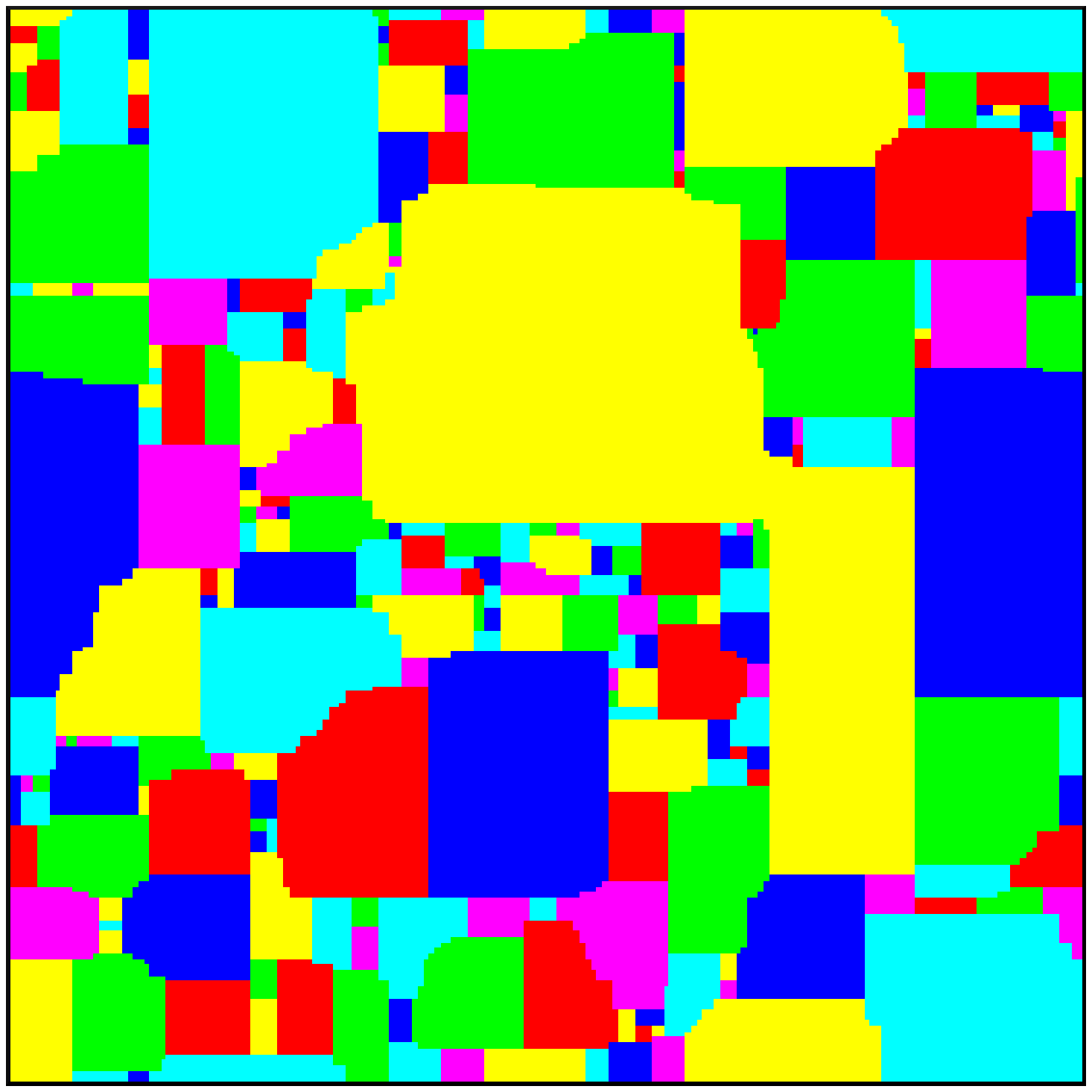}}\,
\subfigure[][t=640000]{\includegraphics[width=0.188\textwidth]{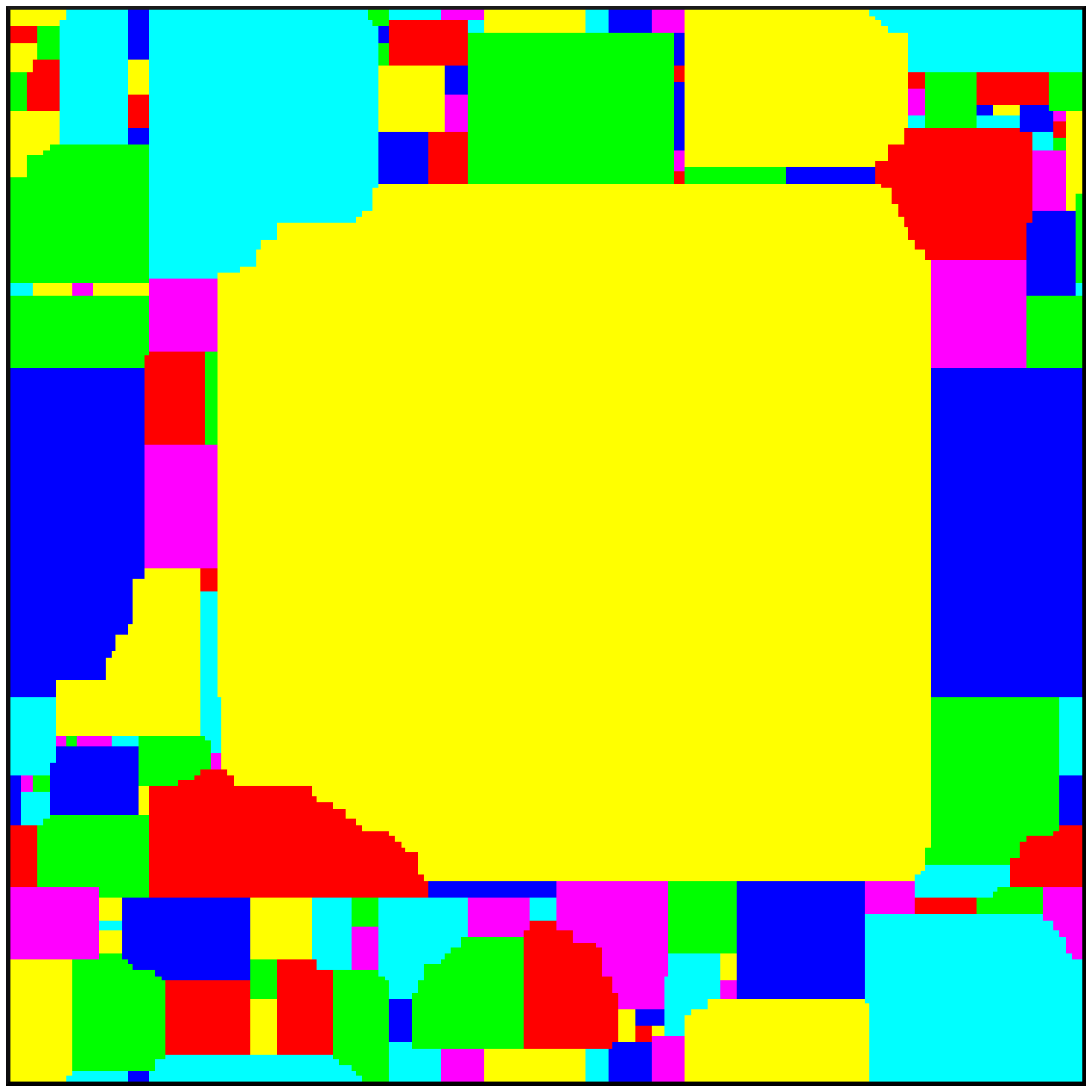}}
\caption{Cluster avalanche in the 6-state Potts model on a lattice of linear
  dimension $L=192$.  See~\cite{SM} for an animation of this evolution. }
\label{fig:av}
  \end{center}
\end{figure}

For $q=3$ pseudo-blinkers are mostly benign, as configurations that contain
pseudo-blinkers typically will undergo several isolated but small decreases
in energy at long times and then get trapped in either a frozen state or a
true blinker state.  For larger $q$, however, macroscopic cluster
rearrangements, or ``avalanches'', can occur, in which clusters can fill
their convex envelopes, leading to successive cluster mergings
(Fig.~\ref{fig:av}).  We shall argue that the mechanism driving cluster
avalanches in the Potts model for large $q$ is an enhanced version of this
filling out of the convex cluster envelopes.

Figure~\ref{fig:av} shows a typical avalanche in the $6$-state Potts model.
In (b), two small yellow clusters merge and this composite cluster begins to
fill its convex envelope.  The resulting cluster continues to grow by
additional mergings and convex envelope filling (c), and then merges with a
large cluster to the southeast (d).  The merged cluster ultimately fills its
convex envelope at $t=640000$ where the lowest-energy state has been reached
(e).  The avalanche occurs at $t\approx 633000$, which is an order of
magnitude greater than the coarsening time.  This avalanche also leads to a
macroscopic decrease of nearly $50\%$ in the energy---from 5051 to 2849!

\begin{figure}[ht]
\begin{center}
\subfigure[][t=634396.5]{\includegraphics[width=0.2\textwidth]{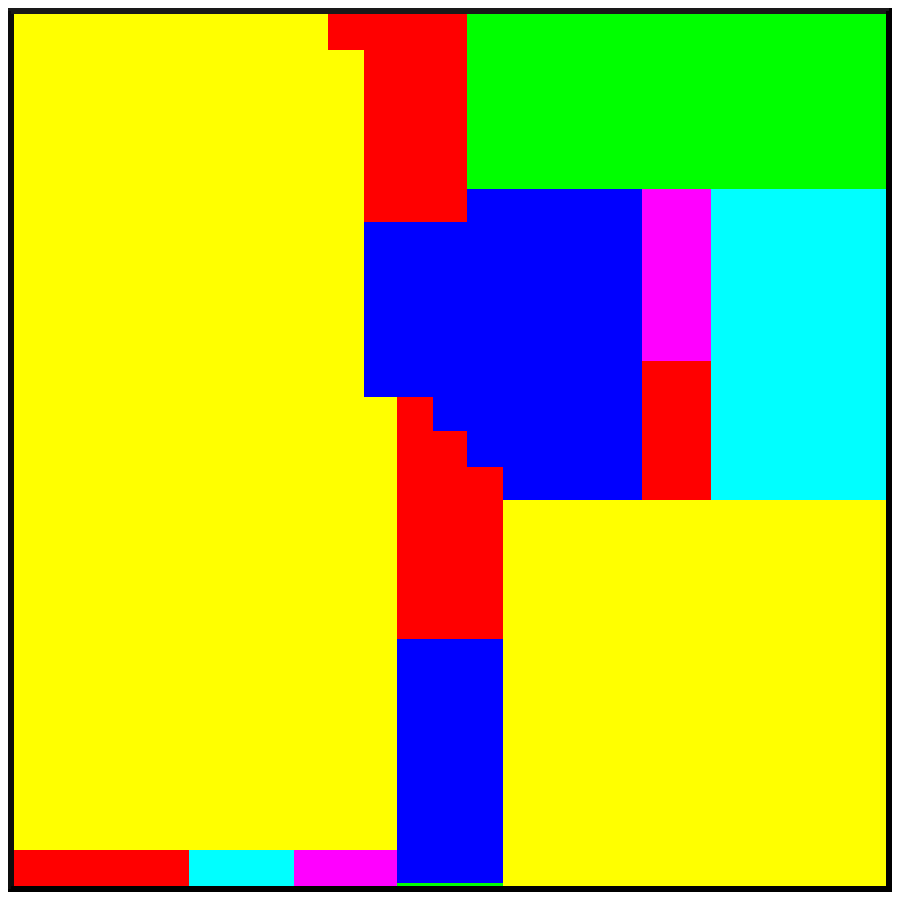}}\quad
\subfigure[][t=634396.8]{\includegraphics[width=0.2\textwidth]{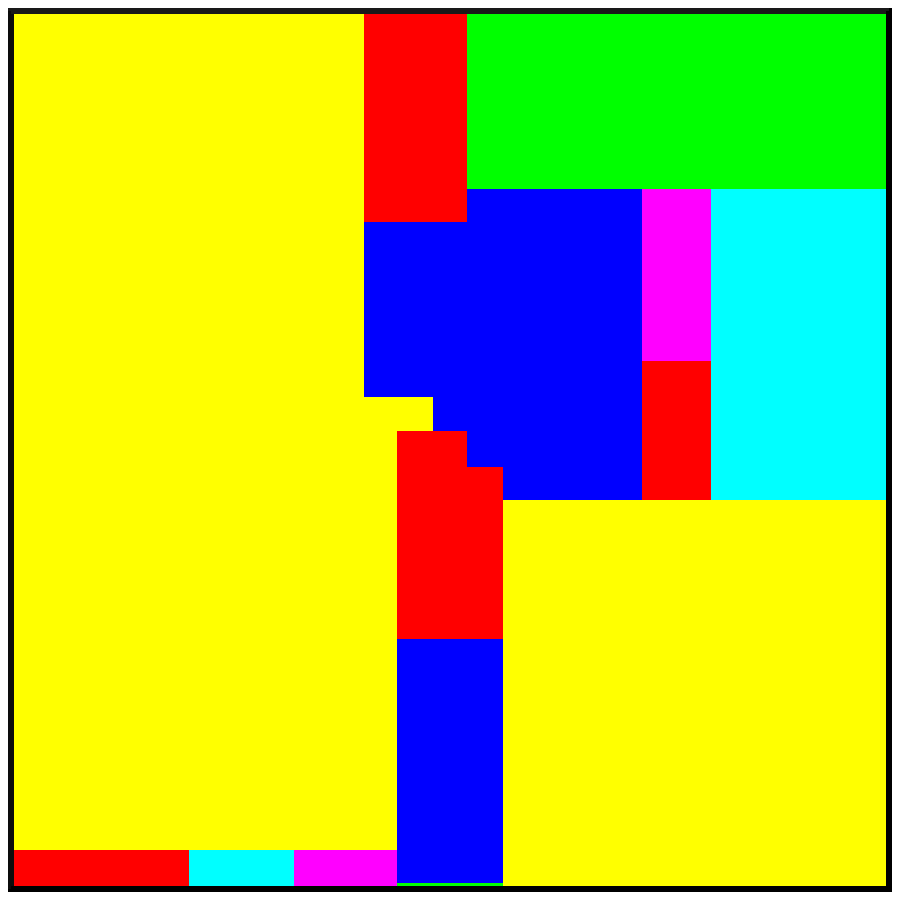}}\quad
\subfigure[][t=634455.0]{\includegraphics[width=0.2\textwidth]{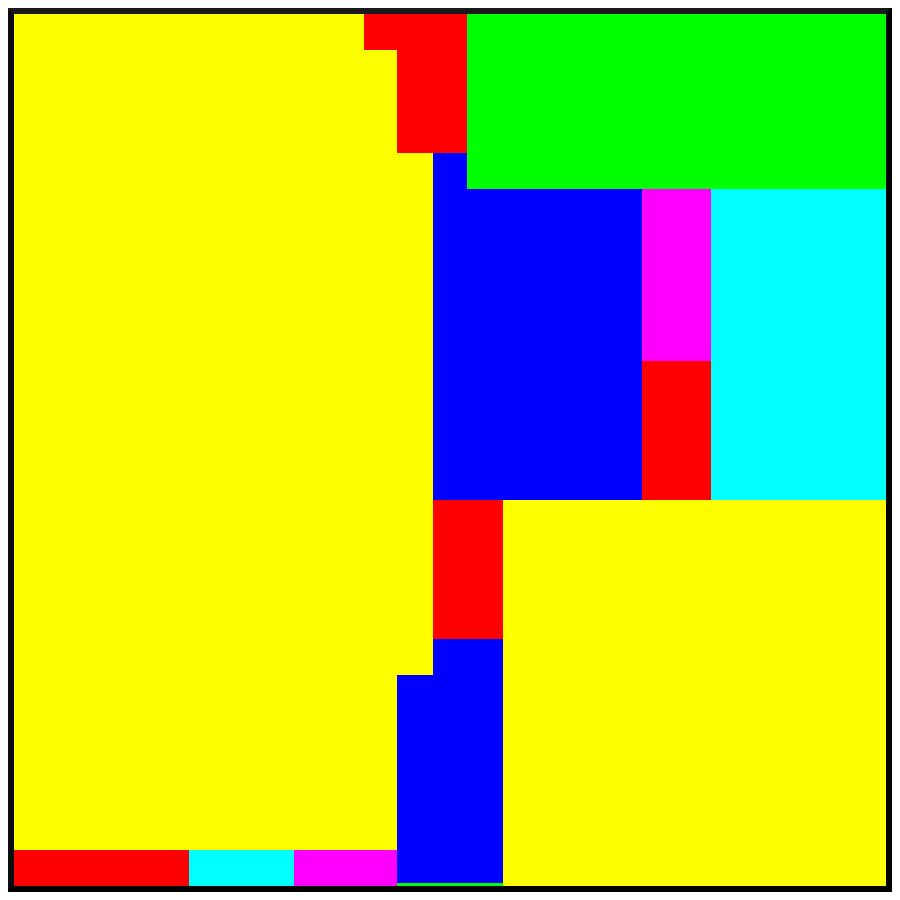}}\quad
\subfigure[][t=634459.4]{\includegraphics[width=0.2\textwidth]{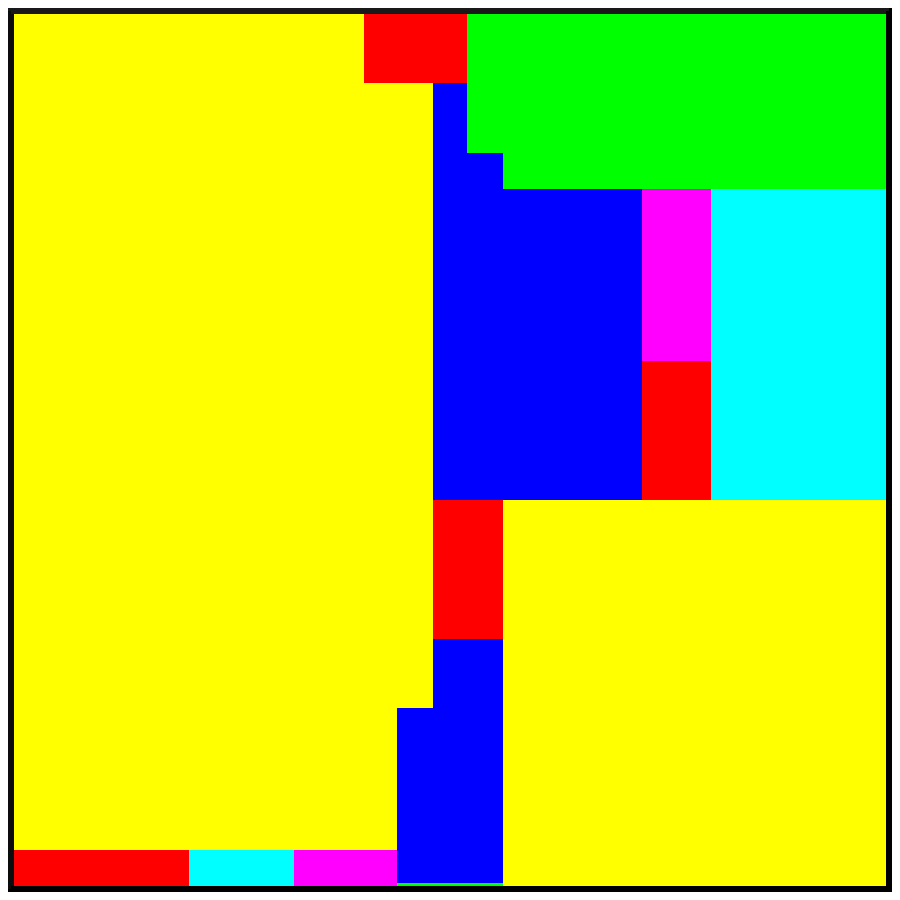}}
\vskip 0.05in
\subfigure[][t=634459.7]{\includegraphics[width=0.2\textwidth]{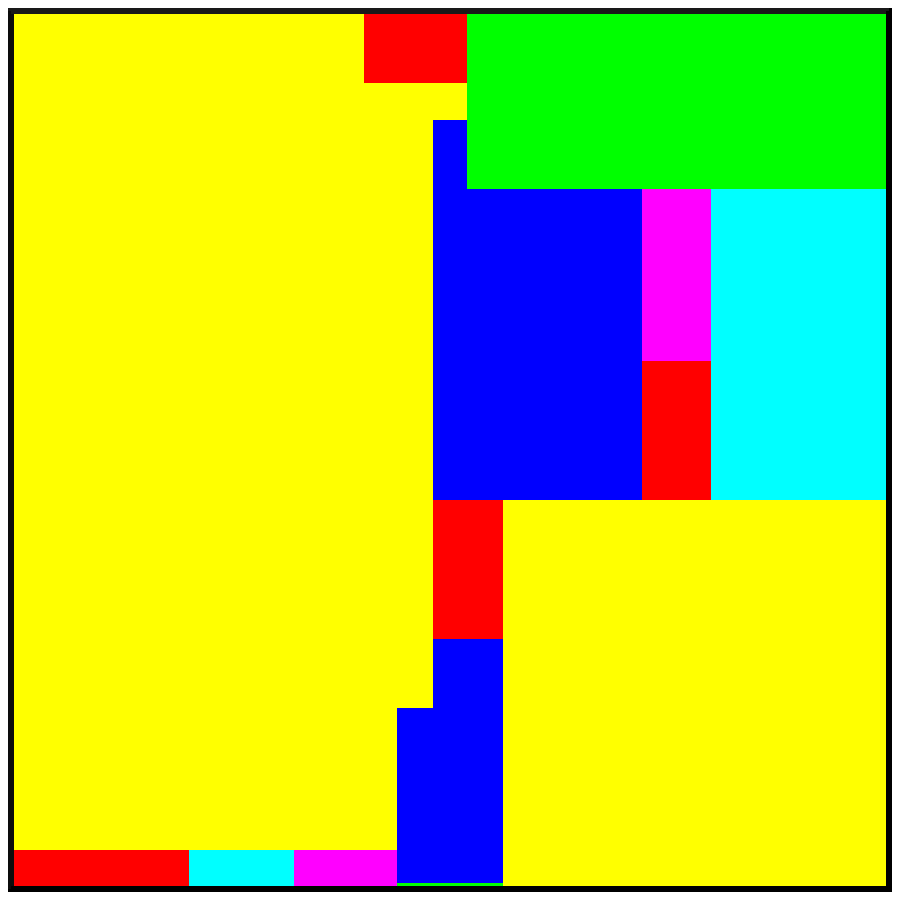}}\quad
\subfigure[][t=634471.5]{\includegraphics[width=0.2\textwidth]{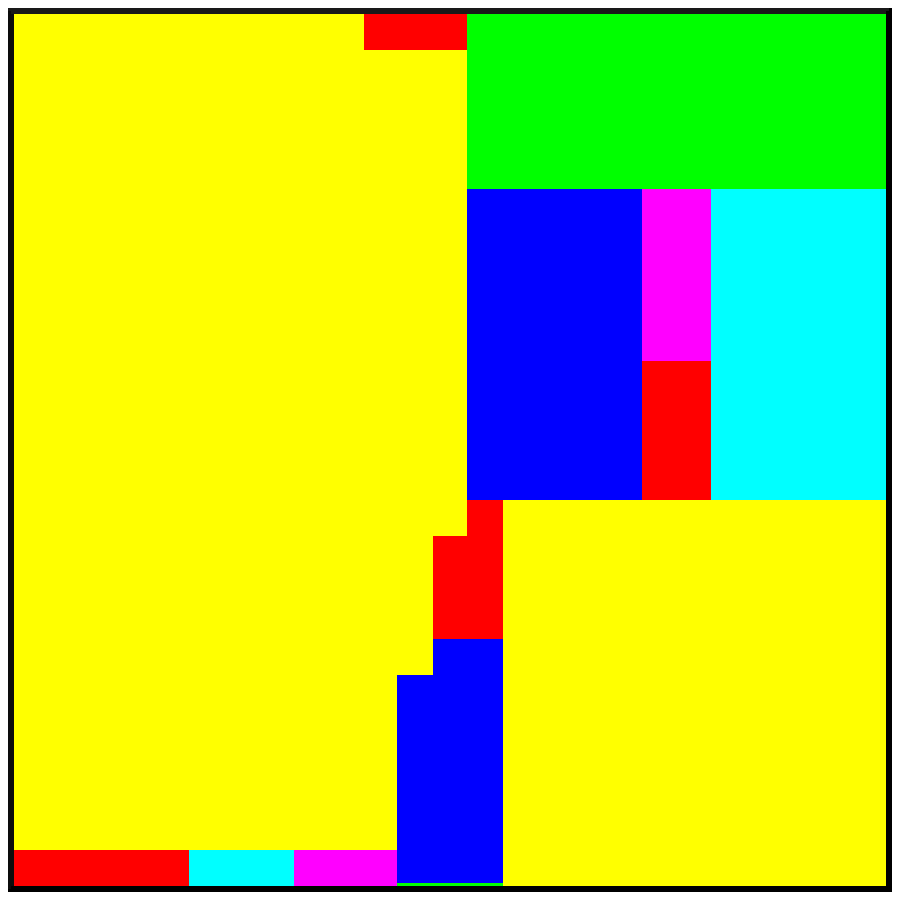}}\quad
\subfigure[][t=634471.7]{\includegraphics[width=0.2\textwidth]{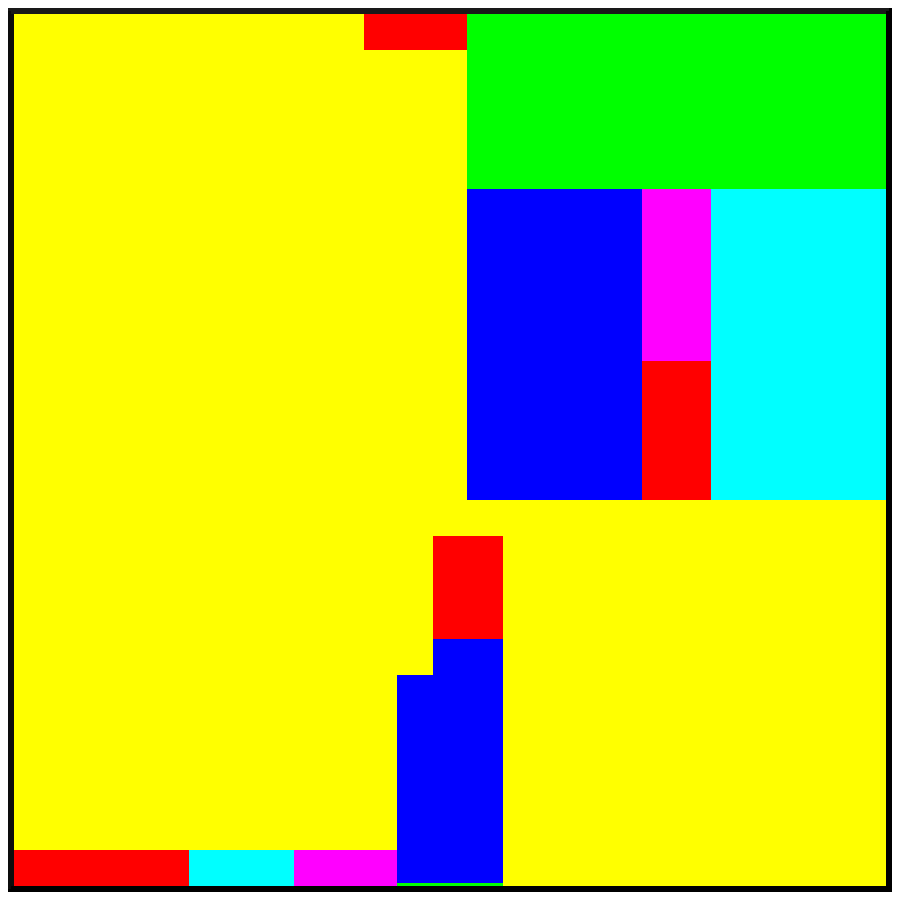}}\quad
\subfigure[][t=634490.0]{\includegraphics[width=0.2\textwidth]{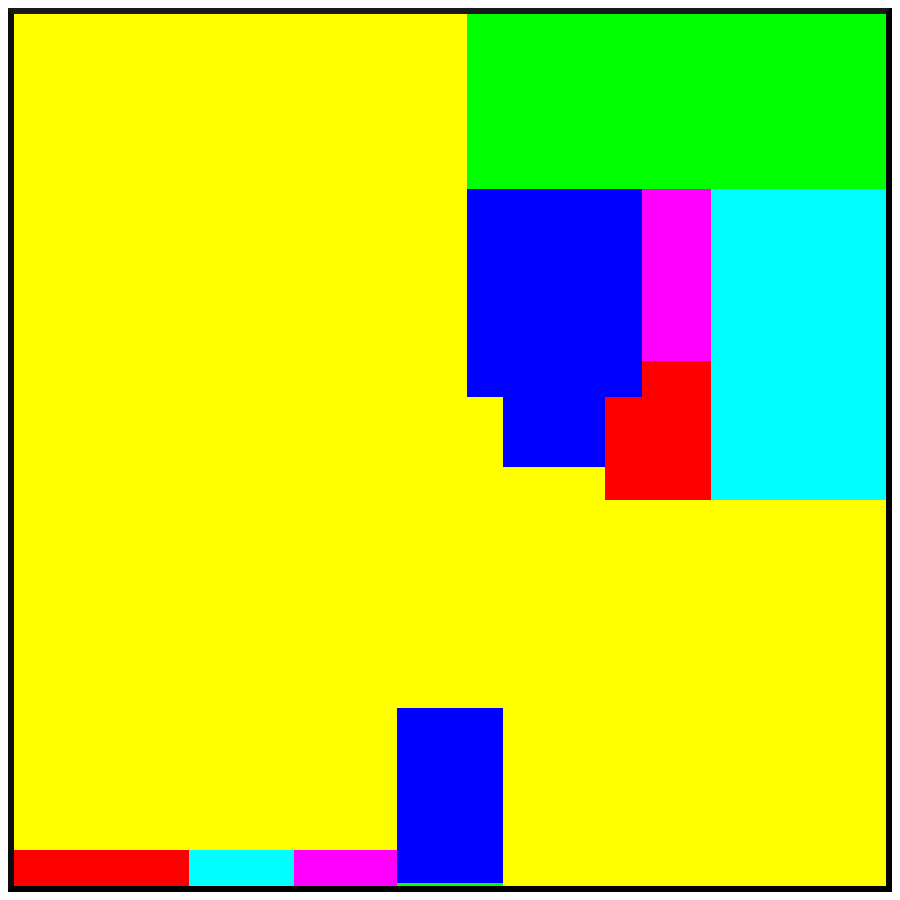}}
\caption{A zoom of Fig.~\ref{fig:av} showing the event sequence that merges
  the two large yellow clusters in \ref{fig:av}(c).  In (b) and in (e), the
  convex envelope of the yellow cluster moves one unit to the right.  In (g),
  a single spin flip merges the two yellow clusters.}
\label{fig:zoom}
  \end{center}
\end{figure}

\begin{figure}[ht]
\begin{center}
\includegraphics[width=0.6\textwidth]{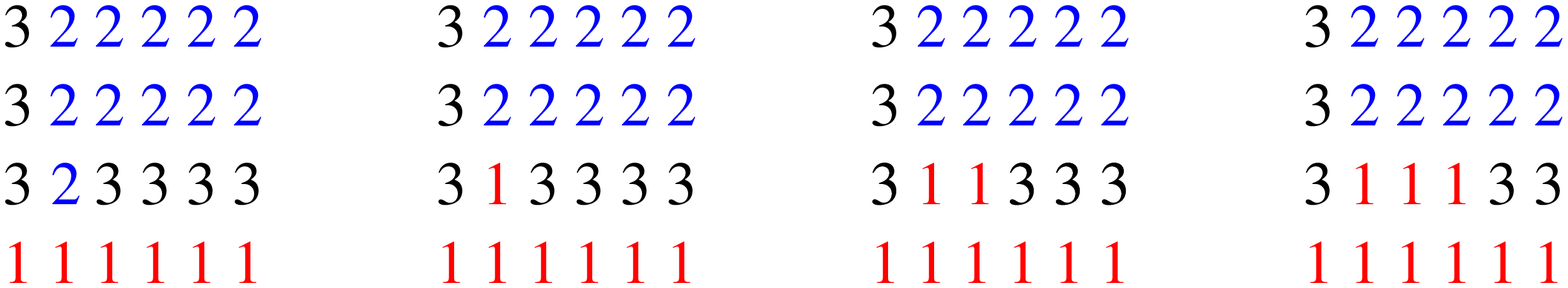}
\caption{Illustration of spin flip events that allow a protrusion of spins in
  the 1 state to form and grow, ultimately leading to the lateral translation
  of a straight interface.}
\label{protrusion}
  \end{center}
\end{figure}

However, if the filling of convex envelopes is the driving mechanism for
avalanches, then the system would never evolve from Fig.~\ref{fig:av}(c) to
\ref{fig:av}(d) because the convex envelopes of the two large yellow clusters
in \ref{fig:av}(c) do not overlap.  These two clusters do merge because
spin-flip events occur that allow a single cluster to expand beyond its
convex envelope.  The sequence of events that illustrate this domain
expansion and subsequent cluster merging are shown in Figure~\ref{fig:zoom}.
The microscopic events that underlie this expansion are illustrated in
Fig.~\ref{protrusion}.  When the number of spin states is three or greater,
then spin-flip events allow a protrusion to form on a straight domain
boundary.  This protrusion can grow and ultimately cause a straight boundary
to translate by one lattice spacing laterally.  When the number of spin
states is greater than or equal to the lattice coordination number, then
there can be spins on domain boundaries that can flip freely to \emph{any}
state, which facilitates the process of domain expansion.

\begin{figure}[ht]
\begin{center}
\subfigure[]{\includegraphics[width=0.22\textwidth]{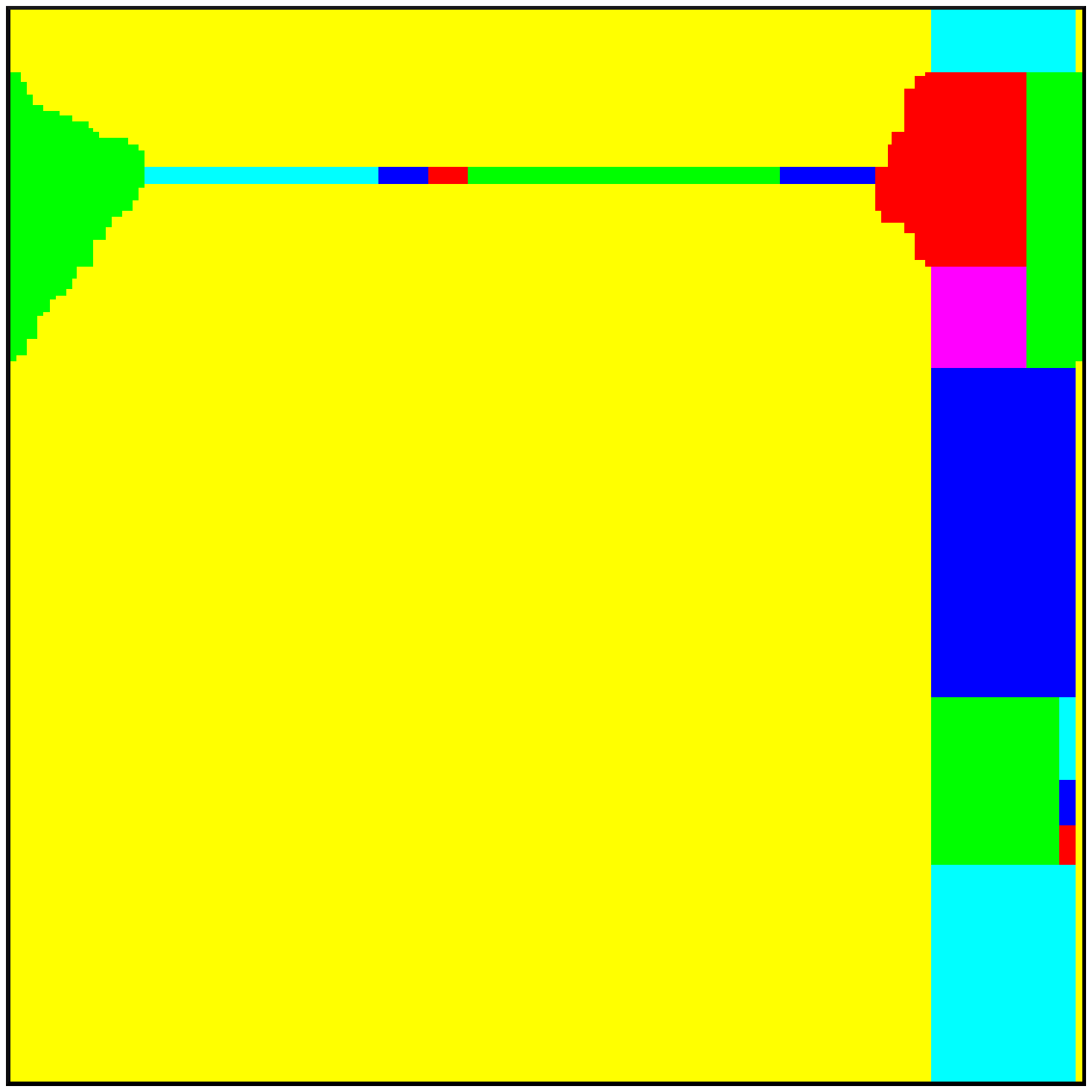}}\quad
\subfigure[]{\includegraphics[width=0.22\textwidth]{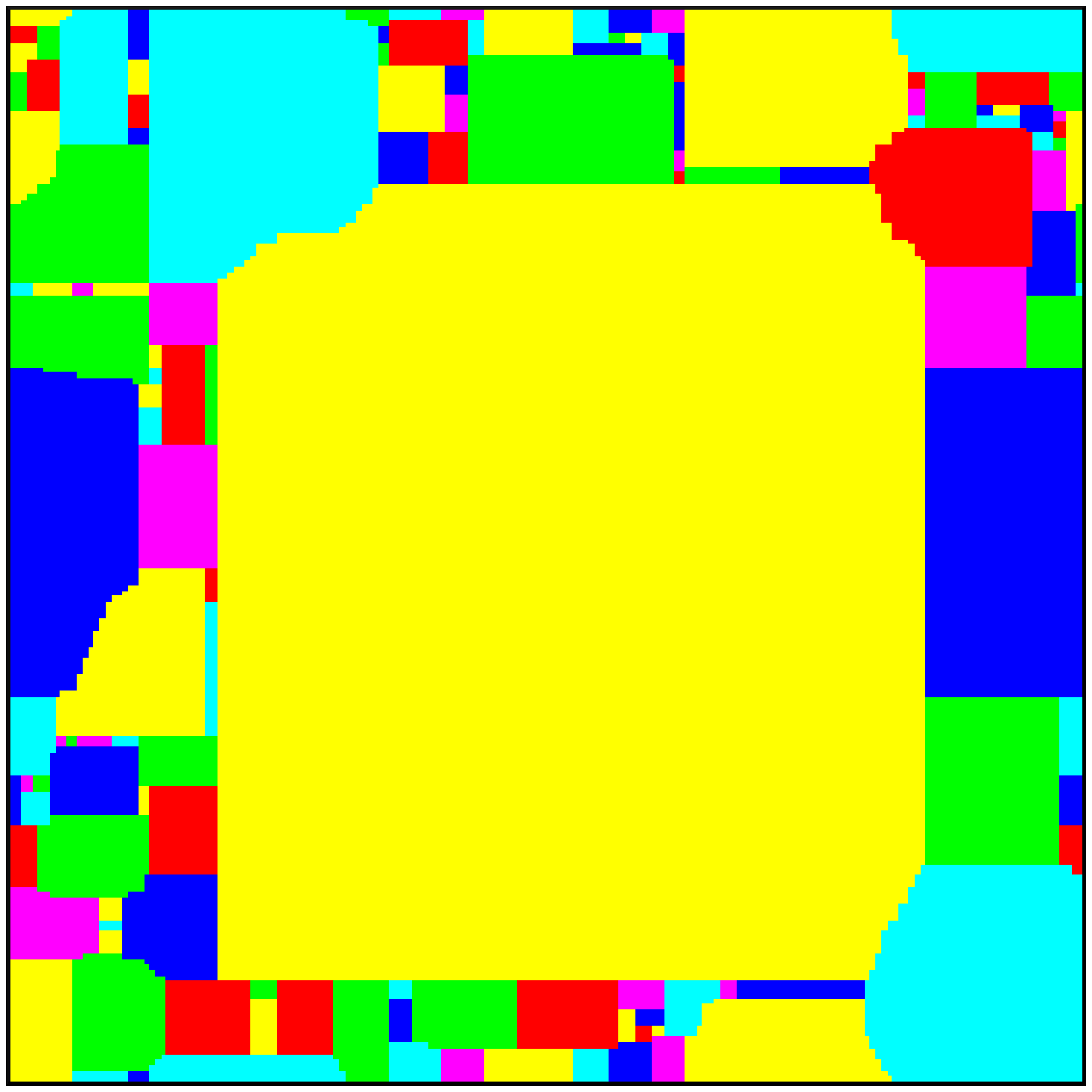}}\quad
\subfigure[]{\includegraphics[width=0.22\textwidth]{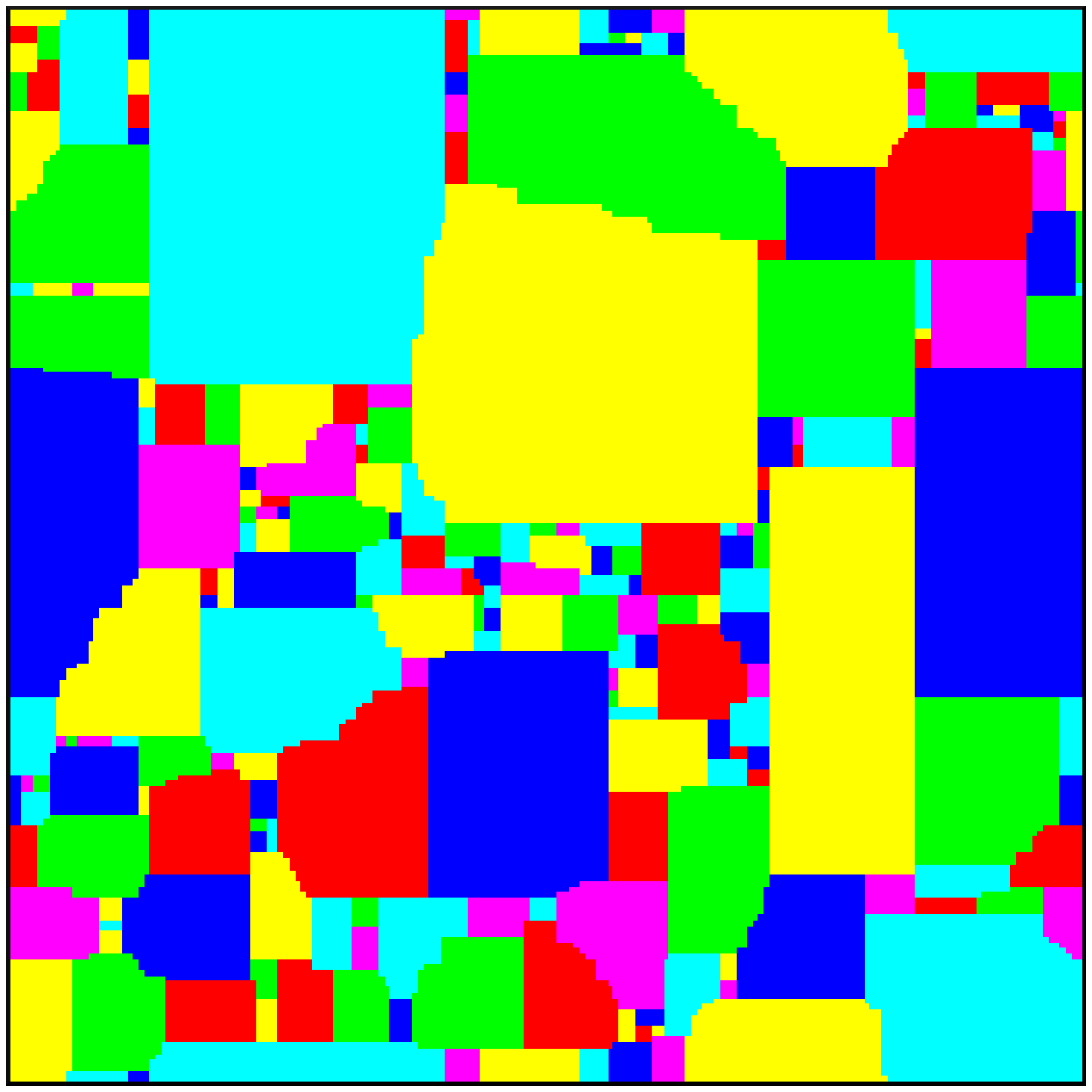}}\quad
\subfigure[]{\includegraphics[width=0.22\textwidth]{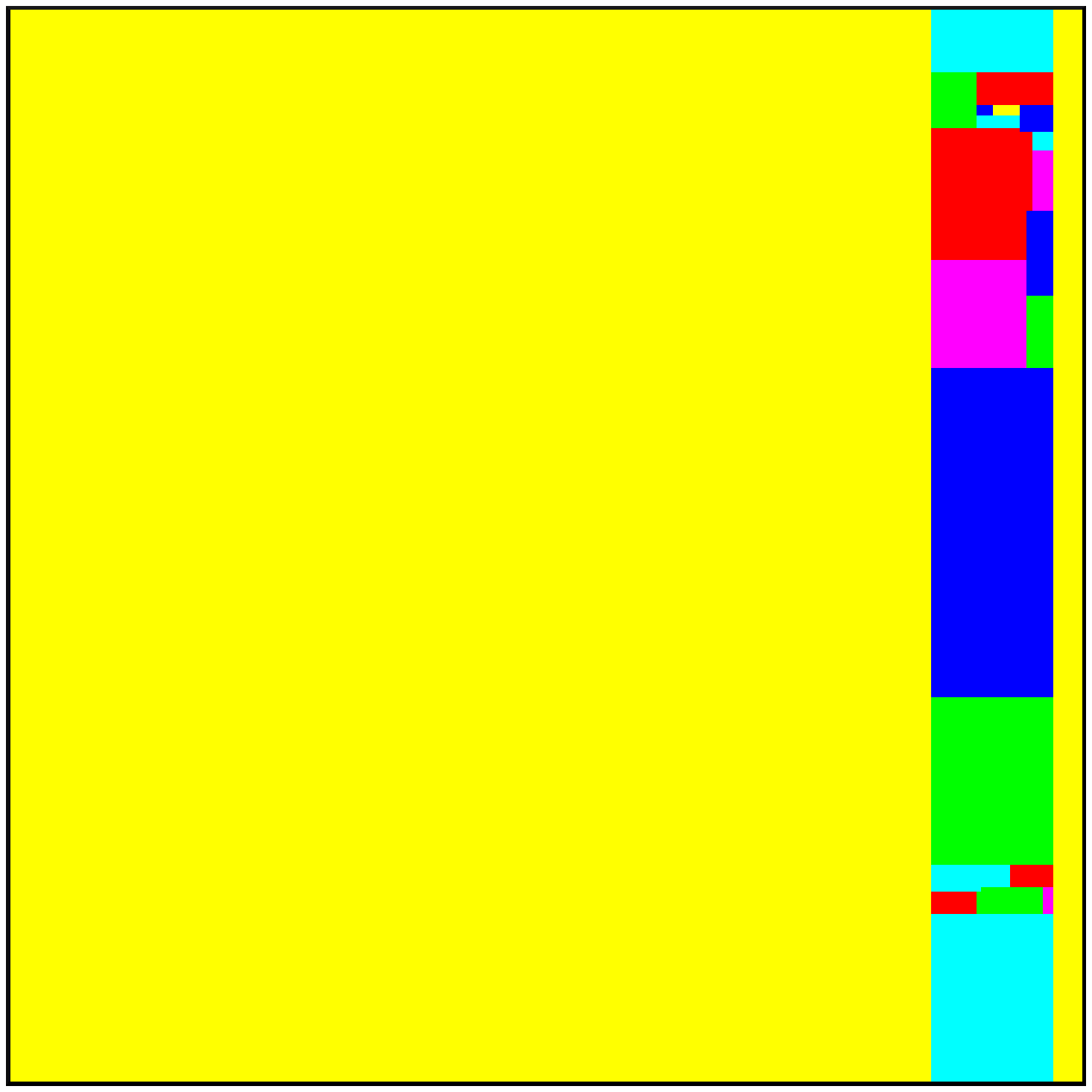}}
\caption{Four long-time states when starting with the configuration of
  Fig.~\ref{fig:av}(b).}
\label{fig:av_diff}
  \end{center}
\end{figure}

These long-time avalanches also evolve unpredictably.  To illustrate this
point, we started with the configuration in Fig.~\ref{fig:av}(b) (immediately
after the two pseudo-blinkers merge) and we simulated $10^4$ independent
realizations of the dynamics.  The final state energies of this subset of
realizations range between $E=490$ and $E=4383$, with $\langle
E\rangle=2827$.  Four representative such final states, with final energies
$E=938$, $2547$, $4103$, and $664$ (a-d) are shown in Fig.~\ref{fig:av_diff}.
The salient point is that avalanching involves an unpredictable sequence of
events, so that it is not possible to infer the final state even quite late
in the evolution.

\begin{figure}[ht]
\begin{center}
\subfigure[][t=30]{\includegraphics[width=0.22\textwidth]{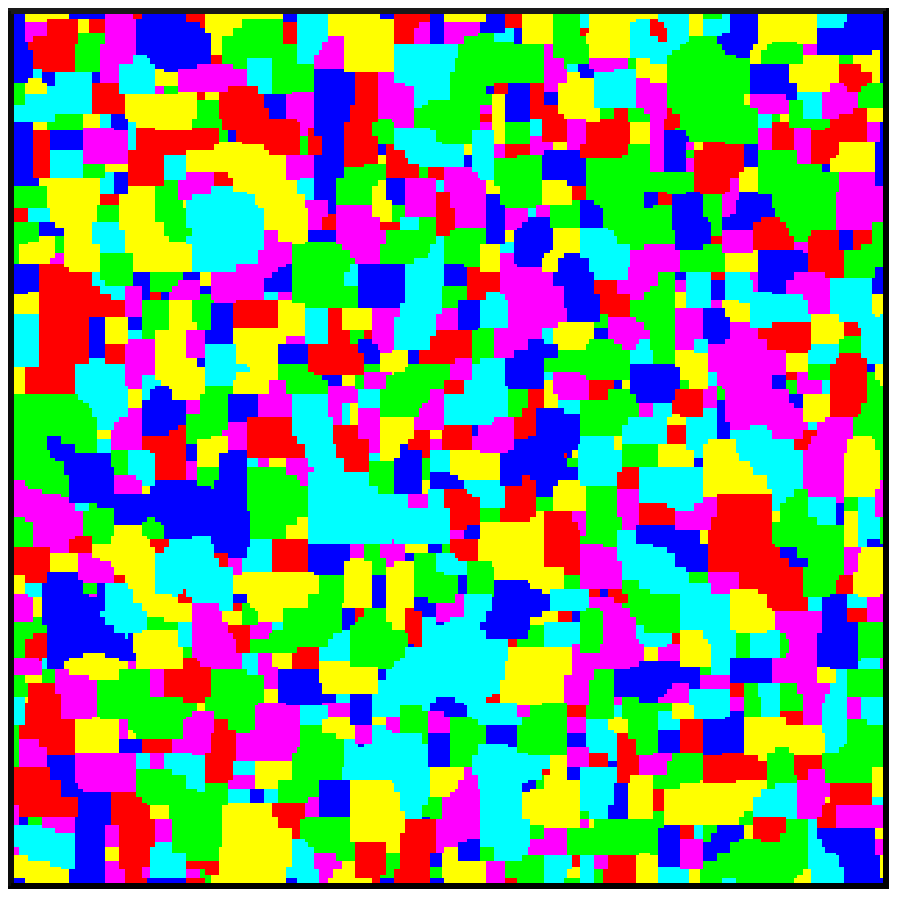}}\quad
\subfigure[][t=100]{\includegraphics[width=0.22\textwidth]{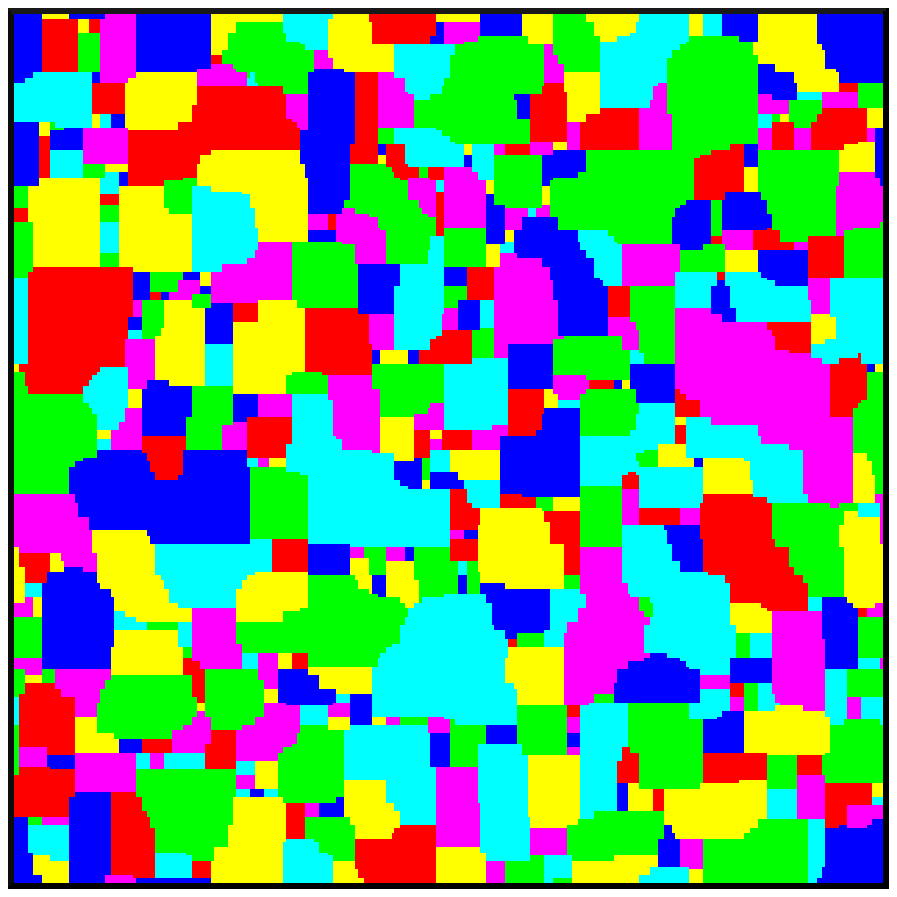}}\quad
\subfigure[][t=1000]{\includegraphics[width=0.22\textwidth]{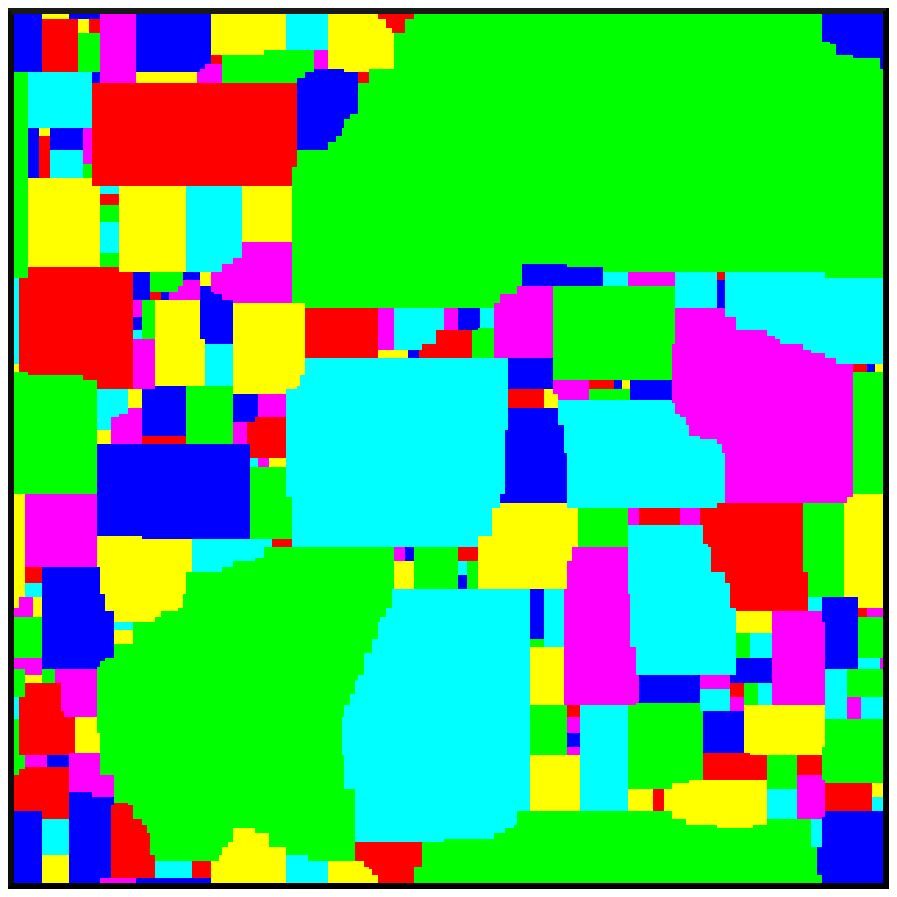}}\quad
\subfigure[][t=7694]{\includegraphics[width=0.22\textwidth]{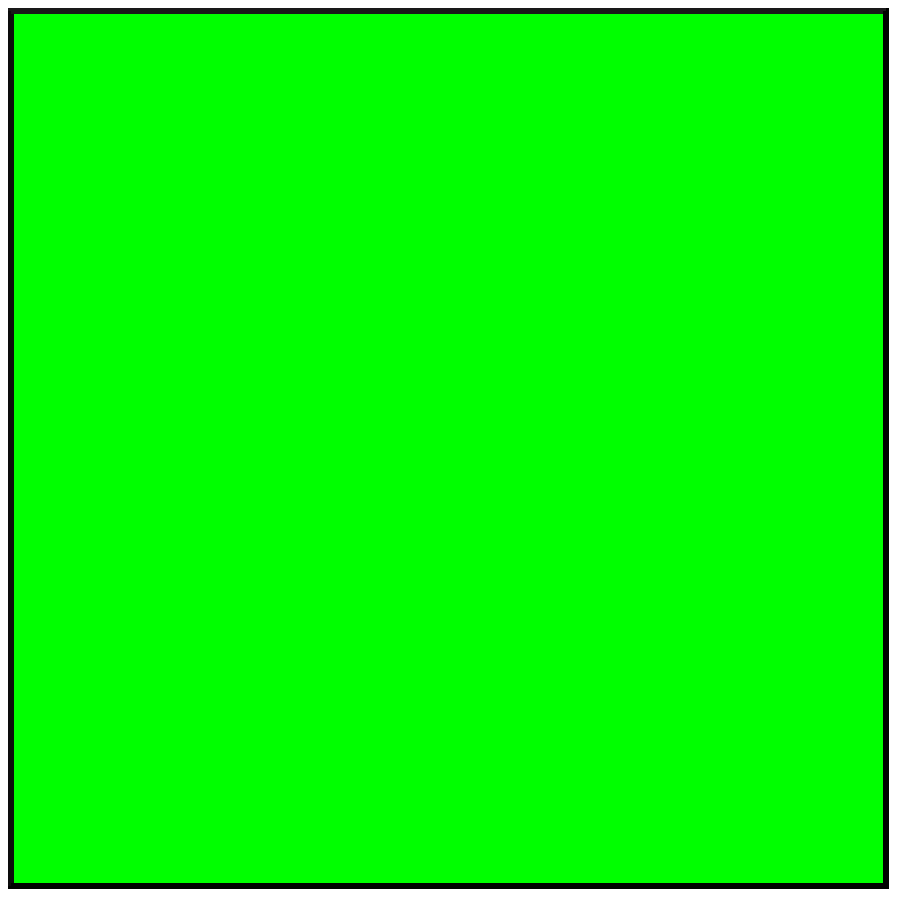}}
\vskip 0.05in
\subfigure[][t=100]{\includegraphics[width=0.22\textwidth]{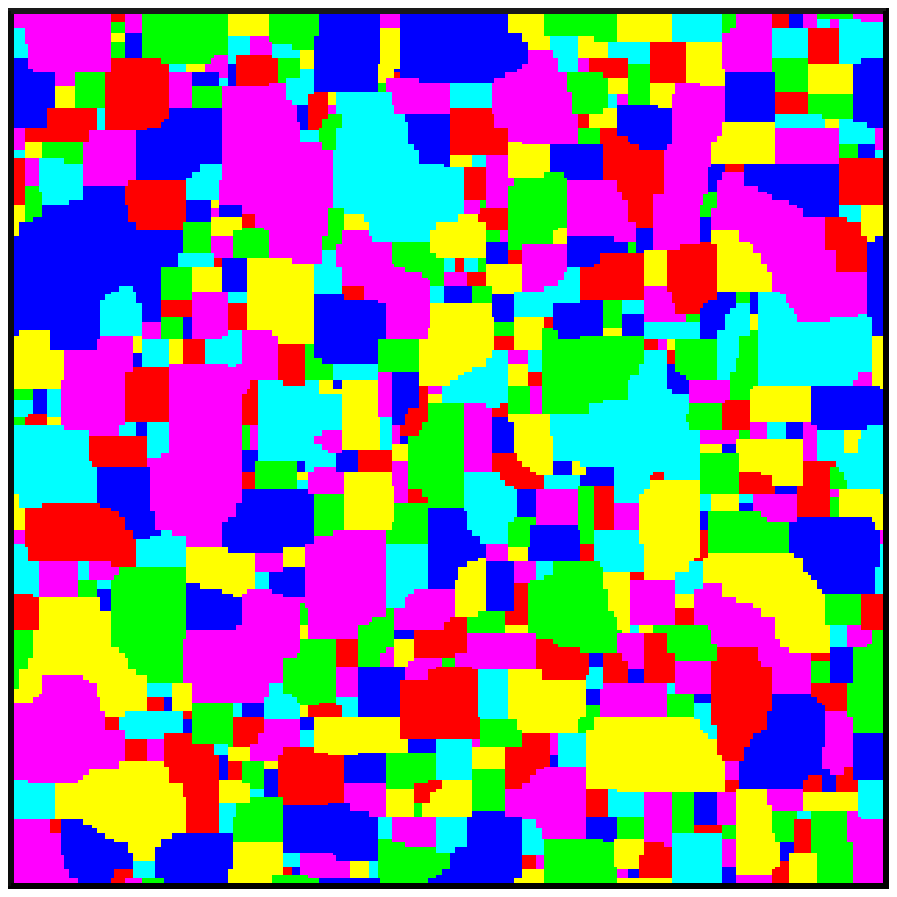}}\quad
\subfigure[][t=330]{\includegraphics[width=0.22\textwidth]{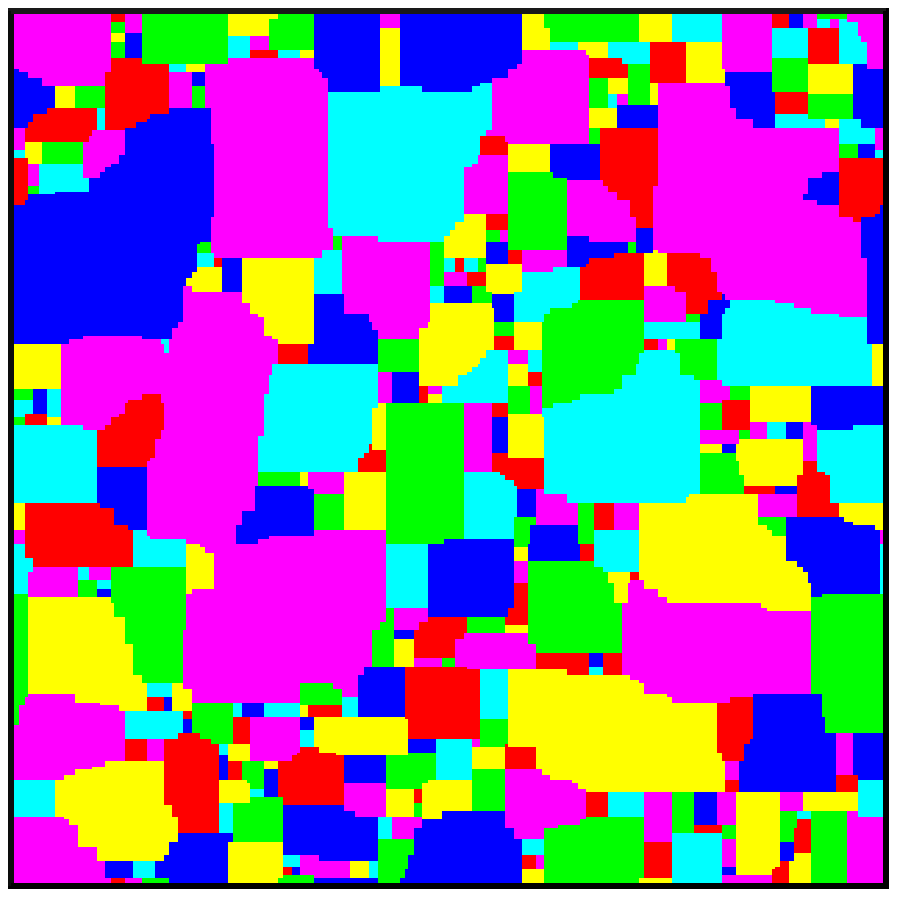}}\quad
\subfigure[][t=1000]{\includegraphics[width=0.22\textwidth]{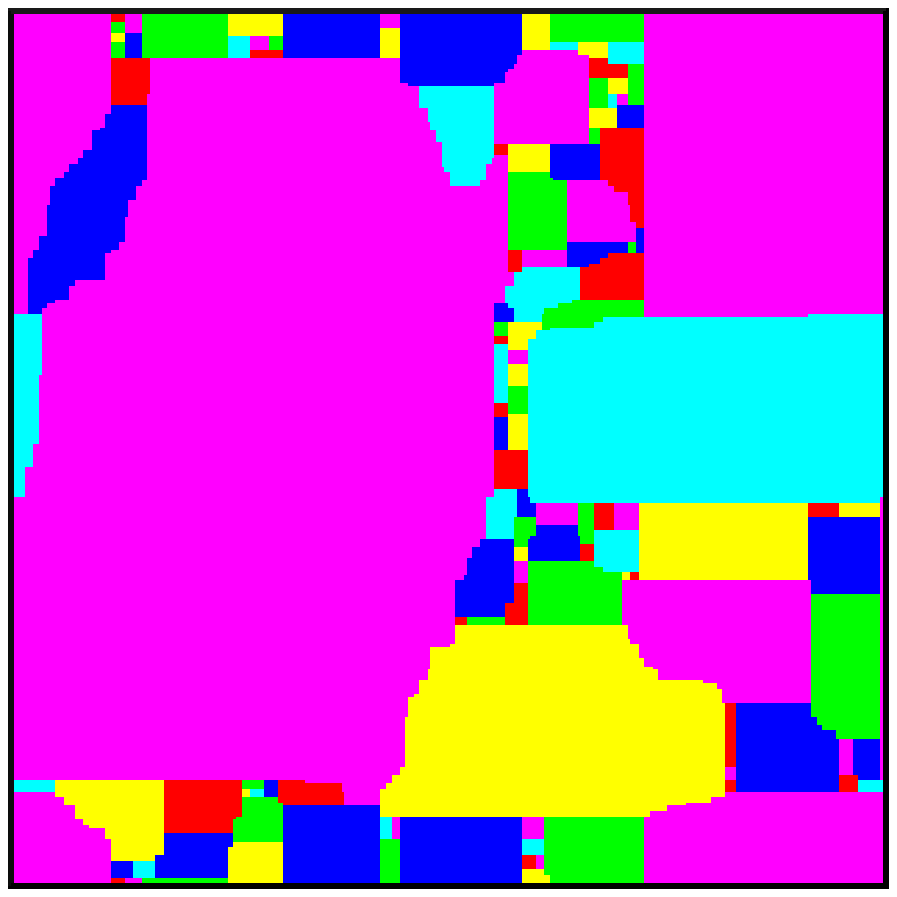}}\quad
\subfigure[][t=6784]{\includegraphics[width=0.22\textwidth]{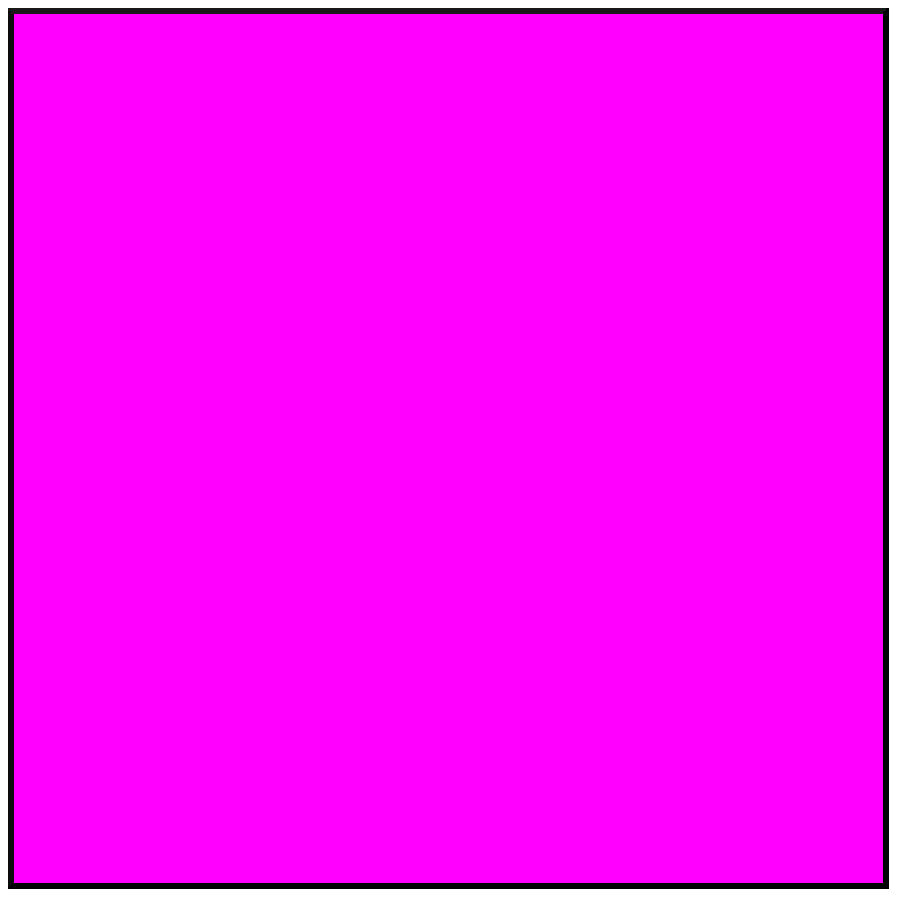}}
\caption{Two examples of a 6-state Potts model on a square lattice of linear
  dimension $L=192$, with equal initial densities of the six states, that
  rapidly evolves to one of the ground states.  See~\cite{SM} for an
  animation of this evolution in the top line.}
\label{fig:quick}
  \end{center}
\end{figure}

Finally, we also find the striking feature that in a small fraction of all
realizations (e.g., with probability approximately 5\% for $q=6$ and $L=384$)
these avalanches can continue to completion, so that the system ultimately
reaches the ground state (Fig.~\ref{fig:quick}).  In this subset of
realizations, expansion of clusters to fill their convex envelopes leads to
merging of clusters and a cascading sequence of merging and growth that
terminates when the ground state is reached. It is striking that these
macroscopic avalanches become more common as the number of Potts states is
increased and that there exist both realization that reach the ground state
and realizations that get stuck in states where the typical cluster size is
quite small.

\section{Discussion}

We studied the following basic problem: how does a system with many
equivalent ground states evolve when suddenly quenched from a disordered
state to zero temperature, and what are the properties of the long-time
states?  In the simpler case of the kinetic Ising model, the ground state is
only sometimes reached in $d=2$ and is never reached for $d \geq 3$.  The
feature that prevents the ground state from being reached in the
three-dimensional Ising case is that the initial state consists of two highly
intertwined spin-up and spin-down domains that cannot disentangle by local
spin-flip dynamics.  Partly because of this entanglement of domains in the
initial state, the evolution is anomalously slow, with the characteristic
time scale varying exponentially in the system size.

In this work, we extended the consideration of this basic question to the
fate of the 2d kinetic $q$-state Potts model that is initially in a
disordered state and is suddenly quenched to zero temperature.  A new feature
of the Potts model with $q\geq 3$ is that domain boundaries no longer have to
be closed loops.  Instead, boundaries can have endpoints, so that three or
more boundaries can meet at a single junction.  These junctions act as
pinning centers that promote freezing of domains.  This effect is readily
apparent in the 3-state Potts model and becomes more prominent as the number
of states $q$ increases (Fig.~\ref{fig:clusters}).  Thus one might expect
that increasing the number of Potts states $q$ would make it more difficult
for a system to reach the ground state after a quench to zero temperature.
Paradoxically, the opposite appears to be the case.

Thus one of our main results is that the Potts model reaches one of its $q$
ground states with a nonzero probability in the thermodynamic limit.  The
mechanism that causes this increased likelihood for reaching the ground state
is the repeated processes of: (i) clusters growing to fill their convex
envelope and (ii) cluster coalescences that lead to sudden increases in the
convex envelopes of the merged cluster.  The combined effect of these
processes can drive an avalanche in which a single expanding cluster engulfs
the entire system so that one of the $q$ ground states is reached.  When the
number of Potts states is large, cluster growth is enhanced by the existence
of spins on the cluster boundary that can flip freely to any spin state and
thereby allow a cluster to grow beyond its convex envelope.

There are additional subtleties in Potts model coarsening.  Perhaps most
prominent is the existence of non-static blinker interfaces that comprise
diagonal (non-vertical or non-horizontal) domain boundaries.  These blinker
interfaces can evolve forever without energy cost.  The slow relaxation
observed in the $q$-state Potts model arises because of the merging of two of
these wandering diagonal interfaces.  The time scale for these
energy-lowering merging events grows exponentially with the typical domain
size.  

To test the universality of our results, we additionally studied a 
continuum system, viz. the time-dependent Ginzburg-Landau (TDGL) equation
with a symmetric 3-well potential to mimic the 3-state Potts model.  The
resulting equations for the evolution of the order parameter are simpler and
more fundamental than the myriad of microscopic models for describing
coarsening of a multi-state discrete spin model.  While many of our results
from integration of the TDGL equations mirror those found for the discrete
Potts models, the former also give rise to unexpected cluster patterns at
long times that, with the benefit of hindsight, resemble the global structure
of domain mosaics in the Potts model at long times.  For the TDGL system with
three degenerate ground states, the ground state is reached in a somewhat
larger fraction of realizations than for the 3-state kinetic Potts model, but
we also found more exotic static states, such as three hexagonal domains, two
squares and two octagonal domains, and even a state that consists of six
hexagonal domains.

The long-time coarsening of the kinetic Potts model and the multi-state TDGL
equations poses many open questions that are ripe for future exploration.
For instance, on the triangular lattice, the domain structure that emerges in
the coarsening regime of the $q$-state kinetic Potts model is quite different
from the corresponding domain structure on the square lattice~\cite{SSGAS83},
namely the lattice effects are strongly suppressed on the triangular
lattice. Thus the kinetic Potts model on the triangular lattice is closer to
the TDGL equation, which is isotropic.  Therefore even universality with
respect to lattice structure is questionable for the kinetic Potts model.  In
particular, it will be interesting to understand if the probability to reach
blinker states remains positive (in the thermodynamic limit).

Our results for TDGL coarsening are much less extensive than for the Potts
model but also suggest intriguing features, such as the existence of stable
multi-cluster states, in addition to the ground state and stripe states.  It
would be also worthwhile to investigate the TDGL equations with $q$
degenerate ground states.  This would involve considering an order parameter
in $q-1$ dimensions with a potential that has minima at the $q$ vertices of a
simplex.  While the correspondence between the discrete spin system and the
TDGL equation results is quite close for the case of two degenerate states
(the Ising model), the precise nature of the correspondence when the number
of degenerate states is three or larger is still to be determined.

\ack{We thank Gary Grest and Alberto Petri for useful comments and
  suggestions.  JO and SR gratefully acknowledge NSF Grant No. DMR-1205797
  for partial financial support of this work.}  \newpage

\end{document}